\documentclass[12pt]{article}

\usepackage{epsf}
\usepackage{latexsym,amssymb,euscript}
\usepackage{graphicx}
\usepackage{amsmath}
\usepackage{subcaption}
\usepackage[]{units}
\usepackage{mwe}
\usepackage{multirow}
\usepackage{verbatim}
\usepackage{geometry}
\usepackage{cite}

\newcommand{\sigmaqq}{\sigma_{q\overline{q}}}
\newcommand{\sigmaqqeff}{\sigma_{q\overline{q}}^{\mathrm{eff}}}
\newcommand{\sigmaww}{\sigma_{\rm ww}}
\newcommand{\sigmaB}{\sigma_{qqq}}
\newcommand{\chiR}{\chi^2_r}

\DeclareMathOperator{\Tr}{Tr}

\begin{document}

\begin{center}
  {\Large \bf Three-quark potentials in an $SU(3)$ effective Polyakov loop
    model}
\end{center}

\vskip 0.3cm
\centerline{O.~Borisenko$^{1\dagger}$, V.~Chelnokov$^{2*}$,
E.~Mendicelli$^{3\ddagger}$, A.~Papa$^{2,3\P}$}

\vskip 0.6cm

\centerline{${}^1$ \sl Bogolyubov Institute for Theoretical Physics,}
\centerline{\sl National Academy of Sciences of Ukraine,}
\centerline{\sl 03143 Kiev, Ukraine}

\vskip 0.2cm

\centerline{${}^2$ \sl Istituto Nazionale di Fisica Nucleare,
Gruppo collegato di Cosenza,}
\centerline{\sl I-87036 Arcavacata di Rende, Cosenza, Italy}

\vskip 0.2cm

\centerline{${}^3$ \sl Dipartimento di Fisica, Universit\`a della
Calabria,}
\centerline{\sl I-87036 Arcavacata di Rende, Cosenza, Italy}

\vskip 0.6cm

\begin{abstract}
  Three-quark potentials are studied in great details in the three-di\-men\-sional
  $SU(3)$ pure gauge theory at finite temperature, for the cases of static
  sources in the fundamental and adjoint representations.
  For this purpose, the corresponding Polyakov loop model in its simplest
  version is adopted. The potentials in question, as well as the conventional
  quark--anti-quark potentials, are calculated numerically both in the
  confinement and deconfinement phases. Results are compared to available
  analytical predictions at strong coupling and in the limit of large
  number of colors $N$. The three-quark potential is tested against the
  expected $\Delta$ and $Y$ laws and the $3q$ string tension entering these
  laws is compared to the conventional $q\bar{q}$ string tension. As a
  byproduct of this investigation, essential features of the critical behaviour
  across the deconfinement transition are elucidated.
\end{abstract}

\vfill
\hrule
\vspace{0.3cm}
$^*$ On leave from Bogolyubov Institute for Theoretical Physics, Kiev

{\it e-mail addresses}:
$^\dagger$oleg@bitp.kiev.ua, \  $^*$Volodymyr.Chelnokov@lnf.infn.it, \\
$^{\ddagger}$emanuelemendicelli@hotmail.it, \ \ $^{\P}$papa@fis.unical.it

\newpage

\section{Introduction}

The interest in studying the interquark potential for a three-quark system
is not a recent issue at all.
It has instead a long history due to its importance in the spectroscopy of
baryons.
The first studies date back to the mid `80s~\cite{sommer_84,sommer_86} and
after more than a decade a new turn of research has
started around the year 2000 which continues till now~\cite{bali,takahashi,forcrand3,forcrand1,forcrand2,polikarpov1,polikarpov2,caselle,sato,Brambilla:2009cd,bakry,Andreev:2015iaa,Andreev:2015riv,koma,Leech:2018lqu}.
New results are somewhat contradictory, which could be reasonably
explained by the difficulty of accurate measurements of the three-quark
potential. But from these discussions spanning many years, two main
\"Ansatze emerged to describe the three-quark potential: the $\Delta$ law and
the $Y$ law. Denoting by $r_1$, $r_2$, $r_3$ the sides of the triangle having
the quarks at its vertices, the $\Delta$ law is defined by
\begin{equation}
\label{delta_law}
V_3=\dfrac{1}{2}\sigma_{q\overline{q}}(r_{1}+r_{2}+r_{3})=\sigma_{q\overline{q}}\;
\Delta \; ,
\end{equation}
which describes a potential linearly rising with half the perimeter of the
triangle.
The $Y$ law describes the three-quark potential as linearly rising with
the minimal total length of the flux lines connecting the three quarks,
\begin{equation}
\label{Y_law}
V_3=\sigma_{q\overline{q}}\; Y \ , 
\end{equation}
where $Y$ is the sum of the distances of the three quarks from the
Fermat-Torricelli point F, which is the point such that this sum is the
least possible. 
When all inner angles of the triangle are less than $2\pi/3$, one has
\begin{equation}
\label{fermat-torr}
Y=\sqrt{\dfrac{r_1^2+r_2^2+r_3^2 + 4\sqrt{3}A}{2}} \ ,
\end{equation}
where $A$ is the area of the triangle; if one of the angles is larger than
$2\pi/3$, we have instead
\begin{equation}
\label{fermat-torr-lambda}
Y={\mathrm{min}} \ (r_1+r_2, r_1+r_3, r_2+r_3 ) \equiv \Lambda \ ,
\end{equation}
which gives rise to the $\Lambda$ law,
\begin{equation}
\label{Lambda_law}
V_3=\sigma_{q\overline{q}}\; \Lambda \ . 
\end{equation}
Some earlier~\cite{takahashi, polikarpov1,polikarpov2} and the most
recent studies~\cite{bakry,koma} in the $SU(3)$ pure gauge theory seem to
support the $Y$-ansatz, while other simulations~\cite{bali,forcrand3,forcrand1,forcrand2} prefer the $\Delta$-ansatz, at least for not too large triangles. 
An even more complicated picture emerged after simulations of the simpler,
$Z(3)$ Potts model in two-dimensions, which is believed to capture the most
essential features of the gauge model~\cite{forcrand2,caselle}. 
Namely, it was conjectured that there might be a smooth crossover between the
$\Delta$ law and the $Y$ law when the size of triangles grows 
(see, however,~\cite{sato} where this scenario has been criticized).  
Also, the paper~\cite{caselle} proposes a new ansatz in which both the $Y$ law
and the $\Lambda$ law are present. 

In this paper we are going to study an $SU(3)$ spin model which is an
effective model for Polyakov loops and can be derived from the original
gauge theory in the strong coupling region. For simplicity, we consider,
following~\cite{caselle}, only its two-dimensional version. Our
primary goal is to get some analytical predictions for the three-point
correlation function of the Polyakov loops and compare them with numerical
simulations. For that we use the $SU(3)$ spins both 
in the fundamental and adjoint representations. The main tool of our analytical
investigation is the large-$N$ expansion. 
Within this expansion we demonstrate that the fundamental three-point
correlator is described by a sum of the $Y$ and $\Lambda$ laws. 
The $\Delta$ contribution is not present. In turn, the connected part of the
adjoint three-point correlation follows the $\Delta$ law in the confinement
phase. In addition, we study the critical region of the model and confirm that
it belongs to the universality class of the two-dimensional $Z(3)$ spin model. 

This paper is organized as follows. In the next section we introduce our
notations, define $SU(N)$ Polyakov loop model and its dual. Certain analytical
predictions for two- and three-point correlation functions are obtained in
the strong coupling expansion and in the large-$N$ limit. Moreover, we
check the restoration of the rotational symmetry for the 3-quark system.
Section~3 outlines some details of our numerical simulations. Here we compare 
numerical data with the strong coupling expansion and study the critical
behaviour of the model using the finite-size scaling analysis. 
Section~4 presents the results of Monte Carlo simulations for the fundamental
and adjoint two- and three-point correlations in the confinement region.
Results for the same quantities above critical temperature are described
in Section~5. In Section~6 we summarize our results.

\section{The model and theoretical expectations}

\subsection{Partition and correlation functions}

We work on a $2d$ Euclidean lattice $\Lambda=L^2$, with sites $x=(x_1,x_2)$,
$x_n\in[0,L-1]$, and denote by $e_n$ the unit vector in the $n$-th direction.
Periodic boundary conditions (BC) are imposed in all directions. Let
$W(x)\in SU(N)$, and ${\rm Tr}W$ be the character of $SU(N)$
in the fundamental representation. Consider the following partition function on
$\Lambda$, which describes the interaction of non-Abelian spins:
\begin{eqnarray}
Z_{\Lambda}(\beta , N)  \ = \ \int \prod_x dW(x)
\exp  \biggl[ \beta \sum_{x,n}  \ \mbox{Re} \ {\rm Tr}W(x) {\rm Tr}W^*(x+e_n)
  \ \biggr] \ .
\label{sunpf}
\end{eqnarray}
The trace of an $SU(N)$ matrix can be parameterized with the help of $N$
angles, {\it e.g.} by taking $W={\rm diag} (e^{i\omega_1}, \cdots, e^{i\omega_N})$,
subject to the constraint
$\prod_k e^{i \omega_k}=1$. In this parameterization the action has the form
\begin{equation}
\mbox{Re} \ {\rm Tr}W(x) {\rm Tr}W^*(x+e_n)   \ = \
\sum_{i,j=1}^N \ \cos \left  [\omega_i(x)-\omega_j(x+e_n) \right ] \ .
\label{sunaction}
\end{equation}
The invariant measure for $SU(N)$ is given by
\begin{equation}
\int dW  \ = \ \int_0^{2\pi} D(\omega)D^*(\omega) \ \delta \left ( \sum_k\omega_k  \right ) \ \prod_{k=1}^N
\frac{d\omega_k}{2 \pi}  \ ,
\label{sunmeasure}
\end{equation}
where
\begin{equation}
D(\omega) \ = \ \prod_{k<l} \left ( e^{i\omega_k} -  e^{i\omega_l}    \right  )
\label{Doperator}
\end{equation}
and $\delta(x)$ is the periodic delta-function.
Due to this constraint, the $SU(N)$ model is invariant only under the global
discrete shift $\omega_k(x)\to\omega_k(x)+\frac{2\pi n}{N}$ for all $k$ and $x$.
This is just the global $Z(N)$ symmetry.

The main subjects of this work are the two- and three-point correlation
functions for the $SU(3)$ model. In the fundamental representation these
correlations are given by
\begin{eqnarray}
\label{corrfunc2_fund_def}
\Gamma_2^{{\rm f}}(\beta , R) \ = \ \left \langle \  {\rm Tr}W(0) {\rm Tr}W^*(R) \  \right \rangle \ ,  \\
\Gamma_3^{{\rm f}}(\beta , \{ x_i \}) \ = \ \left \langle \ {\rm Tr}W(x_1) {\rm Tr}W(x_2) {\rm Tr}W(x_3) \ \right \rangle \ ,
\label{corrfunc3_fund_def}
\end{eqnarray}
while in the adjoint representation the correlations are written as 
\begin{eqnarray}
\label{corrfunc2_adj_def}
\Gamma_2^{{\rm ad}}(\beta , R) \ = \ \left \langle \ \chi_{\rm ad}(W(0)) \chi_{\rm ad}(W(R))  \ \right \rangle \ ,  \\
\Gamma_3^{{\rm ad}}(\beta , \{ x_i \} ) \ = \
\left \langle \ \chi_{\rm ad}(W(x_1)) \chi_{\rm ad}(W(x_2)) \chi_{\rm ad}(W(x_3)) \ \right \rangle \ ,
\label{corrfunc3_adj_def}
\end{eqnarray}
where we use the relation $\chi_{\rm ad}(W)=\rm TrW \rm TrW^* - 1$. 

The partition function~(\ref{sunpf}) can be regarded as the simplest effective
model for the Polyakov loops which can be derived in the strong coupling region
of $3d$ lattice gauge theory (LGT) at finite temperature (see,
{\it e.g.},~\cite{sun_effaction} and references therein).
Namely, the integration over the spatial gauge links on
the anisotropic $(d+1)$-dimensional lattice with two couplings $\beta_s$ and
$\beta_t\equiv\beta$ in the limit $\beta_s =0$ and for $\beta$ sufficiently
small leads to the $d$-dimensional spin model~(\ref{sunpf}). It describes the
deconfinement phase transition of the pure gauge theory, which is of second
order for $SU(3)$ if $d=2$. It is widely assumed that the phase transition is
in the universality class of the two-dimensional $Z(3)$ (Potts) model. The
inverse correlation length (mass gap) is the string tension of the gauge theory.
The correlation length diverges when approaching the critical point with the
critical index $\nu=5/6$. Another important critical index $\eta$, which is a
characteristic of the massless phase, equals $4/15$ exactly at the critical
point. Thus, on the basis of the universality arguments~\cite{svetitsky} we
expect the same values for these indices also in the effective $SU(3)$ Polyakov
loop model. More on the critical behaviour of three-dimensional $SU(N)$ 
LGTs can be found in Refs.\cite{teper_08, 2dun_bkt}.

The model~(\ref{sunpf}) cannot be solved exactly at any finite $N$ and $D>1$.
Therefore, to get some analytical predictions for the behaviour of the
three-point correlation functions we consider the large-$N$ limit of the model.
This limit can in turn be solved exactly by using the dual representation
which we are going to describe shortly. 

\subsection{Dual representation} 

In some applications the dual formulation of the Polyakov loop
model~(\ref{sunpf}) can be useful.
Such formulation for the $SU(3)$ model has been derived in~\cite{spin_flux1}.
Here we use the dual representation obtained by some of us
in~\cite{dual_lgt, dual_sunspin}. This form of dual theory is valid for all
$N$ and can be used both for numerical simulations and for the study of the
large-$N$ limit of the theory. For the $2d$ theory the partition
function~(\ref{sunpf}) on the dual lattice takes the form
\begin{eqnarray}
Z_{\Lambda}(\beta, N) &=& \sum_{\{ r(x) \} =-\infty}^{\infty} \ \sum_{\{ q(l) \} =0}^{\infty} \
\sum_{\{ k(l) \} =-\infty}^{\infty} \ \prod_p  \ Q_N(s(p),\bar{s}(p)) \nonumber   \\
&\times&\prod_l \left [
\frac{\left ( \frac{\beta}{2} \right )^{| r(x) - r(x+e_n) + k(l) N | + 2q(l)}}{(q(l)+| r(x) - r(x+e_n) + k(l) N |)!q(l)!} \right ] \ ,
\label{pf_dual2d}
\end{eqnarray}
where $Q_N(s,\bar{s})$ results from the invariant integration over the $SU(N)$
measure,
\begin{equation}
  Q_N(s,\bar{s}) \ = \  \sum_{\lambda \vdash {\rm min}(s,\bar{s})} \ d(\lambda)
  \ d(\lambda + |k|^N) \ .
\label{QSUN}
\end{equation}
Here $d(\lambda)$ is the dimension of the irreducible representation $\lambda$ 
of the permutation group $S_s$, $s - \bar{s}=kN$ and
\begin{eqnarray}
  s(p) = \frac{1}{2} \  k(p)N + \sum_{l\in p} \ \left ( q(l) \ + \ \frac{1}{2}
  \ |  r(l) |  \right ) \ , 
\label{dual_sp2d}
\end{eqnarray}
\begin{eqnarray}
k(p) \ = \ k(l_1) + k(l_2) - k(l_3) - k(l_4) \ , \ l_i\in p \ , \nonumber \\ 
r(l) \ = \ r(x) - r(x+e_n) +k(l) N \ . 
\label{kp_def}
\end{eqnarray}
$\lambda$ is enumerated by the partition
$\lambda=(\lambda_1,\lambda_2,\cdots,\lambda_{l(\lambda)})$
of $s$, {\it i.e.} $\sum_{i=1}^{l(\lambda)} \lambda_i = s$,
where $l(\lambda)$ is the length of the partition and
$\lambda_1\geq\lambda_2\cdots\lambda_{l(\lambda)}> 0$.
The sum in~(\ref{QSUN}) is taken over all $\lambda$'s such that
$l(\lambda)\leq N$ and the convention
$\lambda+q^N \equiv (\lambda_1+q,\cdots,\lambda_N+q)$ has been adopted.
For the exact expressions of the different correlation functions we refer the
reader to the paper~\cite{dual_sunspin}.

\subsection{Large-$N$ solution} 

Using the dual representation~(\ref{pf_dual2d}), one can construct an exact
solution of the model in the large-$N$ limit~\cite{dual_largeN_exact} and even
estimate the first non-trivial corrections specific for the $SU(N)$ group. As
an example, we give here the expression for the most general correlation
function and for the partition function in the confinement region in the
presence of sources
\begin{equation}
\langle \ \prod_x \left ( \mbox{Tr} W(x)  \right )^{\eta(x)} 
\left ( \mbox{Tr} W^*(x)  \right )^{\bar{\eta}(x)} \ \rangle \ = \ \frac{Z(\eta,\bar{\eta})}{Z(0,0)} \ ,  
\label{gencorr_largeN}
\end{equation}
\begin{eqnarray}
Z(\eta,\bar{\eta}) \ = \ \int_{-\infty}^{\infty} \ \prod_x d\alpha(x)d\sigma(x) \ 
(\alpha(x)+i\sigma(x))^{\eta(x)} \ (\alpha(x)-i\sigma(x))^{\bar{\eta}(x)} \nonumber  \\ 
e^{-\sum_{x,x^{\prime}} G_{x,x^{\prime}} (\alpha(x)\alpha(x^{\prime}) + \sigma(x)\sigma(x^{\prime}))} \ 
\prod_x \left ( 1 + \frac{2}{N!} \mbox{Re} (\alpha(x)-i\sigma(x))^N  \right )
\;.
\label{PF_largeN}
\end{eqnarray}
The Gaussian part describes the solution in the large-$N$ limit, while the
product over $x$ in the second line presents the first correction due to
$SU(N)$. $G_{x,x^{\prime}}$ is the massive two-dimensional Green function 
for the scalar field. 

This solution, together with a similar one in the deconfinement phase, enables
one to calculate both fundamental and adjoint two- and three-point correlations
in that limit. 
Different results are obtained in the small and large $\beta$ regions separated
by the deconfinement phase transition. If we take $N=3$, then for the
confinement phase we get
\begin{eqnarray}
\label{corrfunc2_fund_largeN_small_beta}
\Gamma_2^{{\rm f}}(\beta , R) \ \sim \ G(\beta, R) \ ,
\end{eqnarray}
\begin{eqnarray}
\Gamma_3^{{\rm f}}(\beta , \{ x_i \} ) \ \sim \   \sum_y \prod_{i=1}^3 \ G(\beta, |x_i-y|) \ ,
\label{corrfunc3_fund_largeN_small_beta}
\end{eqnarray}
\begin{eqnarray}
\label{corrfunc2_adj_largeN_small_beta}
\Gamma_2^{{\rm ad}}(\beta , R) \ \sim \ G(\beta, R)^2 + M_{\rm ad}(\beta)^2 \ ,
\end{eqnarray}
\begin{eqnarray}
\Gamma_3^{{\rm ad}}(\beta , \{ x_i \} ) \ &\sim& \  \prod_{i=1}^3 G(\beta, |x_i - x_{i+1}|)
   \nonumber \\
   &&  {} + M_{\rm ad}(\beta) \sum_{i=1}^3 G(\beta, |x_i - x_{i+1}|)^2 + M_{\rm ad}(\beta)^3
\label{corrfunc3_adj_largeN_small_beta}
\end{eqnarray}
and in the deconfinement phase we get
\begin{eqnarray}
\label{corrfunc2_fund_largeN_large_beta}
\Gamma_2^{{\rm f}}(\beta , R) \ \sim \ M_{\rm f}(\beta)^2 \exp \left[ \alpha(\beta) G(\beta, R) \right] \ ,
\end{eqnarray}
\begin{eqnarray}
\Gamma_3^{{\rm f}}(\beta , \{ x_i \} ) \ \sim \ M_{\rm f}(\beta)^3 \exp \left[ \alpha(\beta) 
\sum_{i=1}^3 G(\beta, |x_i - x_{i+1}|) \right]  \ ,
\label{corrfunc3_fund_largeN_large_beta}
\end{eqnarray}
\begin{eqnarray}
\label{corrfunc2_adj_largeN_large_beta}
\Gamma_2^{{\rm ad}}(\beta , R) \ \sim \ (M_{\rm ad}(\beta) + 1)^2 
\exp \left[4 \alpha(\beta) G(\beta, R) \right] - 2 M_{\rm ad}(\beta) - 1 \ ,
\end{eqnarray}
\begin{eqnarray}
\Gamma_3^{{\rm ad}}(\beta , \{ x_i \} ) \ & \sim & \  (M_{\rm ad}(\beta) + 1)^3 
\exp \left[ 4 \alpha(\beta) \sum_{i=1}^3 G(\beta, |x_i - x_{i+1}|) \right]
 \nonumber \\
 && - \sum_{i=1}^3 (M_{\rm ad}(\beta) + 1)^2 \exp \left[4 \alpha(\beta) G(\beta, |x_i - x_{i + 1}|) \right]
 \nonumber \\
 && + 3 M_{\rm ad}(\beta) + 2 \ . 
\label{corrfunc3_adj_largeN_large_beta}
\end{eqnarray}
In the equations above the Green function in the thermodynamical limit is given
by
\begin{equation}
G(\beta, x) \ = \ \int_0^{2\pi} \ \frac{d\omega_1 d\omega_2}{(2\pi)^2} \
\frac{e^{i\sum_n\omega_n x_n}}{m(\beta)+2- \sum_{n=1}^2\cos\omega_n} \ \sim \ 
\frac{e^{-m(\beta)R}}{\sqrt{R}} \ , 
\label{Greenfunc_largeN}
\end{equation}
where $R^2=x_1^2+x_2^2$ and the functional dependence of the mass $m$ on
$\beta$ is different in the confined and deconfined phases. 
In the confinement phase the mass $m(\beta)$ coincides with the $q\bar{q}$
string tension, while in the deconfinement phase this quantity has the meaning
of screening mass.  
$M_{\rm f}(\beta)$ and $M_{\rm ad}(\beta)$ define the fundamental and adjoint
magnetizations at a given $\beta$, correspondingly. They also depend on the
considered phase. For example, $M_{\rm f}(\beta)=0$ in the confined phase.  
$\alpha(\beta)$ is another $R$-independent quantity which appears due to
Gaussian integration around the large-$N$ solution. 
All four quantities - $m(\beta)$, $\alpha(\beta)$, $M_{\rm f}(\beta)$,
$M_{\rm ad}(\beta)$ - are known exactly in the large-$N$ expansion. 

\subsection{$3q$ potential} 

In what follows our strategy relies on the assumption that the large-$N$
formulae~(\ref{corrfunc2_fund_largeN_small_beta})-(\ref{corrfunc3_adj_largeN_large_beta}) 
remain qualitatively valid (up to one correction explained below) at
finite $N$, in particular for $N=3$. We expect that the most essential difference 
between the large-$N$ limit and the $N=3$ case exhibits itself in the vicinity of the critical point. 
Indeed, both our solution~\cite{dual_largeN_exact} and the mean-field solution of 
Ref.~\cite{largeN_mean-field} reveal the existence of a third order phase transition 
at large $N$. Meanwhile, as described above, the $SU(3)$ Polyakov loop model belongs 
to the universality class of the two-dimensional $Z(3)$ model. It means, in particular that 
the critical behaviour of two- and three-point correlation functions is described by a different 
set of the critical indices $\nu$ and $\eta$. 
Therefore, we shall use (\ref{corrfunc2_fund_largeN_small_beta})-(\ref{corrfunc3_adj_largeN_large_beta})
as fitting functions, where the quantities $m(\beta)$,
$\alpha(\beta)$, $M_{\rm f}(\beta)$, $M_{\rm ad}(\beta)$ are unknown parameters to
be found from fits of numerical data. In most cases, we use the asymptotic
expansion for the Green function $G(\beta, x)$ given on the right-hand side of
Eq.~(\ref{Greenfunc_largeN}). As we explained above, only the critical indices 
appearing in these quantities can vary with $N$. 
Also, we introduce here another quantity, namely the index $\eta$, in order to
describe the power dependence of the correlation function, $R^{-\eta}$, on the
distance. This could again be important in the vicinity of the critical point. 
In general, this introduces a correction to the potential of the form 
\begin{equation}
V_{\mathrm{Coulomb}} \ = \ \eta \ln D \ , \ \ \ \ \ D=R, Y ,\Delta , \Lambda \;,
\label{coulombpart}
\end{equation}
and is interpreted as the Coulomb part of the full potential in the
two-dimensional theory.

Since the asymptotic behaviour of $G(\beta, x)$ is known, it follows that we
actually know the large-distance behaviour of all two- and three-point
functions listed above, but $\Gamma_3^{{\rm f}}(\beta , \{ x_i \} )$. 
The behaviour of the latter can be analyzed by the saddle-point method when
at least one side of the triangle is large enough. We find two types of the
behaviour: 
\begin{enumerate}
\item 
  All inner angles of the triangle are less than $2\pi/3$. The three-point
  fundamental correlation function is given by the sum of two terms
  corresponding to $Y$ and $\Lambda$ laws 
\begin{equation}
\Gamma_3^{{\rm f}}(\beta , \{ x_i \} ) \ \approx \ A \ \frac{e^{-\sigma_{qqq} Y}}{\sqrt{Y}} + 
B \ \frac{e^{-\sigma_{qqq} \Lambda}}{\sqrt{\Lambda}} \ ,   
\label{YandLambda_LargeN}
\end{equation}
where $A,B$ are constants and $\sigma_{qqq}=\sigma_{q\bar{q}}$. This behaviour
resembles the behaviour of the three-point correlation function in the $Z(3)$
spin model~\cite{caselle}. 
\item 
  One of the angles is larger than $2\pi/3$. In this case the asymptotics is
  described by the above formula with $A=0$. Thus, only the $\Lambda$ law is
  present. This again agrees with the $Z(3)$ spin model. 
\end{enumerate}
Let us also emphasize that we could not find the $\Delta$ law contribution in
our large-$N$ approach. 
Nevertheless, we attempt to fit numerical data both to $Y$ and $\Delta$ laws in
the following.  
Finally, let us stress that the connected part of the three-point adjoint
correlation follows the $\Delta$ law in the confinement phase, as is seen
from Eq.~(\ref{corrfunc3_adj_largeN_small_beta}).

\subsection{Strong coupling expansion} 

When $\beta$ is sufficiently small, one can use the conventional strong coupling
expansion to demonstrate the exponential decay of the fundamental two- and
three-point correlation functions.
Instead, adjoint correlations stay constant over large distance.
To check our codes we have calculated the leading orders of the strong-coupling
expansion for the two-point correlator at distance $R=1$ and for the
three-point correlator in the isosceles-triangle geometry $T$ with base
$b=2$ and height $h=1$. The results read 
\begin{equation}
\label{Strong_coupling_Gamma2_f}
\Gamma_2^{{\rm f}}(\beta , 1) =\
\frac{1}{2} \beta +
\dfrac{1}{8}\beta^2+
\dfrac{9}{8}\beta^3 +
\dfrac{385}{384}\beta^4 + {\cal O}(\beta^5) \ ,
\end{equation}
\begin{equation}
\label{Strong_coupling_Gamma3_f}
\Gamma_3^{{\rm f}}(\beta , T) =\
\dfrac{1}{8}\beta^3 +
\dfrac{1}{2}\beta^4 +
\dfrac{145}{128}\beta^5 +
\dfrac{29}{8}\beta^6 + {\cal O}(\beta^7) \ ,
\end{equation}
\begin{equation}
\label{Strong_coupling_Gamma2_ad}
\Gamma_2^{{\rm ad}}(\beta , 1) =\ 
\frac{1}{4} \beta^2 +
\frac{1}{6} \beta^3 +
\frac{17}{8} \beta^4 + {\cal O}(\beta^5) \ ,
\end{equation}
\begin{equation}
\label{Strong_coupling_Gamma3_ad}
\Gamma_3^{{\rm ad}}(\beta , T) = \
\frac{27}{16} \beta^6 +
\frac{487}{192} \beta^7 + {\cal O}(\beta^8) \ .
\end{equation}
For arbitrary isosceles triangle $T$ with base $b$ and height $h$ one obtains 
\begin{equation}
\Gamma_3^{{\rm f}}(\beta , T) \sim \ \beta^{h+b} \equiv \beta^{Y_l} \ .
\end{equation}
On a cubic lattice $Y_l=h+b$ is the minimal sum of the lattice distances from
the triangle vertices to an arbitrary lattice point. Then, according to
Eq.~(\ref{Y_law}),
\begin{equation}
\label{Ylat_law}
V_3 \ =  \ \sigma_{q\overline{q}}\; Y_l \ , \ \ \ \ \ \sigma_{q\overline{q}} \
\approx \ \ln \beta 
\end{equation}
in the strong coupling region on the finite lattice. Thus, strictly speaking the
strong coupling expansion predicts not an exact $Y$ law, as it is often stated
in the literature, but rather a $Y_l$ law. 
In general, $Y_l>Y$ and we expect that the rotational symmetry will be restored
quickly with  $\beta$ and the triangle sides increasing. This should result in
the restoration of the genuine $Y$ law. 
\begin{figure}[tb]
\centering
\includegraphics[width=0.5\textwidth]{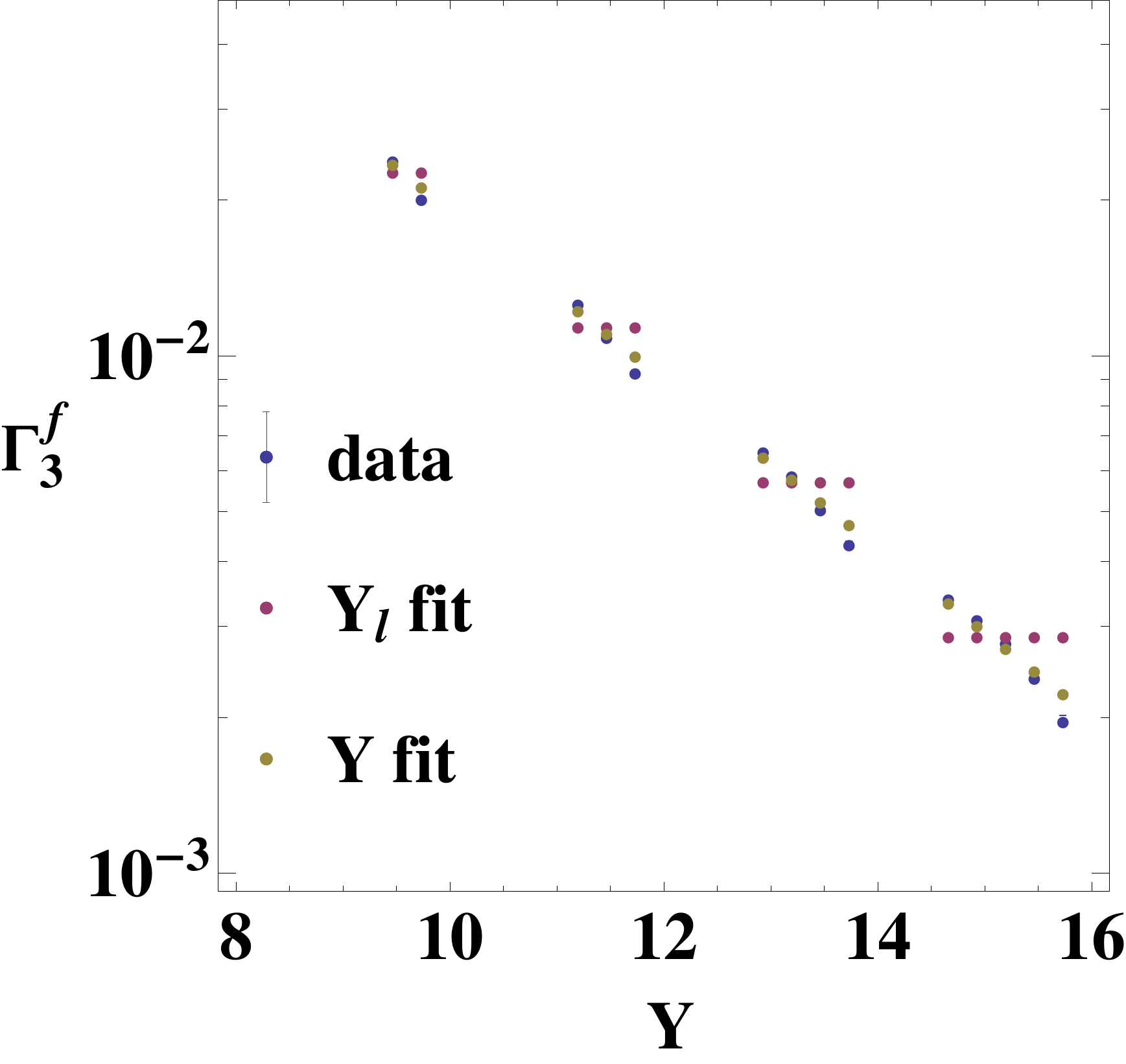}
\caption{Comparison of the three point correlation function $\Gamma_3^f$ at
  $\beta = 0.41$ with the fit using $Y_l$ and $Y$ laws.}
\label{fig:compareYL}
\end{figure} 
To demonstrate that such a restoration really takes place, we have studied 
the three-point correlation function for triangles with $2 \leq b \leq
10$ and $6 \leq h \leq 14$  at $\beta = 0.41$. 
The fact that the rotational symmetry is already restored at this value of
$\beta$ is shown on Fig.~\ref{fig:compareYL}, where we compare numerical data
with the fits of the form $e^{-\sigma_{qqq}D}/D^{\eta}$ for $D=Y_l$ and $D=Y$. 
Clearly, $D=Y$ describes data better than $D=Y_l$.

\section{Details of numerical simulations}

To calculate the correlation functions from numerical simulations we used two
different approaches.
The first is the simulation of the model in terms of the eigenvalues
$\omega_i(x)$ of the $SU(3)$ spins, described in more detail in~\cite{2dun_bkt}.
In this approach (denoted as {\em standard} in the following), an updating
sweep consisted in the combination of a local Metropolis update of each lattice
site, followed by two updates by the Wolff algorithm, consisting in $Z(3)$
reflections of the clusters.
An alternative approach is the simulation of the dual model~(\ref{pf_dual2d}),
using the heatbath update for the link variables $k$, $q$ and the dual site
variables $r$. In this case, we can measure only observables invariant under
the global $Z(3)$ symmetry. 

In both approaches we measured two- and three-point correlation functions
in the fundamental and adjoint representations, taking for the two point
correlation function pairs of points separated by $R$ in one of the two
lattice directions, with $R=2,4,\ldots,L/2$.
For the three-point correlation functions two geometries were studied:
isosceles triangles with base $b$ and height $h$, and right-angled triangles
with the catheti (of lengths $a_1$ and $a_2$) along the two lattice directions.
In both cases, $b$ and $h$, and $a_1$ and $a_2$, took independent values in
the set $\{2, 4, \ldots, L/2\}$.

In addition to the two- and three-point correlations~(\ref{corrfunc2_fund_def})-(\ref{corrfunc3_adj_def}), the magnetizations and their susceptibilities were
measured:
\begin{eqnarray}
\label{magnetization_def}
M_{\mathrm{f}} & = & \left\langle \chi_{\mathrm{f}}(W_x) \right\rangle = \left\langle \Tr W_x \right\rangle \ , \\
\label{magnetization_ad_def}
M_{\mathrm{ad}} & = & \left\langle \chi_{\mathrm{ad}}(W_x) \right\rangle = \left\langle \Tr W_x \Tr W_x^* - 1 \right\rangle \ , \\
\label{susceptibility_def}
\chi_L^{(M_\mathrm{f})} & = &
L^2 \left( \left\langle \left(\chi_{\mathrm{f}}(W_x)\right)^2 \right\rangle -
\left\langle \chi_{\mathrm{f}}(W_x) \right\rangle^2 \right) \ ,\\
\chi_L^{(M_\mathrm{ad})} & = &
L^2 \left( \left\langle \left(\chi_{\mathrm{ad}}(W_x)\right)^2 \right\rangle -
\left\langle \chi_{\mathrm{ad}}(W_x) \right\rangle^2 \right) \ .
\label{susceptibility_ad_def}
\end{eqnarray}
The $x$-dependent values are averaged over all sites of the lattice.

For each simulation we performed $10^4$ thermalization updates, and then
made measurements every ten whole lattice updates (sweeps), collecting
a statistics of $10^5$ -- $10^6$. To estimate statistical errors a
jackknife analysis was performed at different blocking over bins
with size varying from 500 to 10000.

A comparison of the two simulation methods showed that the dual code
performs better at small values of $\beta$, while giving much larger
fluctuations than the standard one when $\beta$ is close to its critical value.
What is more important -- at larger values of $\beta$ for the fundamental
correlation function the fluctuations rapidly increase with the distance
between the points.
Due to this, most of the results presented here have been obtained in the
standard approach, and the dual code was used only for cross-check purposes. 

\subsection{Comparison with strong coupling}

\begin{figure}[tb]
\centering
\includegraphics[width=0.485\textwidth]{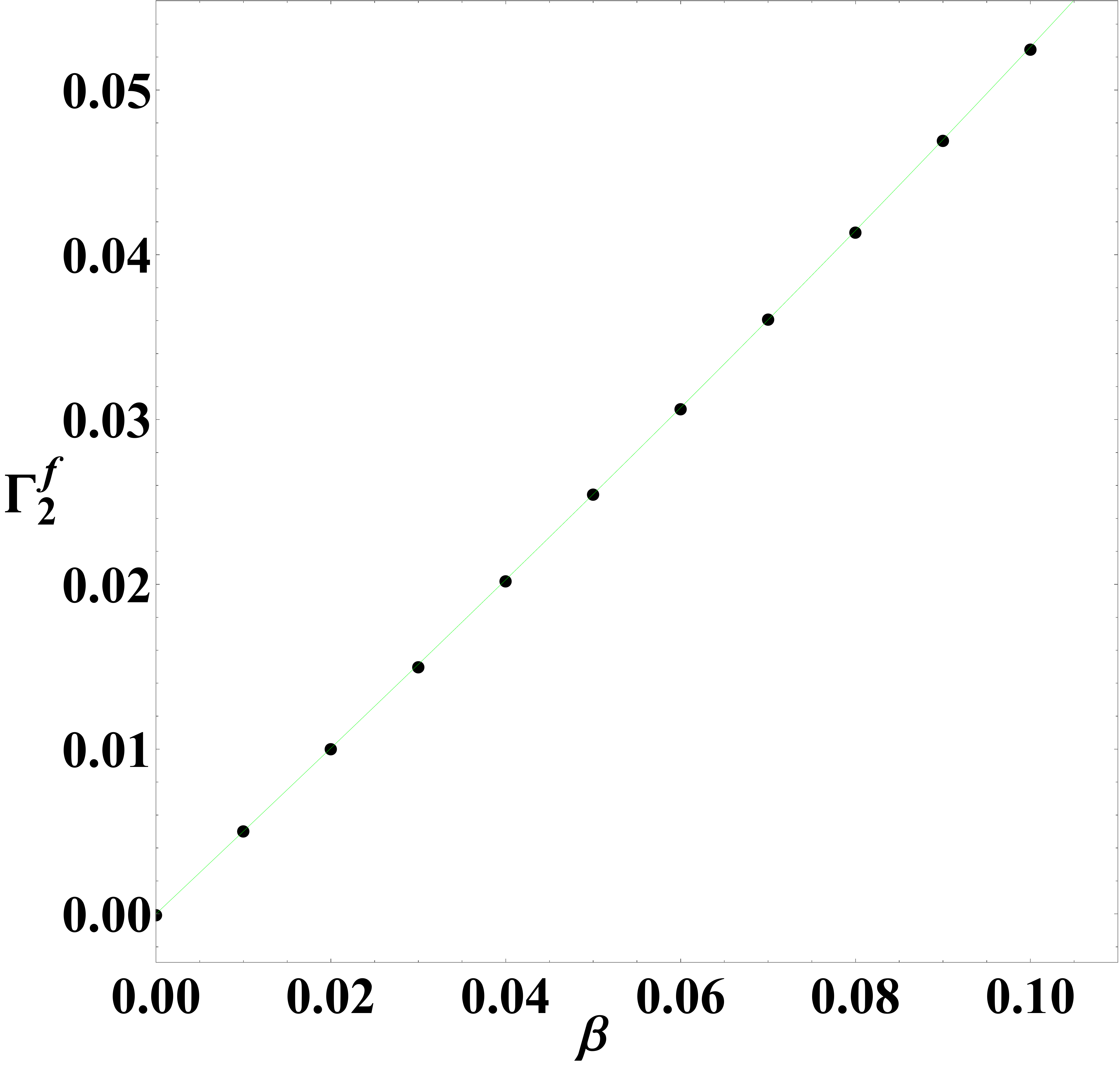}
\includegraphics[width=0.505\textwidth]{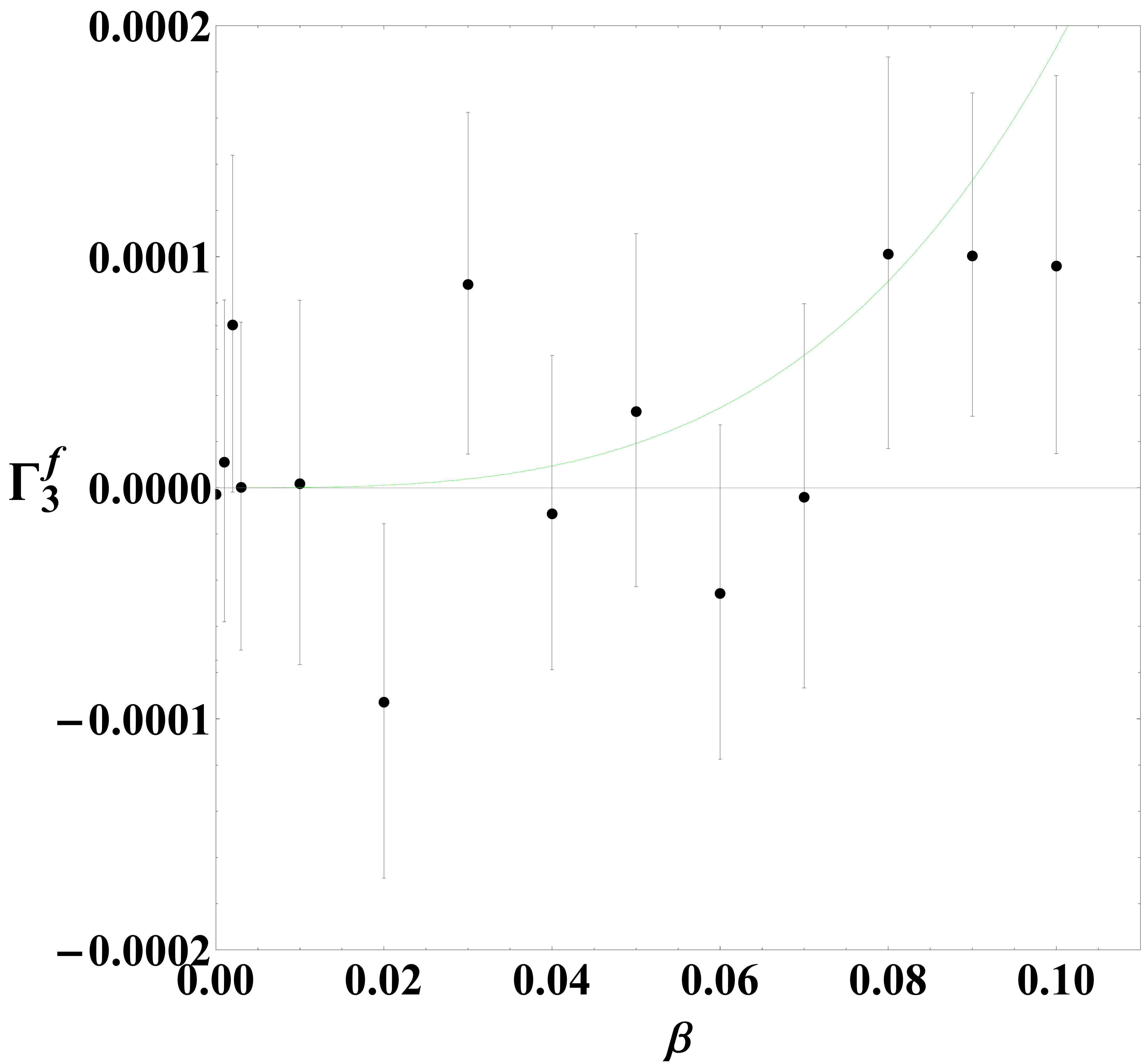}
\caption{Two-point (left) and three-point (right) correlation functions
  in the fundamental representation versus $\beta$. The green solid lines
  represent the strong coupling expansions,
  given respectively in Eqs.~(\ref{Strong_coupling_Gamma2_f}) and
  (\ref{Strong_coupling_Gamma3_f}).}
\label{fig:Gamma_strong}
\end{figure}

\begin{figure}[tbh!]
\centering
\includegraphics[width=0.48\textwidth]{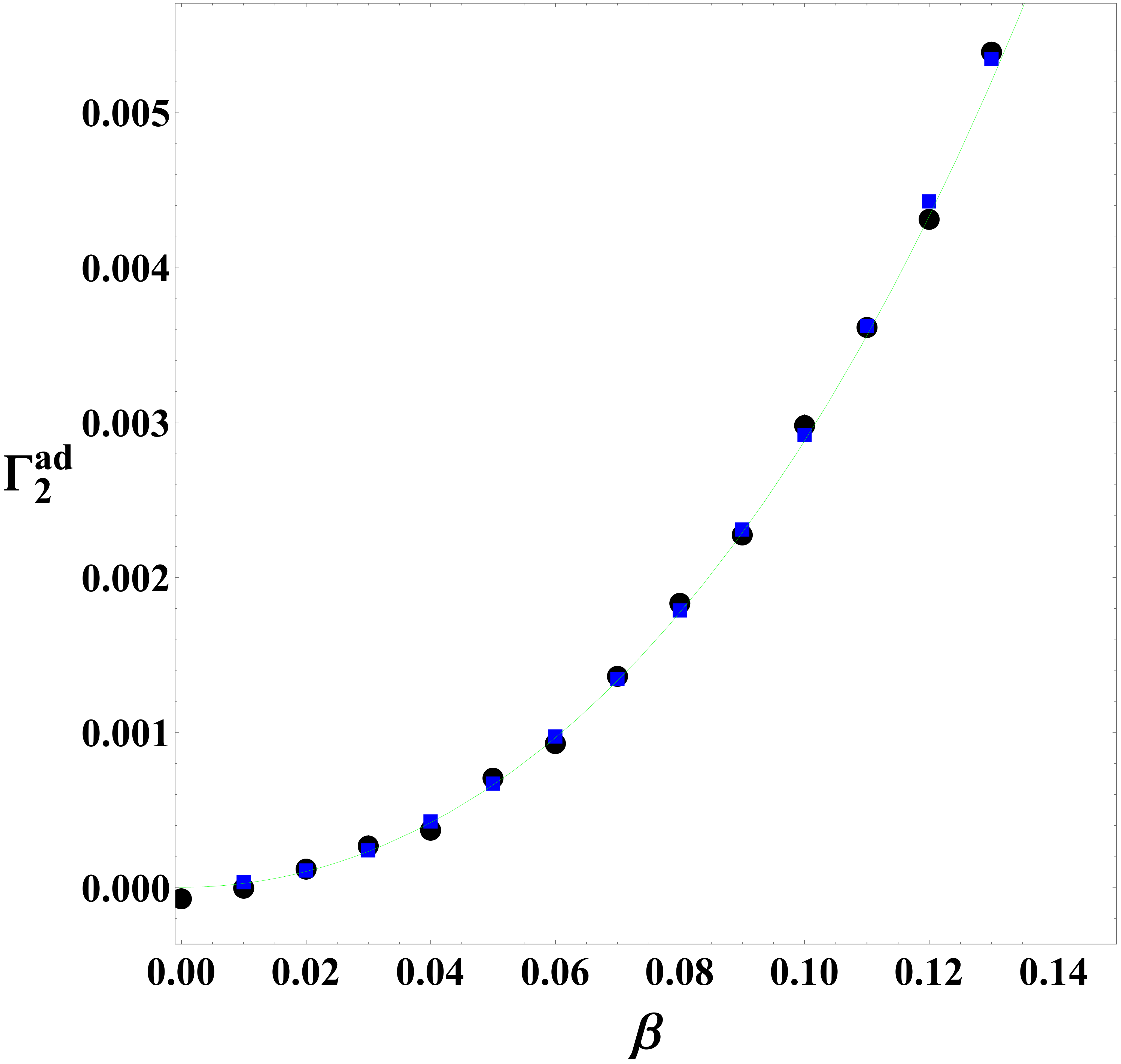}
\includegraphics[width=0.51\textwidth]{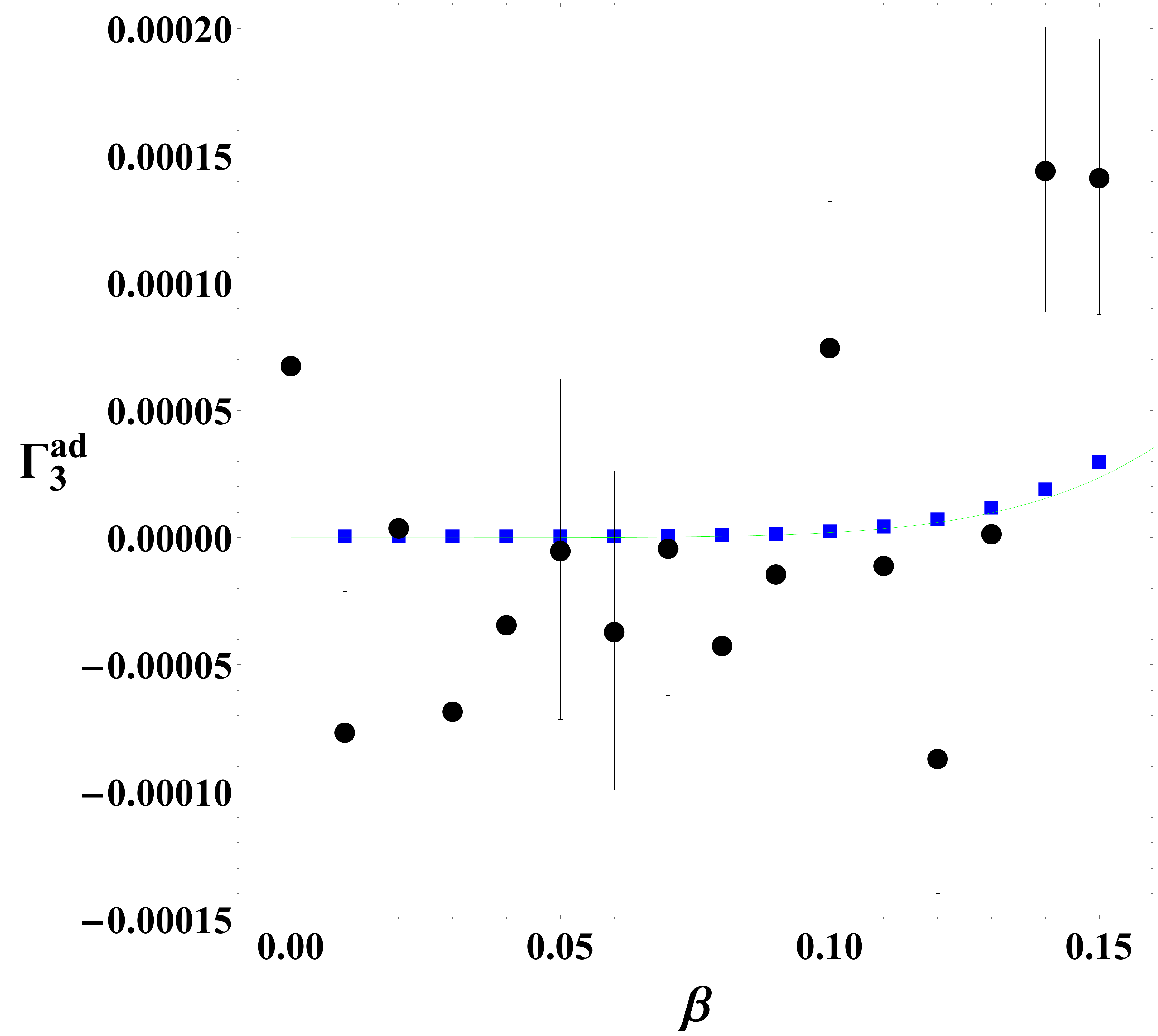}
\caption{Two-point (left) and three-point (right) correlation functions
  in the adjoint representation versus $\beta$.
  The solid green lines represent the strong coupling expansions,
  given respectively in Eqs.~\eqref{Strong_coupling_Gamma2_ad} and~\eqref{Strong_coupling_Gamma3_ad}.
  The round (square) symbols refer to simulations in the standard (dual)
  formulation.}
\label{fig:Gamma_AD_strong}
\end{figure}

To test our algorithms we performed a set of simulations at small values of
$\beta$ ($\beta < 0.15$), and compared the obtained values of
$\Gamma_2^{\mathrm{f}}$, $\Gamma_3^{\mathrm{f}}$, 
$\Gamma_2^{\mathrm{ad}}$ and $\Gamma_3^{\mathrm{ad}}$ with the corresponding
determinations in the strong coupling expansion (Eqs.~(\ref{Strong_coupling_Gamma2_f})-(\ref{Strong_coupling_Gamma3_ad})).
The results of the comparison are shown in Fig.~\ref{fig:Gamma_strong} for the
correlations in the fundamental representation, and in Fig.~\ref{fig:Gamma_AD_strong}
for the correlations in the adjoint representation. It can be seen that the
two-point correlation, both in the fundamental and adjoint representation,
is in good agreement with the strong coupling expansion. For the three-point
correlation, due to its small absolute value, statistical errors 
in the standard simulation are too large to make any statement about
agreement with the strong coupling prediction. The results for the adjoint
correlation from the dual code are compatible with the strong coupling
expansion up to $\beta = 0.1$.
Since the results of the two simulation codes agree in the region around
$\beta_{\mathrm{c}}$, where most of our simulations were carried out, we
are confident in the reliability of our measurements.

\subsection{Critical behaviour}

\begin{figure}[h!]
\includegraphics[width=.310\textwidth]{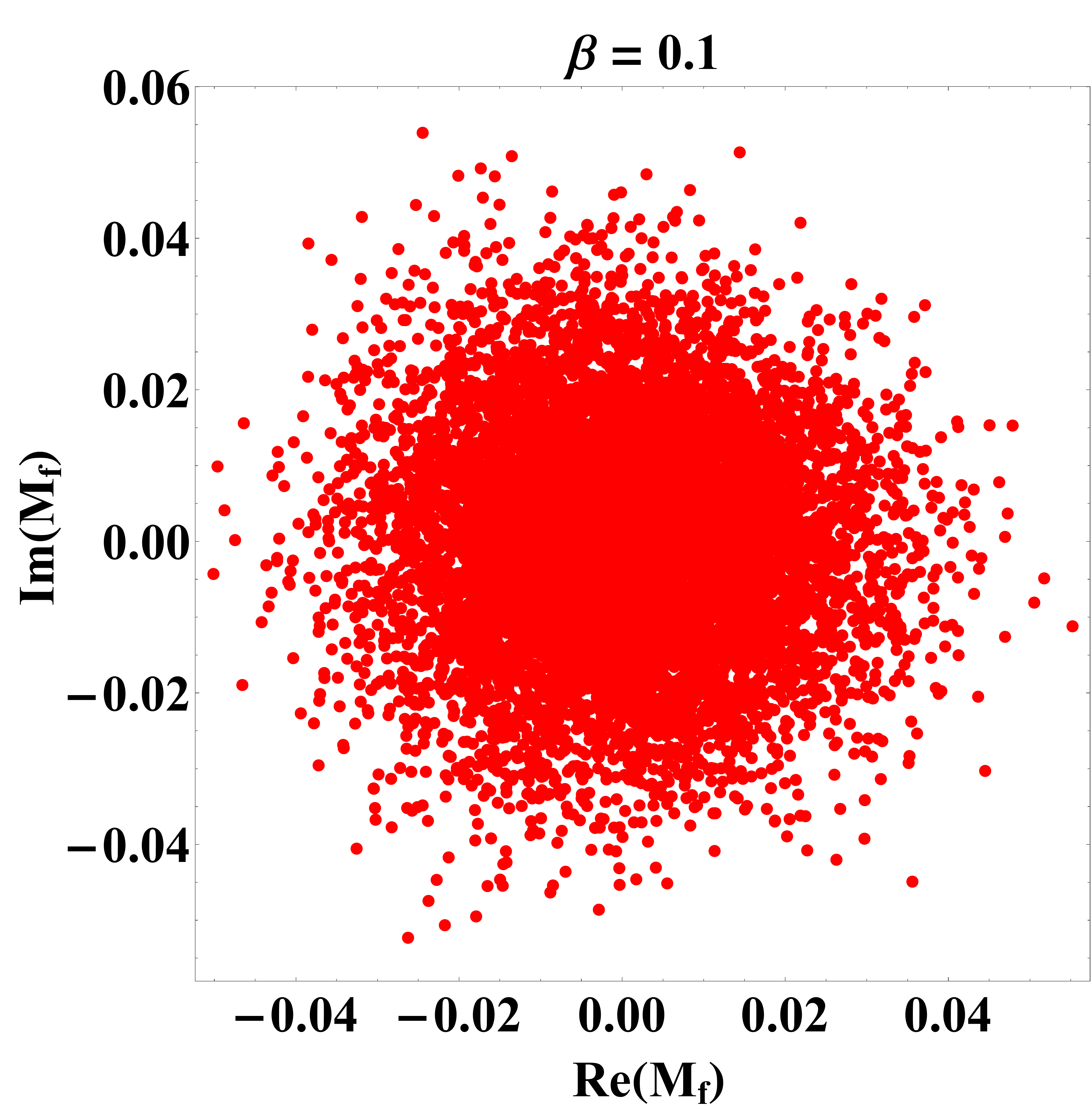}
\quad
\includegraphics[width=.310\textwidth]{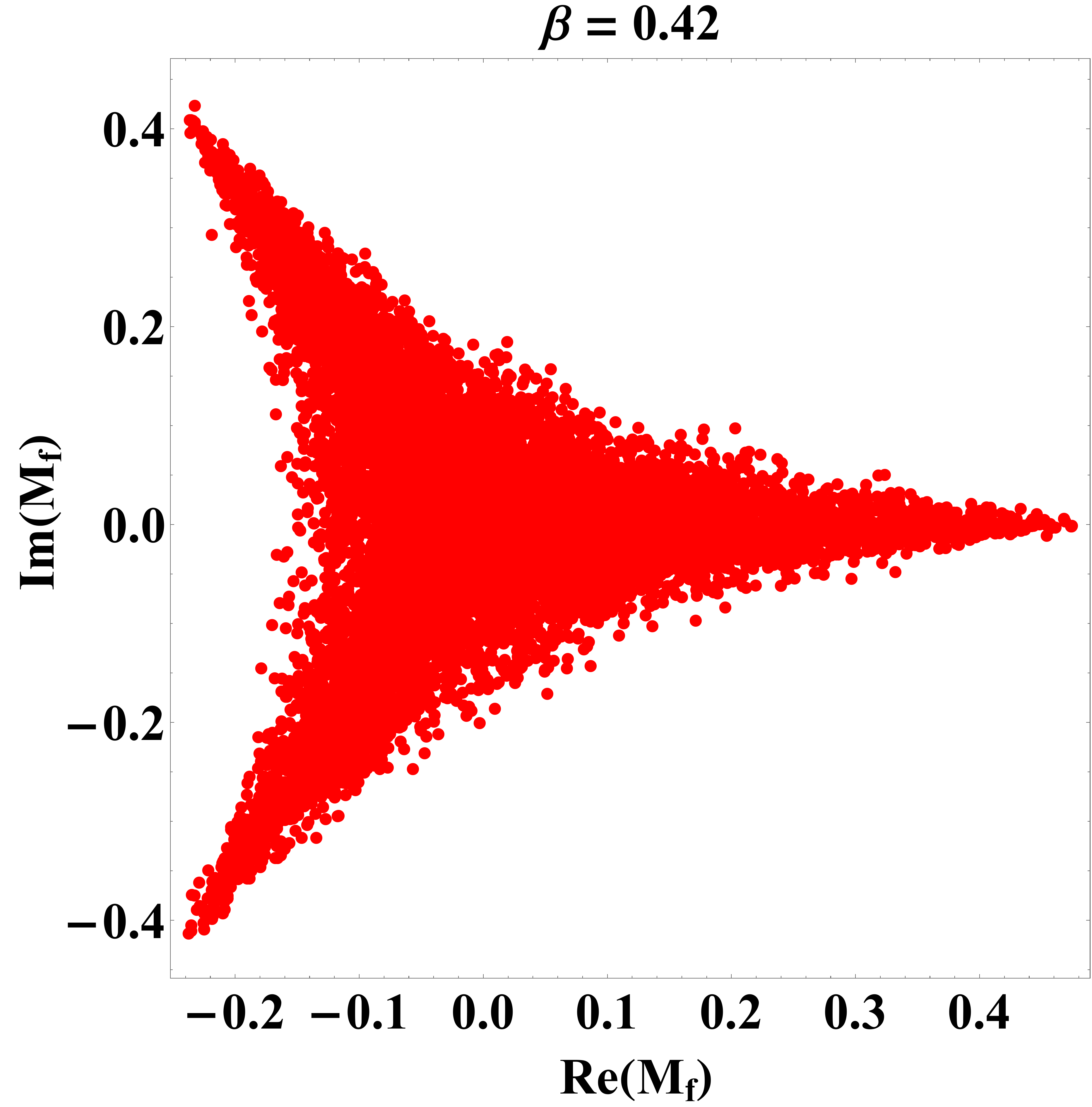}
\quad
\includegraphics[width=.310\textwidth]{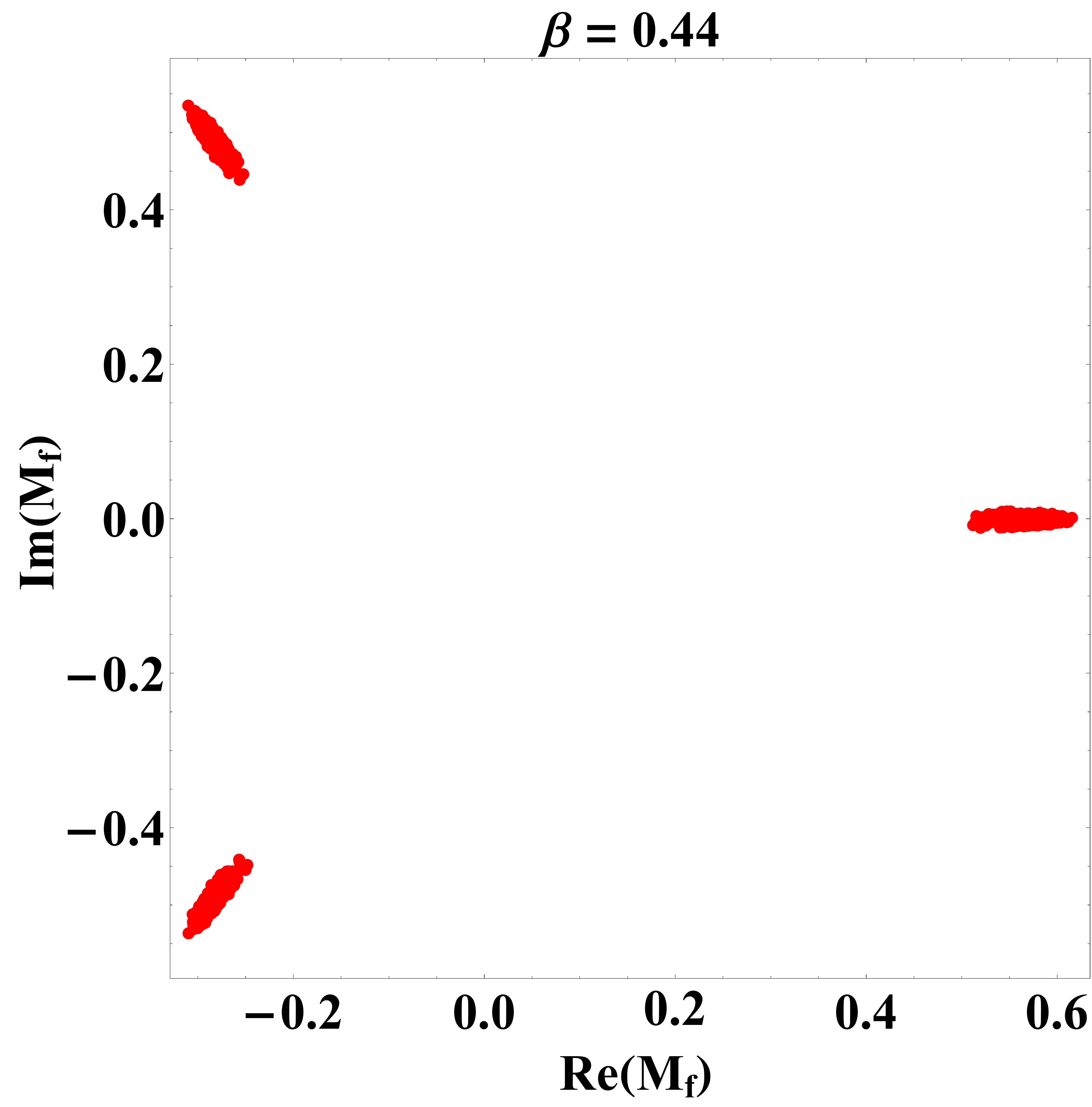}
\caption{Scatter plots of the complex magnetization $M_{\mathrm{f}}$ at $\beta=0.1, 0.42, 0.44$ on a $32\times 32$ lattice. }
\label{fig:magnetization}
\end{figure}

A clear indication of the two-phase structure of the model is provided
by the scatter plots of the complex magnetization at different values of
$\beta$, shown in Fig.~\ref{fig:magnetization}.

\begin{figure}[h!]
\centering
	\includegraphics[width=0.475\textwidth]{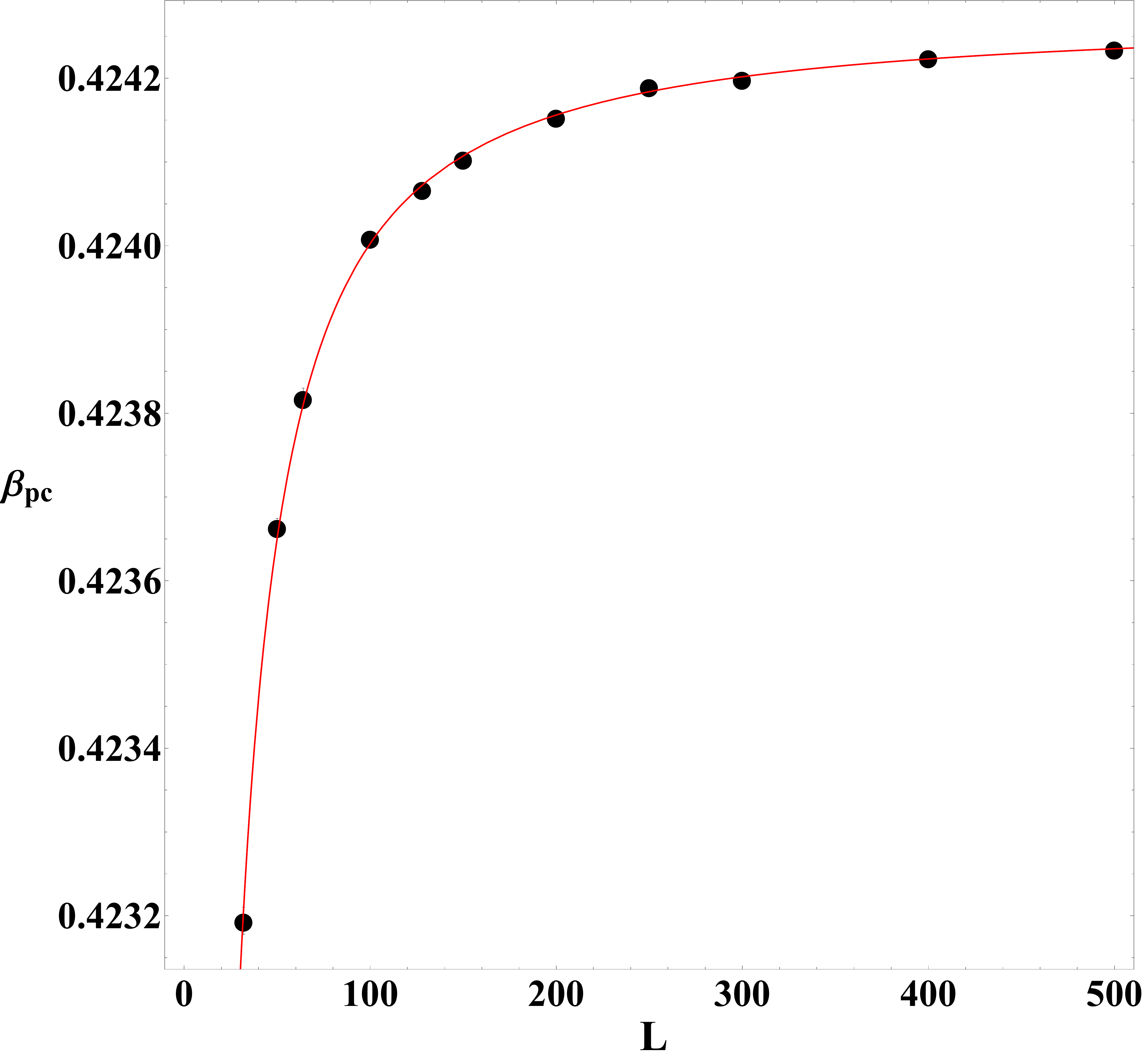}
	\includegraphics[width=0.515\textwidth]{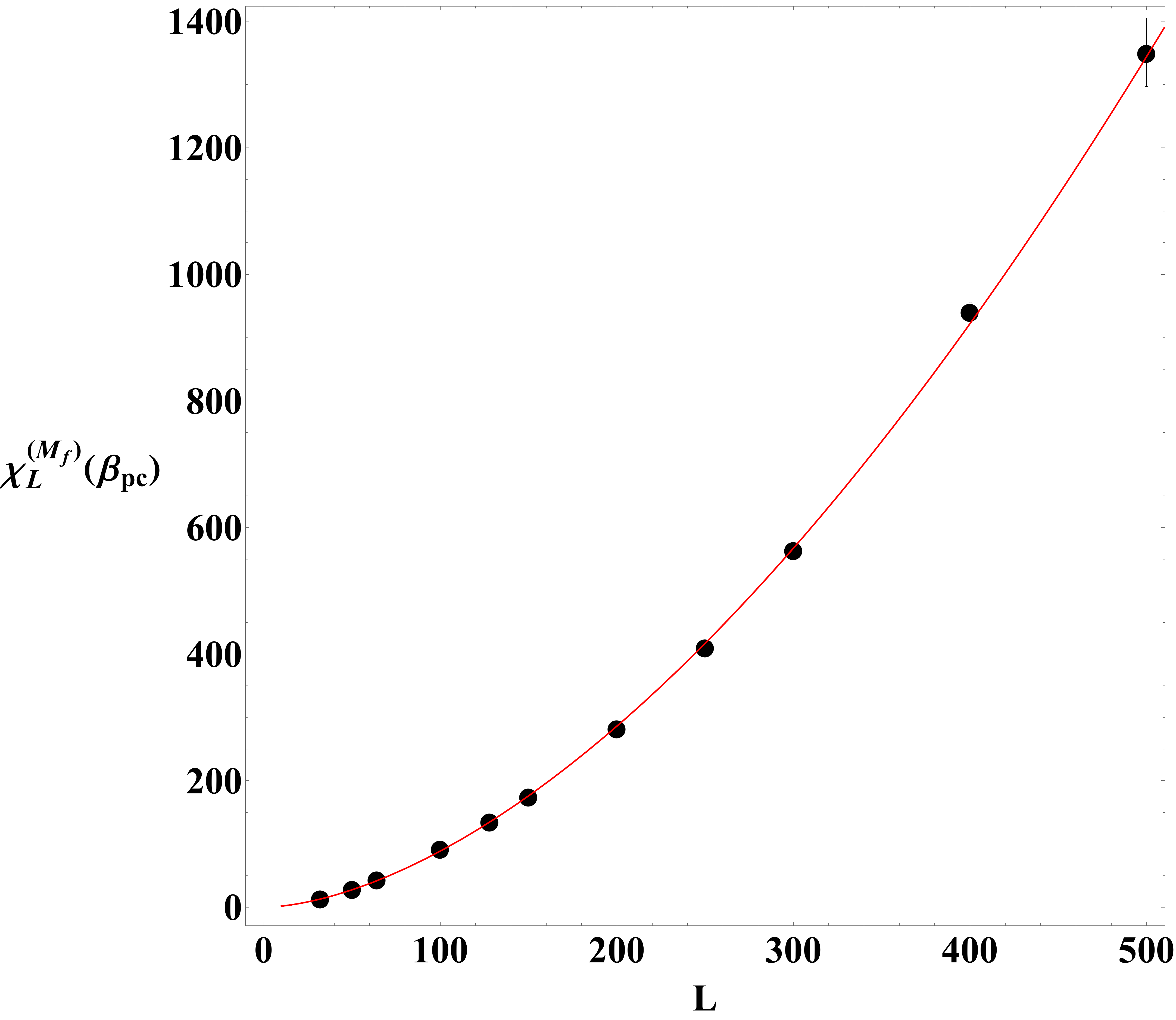}
        \caption{Fits of the $\beta_{\rm pc}(L)$ values determined from the
          magnetization susceptibility $\chi_L^{(M_{\rm f})}$ (left) and of the
          peak value of the magnetic susceptibility (right) versus the lattice
          size $L$.
          The solid red lines give the result of the fits with the 
          scaling functions in Eqs.~\eqref{beta_pc_scaling}
          and~\eqref{sp_Height_scaling}, respectively.}\label{fig:crit_fit}
\end{figure}

To precisely locate the $\beta_{\mathrm{c}}$ at which the phase transition occurs,
we have studied the magnetization susceptibility for different lattice sizes
$L$, extracting the value of $\beta_{\rm pc}(L)$ from a fit of the peak
of the susceptibility with a Lorentzian function. The obtained values
of $\beta_{\rm pc}(L)$ have been fitted with the scaling law for a second
order transition (see the left panel of Fig.~\ref{fig:crit_fit})
\begin{equation}
\label{beta_pc_scaling}
\beta_{\rm pc}(L) = \beta_{\rm c} + \dfrac{A}{L^{1/\nu}}
\end{equation}
with the following resulting parameters:
\begin{equation*}
  A =-0.0675(62),\ \ \ \beta_{\rm_c} =0.4242748(39),\ \ \ \nu =  0.835(17),
  \ \ \ \chi^2_{\rm r} = 1.18\ .
\end{equation*}
The value for the critical index $\nu$ is in agreement with the critical index
$\nu = 5/6$ of the two-dimensional three-state Potts model, to whose
universality class our model is believed to belong.
A direct extraction of the critical exponent $\nu$, performed in the
subsection~\ref{nu_from_sigma}, gives a compatible result, which is 
sensitive to the choice of the region of $\beta$ values where critical
scaling is supposed to hold.

As a second check of the order of the phase transition and of the universality
class, we studied the dependence of the peak value of the magnetic
susceptibility for different lattice sizes using the scaling law
(see the right panel of Fig.~\ref{fig:crit_fit})
\begin{equation}
\label{sp_Height_scaling}
\chi_L^{(M)}(\beta_{\rm pc})(L) = B L^{\gamma/\nu}\ .
\end{equation}
We found
\begin{equation*}
 B = 0.0282(27),\ \ \ \gamma/\nu = 1.737(17),\ \ \ \chi^2_{\rm r} = 0.30\ .
\end{equation*}
The obtained value for $\gamma/\nu$ is in agreement with the hyperscaling
relation $2 - \eta = \gamma/\nu$ for the three-state Potts model, which
gives $\eta\approx 0.263$. The expected value of $\eta$ is 4/15.
These findings support the $Z(3)$ universality class of the present Polyakov
loop spin model.  

\section{Correlation functions in confinement phase}

\subsection{Extraction of $\sigma_{q\overline{q}}$ from $\Gamma_2^{\rm f}$}

The potential parameter $\sigma_{q\overline{q}}$ is extracted from the measurements
of the observable $\Gamma_2^{\rm f}$. Following
Eq.~(\ref{corrfunc2_fund_largeN_small_beta}) and the explanation in
subsection~2.4, we expect
\begin{equation}
\label{gamma_2_R}
\Gamma_2^{\rm f}(R) = A\, \frac{e^{-\sigma_{q\overline{q}}\, R}}{R^\eta} \ .
\end{equation}
One can extract $\sigmaqq$ from the following ratio:
\begin{equation}
\label{sigmacalca}
\sigma_{q\overline{q}}^{\rm eff}(R) \equiv -\dfrac{1}{2} \ln \left[\dfrac{\Gamma_2^{\rm f}(R)}{\Gamma_2^{\rm f}(R-2)}\right]
  = \sigmaqq + \dfrac{\eta}{2} \ln \left[ \dfrac{R}{R-2}\right] \ .
\end{equation}
We have compared our Monte Carlo data for $\sigma_{q\overline{q}}^{\rm eff}(R)$ with
the formula $\sigmaqq+\eta/2\; \ln \left[R/(R-2)\right]$.
The interval of $\beta$ values that we considered for the extraction
of $\sigma_{q\overline{q}}$ was $[0.41,0.42]$, since for $\beta < 0.41$ the
two-point correlation drops too fast to be of any use. The values of 
$\sigmaqqeff$ obtained in the selected range of $\beta$ do not show
any significant difference when moving from a $64\times 64$ to
a $128\times 128$ lattice, thus making unnecessary to perform simulation
on even larger lattices.

As an alternative method for extracting $\sigmaqq$, we measured the wall-wall
correlation function,
\begin{equation}
\label{wall_wall_corr}
\Gamma_2^{\rm ww}(R) = \left\langle \frac{1}{L^3} \sum_{x, y_1, y_2 = 1}^L
\Tr W(x, y_1) \Tr W^\dagger(x + R, y_2) \right \rangle \ ,
\end{equation}
which is known to obey the exponential decay law, with no power corrections,
\begin{equation}
\label{Wall-Wall_R}
\Gamma_2^{\rm ww}(R)=A\,e^{-\sigma_{q\overline{q}}\, R}\ .
\end{equation}

Introducing, similarly to~(\ref{sigmacalca}),
\begin{equation}
\label{sigma_wall_wall}
\sigmaww^{\mathrm{eff}}(R)\equiv-\dfrac{1}{2} \ln \left[\dfrac{\Gamma_2^{\rm ww}{(R)}}{\Gamma_2^{\rm ww}(R-2)}\right] \ ,
\end{equation}
we found that $\sigmaww^{\mathrm{eff}}(R)$ exhibits a long plateau at each
considered $\beta$ value; we took as plateau value $\sigmaww$ the value
of $\sigmaww^{\mathrm{eff}}(R)$ at the smallest value of $R$ after which
all values of $\sigmaww^{\mathrm{eff}}(R)$ agree within statistical
uncertainties. Results for the $128\times 128$ lattice are summarized in
Table~\ref{table:128_sigma2_new}: we can see that results of
$\sigmaqq$ obtained from the fitting of $\sigmaqqeff(R)$ according
to~(\ref{sigmacalca}) are in good agreement with the plateau values of
$\sigmaww$. We ascribe the difference between the $\sigmaqq$ values 
obtained for different choices of $R_{\min}$ to possible systematic effects
arising from the difficulty to extract the parameters of the exponential decay
corrected by a power law and to our treating of the correlation function
errors as independent values. 
These systematic errors seem also to result in a value of $\sigmaww$ being 
in most cases slightly higher than the estimates of $\sigmaqq$.

\begin{table}[]
\centering
\begin{tabular}{ | c | c | c | c | c | c | c |}
\hline
\hline
 $\beta$
 & $R_{\min}$ & $R_{\max}$ & $\eta$   &  $\sigmaqq$   & $\chiR$ & $\sigmaww$ \\
\hline
\multirow{4}{*}{0.41}
 & 4  & 20 & 0.6242(36) & 0.3309(12) & 0.60 & \multirow{4}{*}{0.3436(49)}\\
 & 6  & 20 & 0.659 (14) & 0.3245(28) & 0.35 & \\
 & 8  & 20 & 0.735 (50) & 0.3144(69) & 0.28 & \\
 & 10 & 20 & 0.61  (18) & 0.327 (19) & 0.31 & \\
\hline
\multirow{4}{*}{0.412}
 & 4  & 20 & 0.6274(33) & 0.29939(10) & 0.70 & \multirow{4}{*}{0.3101(32)}\\
 & 6  & 20 & 0.601 (14) & 0.3044 (27) & 0.49 & \\
 & 8  & 20 & 0.595 (54) & 0.3051 (72) & 0.59 & \\
 & 10 & 20 & 0.62  (18) & 0.303  (19) & 0.74 & \\
\hline
\multirow{4}{*}{0.414}
 & 4  & 26 & 0.6439(34)   & 0.2620 (10)  & 0.75 & \multirow{4}{*}{0.2733(16)}\\
 & 6  & 26 & 0.620 (13)   & 0.2663 (25)  & 0.60	& \\
 & 8  & 26 & 0.600 (44)   & 0.2687  (57) & 0.66 & \\
 & 10 & 26 & 0.60  (0.12) & 0.269  (13)  & 0.75 & \\
\hline
\multirow{4}{*}{0.415}
 & 4  & 22 & 0.6480(25) & 0.2434(16) & 0.58 & \multirow{4}{*}{0.2501(20)}\\
 & 6  & 22 & 0.6330(93) & 0.2460(17) & 0.47 & \\
 & 8  & 22 & 0.611 (28) & 0.2487(38) & 0.50 & \\
 & 10 & 22 & 0.560 (83) & 0.2537(84) & 0.55 & \\
\hline
\multirow{4}{*}{0.416}
 & 4  & 26 & 0.6553(43) & 0.2242(13) & 1.63 & \multirow{4}{*}{0.2356(14)}\\
 & 6  & 26 & 0.623 (13) & 0.2296(24) & 1.04	& \\
 & 8  & 26 & 0.640 (61) & 0.2305(53) & 1.16 & \\
 & 10 & 26 & 0.57  (10) & 0.235 (11) & 1.29 & \\
\hline
\multirow{4}{*}{0.417}
 & 4  & 28 & 0.6649(31) & 0.20343(94) & 1.19 & \multirow{4}{*}{0.2144(10)}\\
 & 6  & 28 & 0.6349(69) & 0.2084 (12) & 0.43 & \\
 & 8  & 28 & 0.627 (21) & 0.2093 (27) & 0.47 & \\
 & 10 & 28 & 0.598 (52) & 0.2121 (86) & 0.51 & \\
\hline
\multirow{4}{*}{0.418}
 & 4  & 28 & 0.6747(28) & 0.18187(83) & 1.12 & \multirow{4}{*}{0.1927(13)}\\
 & 6  & 28 & 0.6515(75) & 0.1857 (13) & 0.61 & \\
 & 8  & 28 & 0.647 (22) & 0.1863 (28) & 0.67 & \\
 & 10 & 28 & 0.598 (49) & 0.1909 (50) & 0.66 & \\
\hline
\multirow{4}{*}{0.419}
 & 4  & 36 & 0.6835(35) & 0.16037(97) & 1.62 & \multirow{4}{*}{0.1700(21)}\\
 & 6  & 36 & 0.668 (10) & 0.1628 (18) & 1.48 & \\
 & 8  & 36 & 0.667 (27) & 0.1630 (34) & 1.59 & \\
 & 10 & 36 & 0.690 (58) & 0.1609 (57) & 1.69 & \\
\hline
\multirow{4}{*}{0.42}
 & 4  & 36 & 0.6938(38) & 0.13758(53) & 0.69 & \multirow{4}{*}{0.1476(13)}\\
 & 6  & 36 & 0.6841(54) & 0.13900(90) & 0.59 & \\
 & 8  & 36 & 0.670 (13) & 0.1405 (15) & 0.58 & \\
 & 10 & 36 & 0.652 (26) & 0.1421 (25) & 0.59 & \\
\hline		 		
\multirow{4}{*}{0.423}
 & 4  & 56 & 0.7756(16) & 0.05295(40) & 1.33 & \multirow{4}{*}{0.05809(56)}\\ 
 & 6  & 56 & 0.7887(37) & 0.05143(52) & 0.87 & \\
 & 8  & 56 & 0.7841(68) & 0.05185(73) & 0.89 & \\ 	
 & 10 & 56 & 0.781 (11) & 0.0521(90)  & 0.92 & \\ 
\hline			 		
\multirow{4}{*}{0.424}
 & 4  & 52 & 0.8844(44) & 0.02306(68) & 3.84 & \multirow{4}{*}{0.01612(30)}\\ 
 & 6  & 52 & 0.9089(44) & 0.02073(53) & 1.30 & \\ 
 & 8  & 52 & 0.9132(71) & 0.02040(69) & 1.33 & \\ 		
 & 10 & 52 & 0.903 (11) & 0.02106(90) & 1.31 & \\ 	
\hline
\end{tabular}
\caption{Best-fit parameters $\eta$ and $\sigmaqq$,
obtained from fits of the Monte Carlo values for $\sigmaqqeff(R)$ with the
function $\sigmaqq + \eta/2\; \ln \left[R/(R-2)\right]$ on a $128\times 128$
lattice. The second and third columns give the minimum and maximum values of
the distance $R$ considered in the fit. The last column gives the determination
of $\sigmaww$.}
\label{table:128_sigma2_new}
\end{table}

\subsection{Extraction of $\sigmaB$ from $\Gamma_3^{\rm f}$}

\begin{figure}[tbp]
\centering
{\includegraphics[width=.48\textwidth]{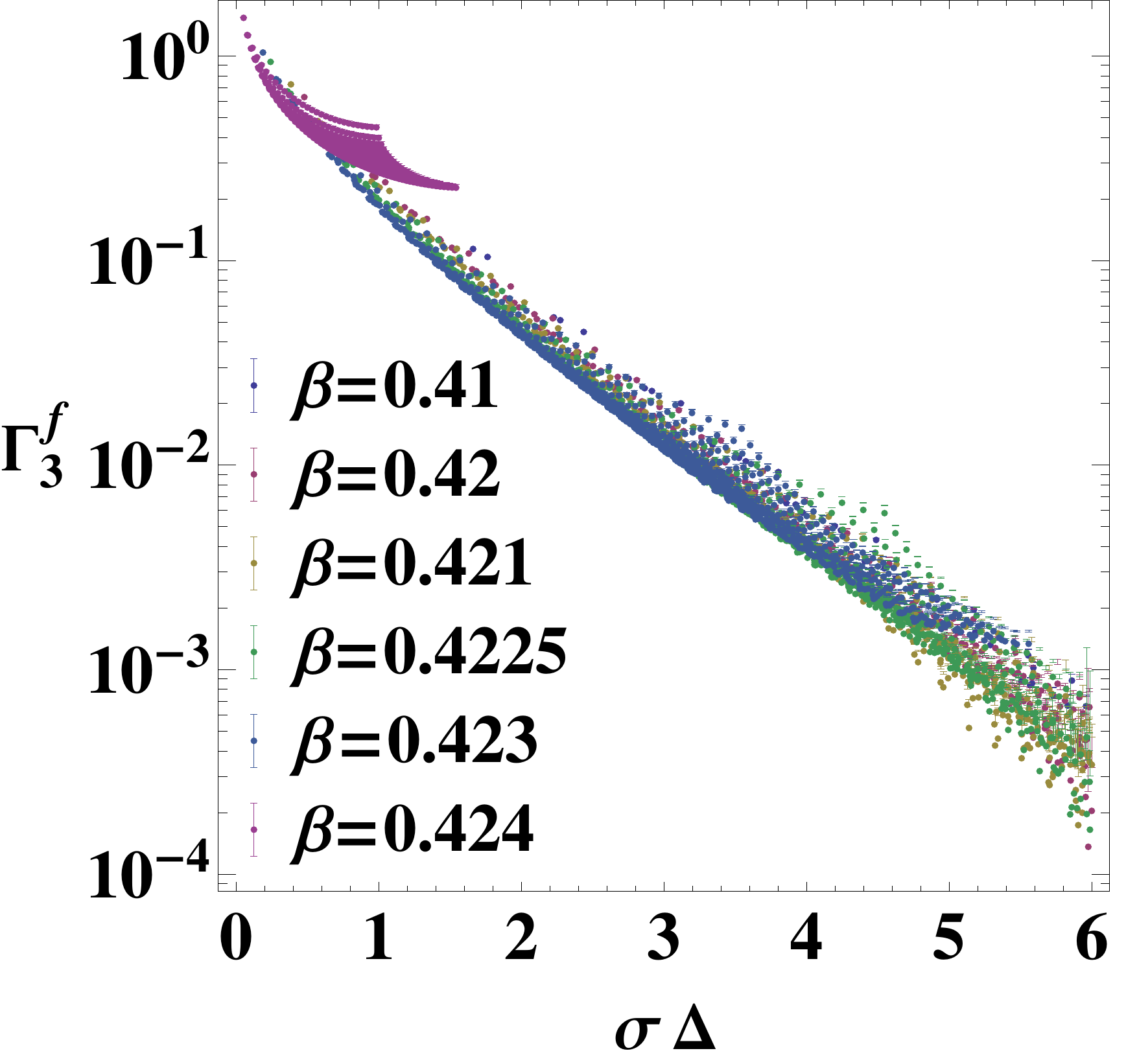}}
\quad
{\includegraphics[width=.48\textwidth]{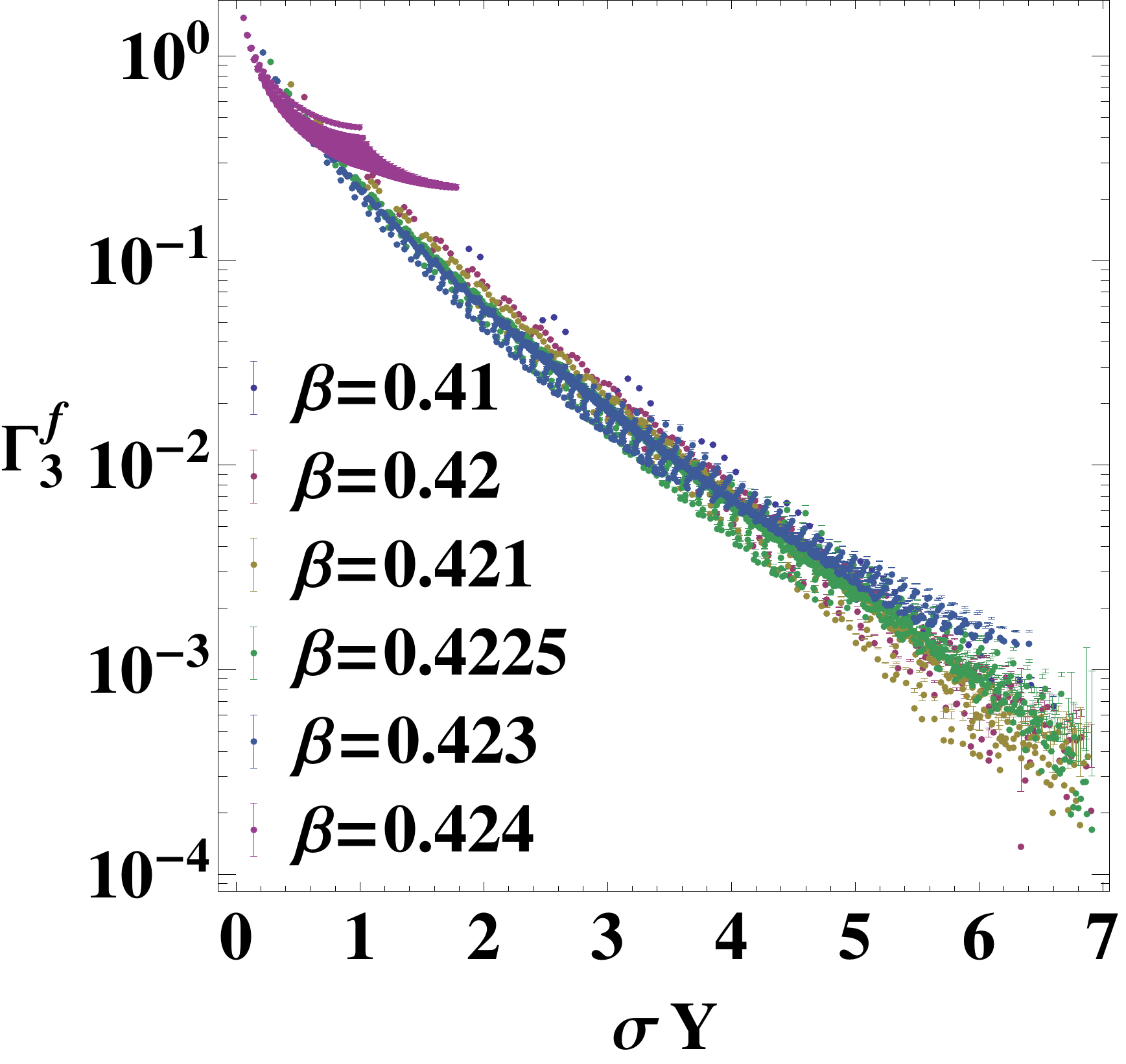}}
\caption{$\Gamma_3^{\rm f}$ versus $\sigma \Delta$ (left) and $\sigma Y$ (right), for the isosceles geometry.
In both cases the value of $\sigmaww$ is used for $\sigma$.}
\label{fig:scaled_Gamma3f}
\end{figure} 

\begin{figure}[tbp]
\centering
{\includegraphics[width=.48\textwidth]{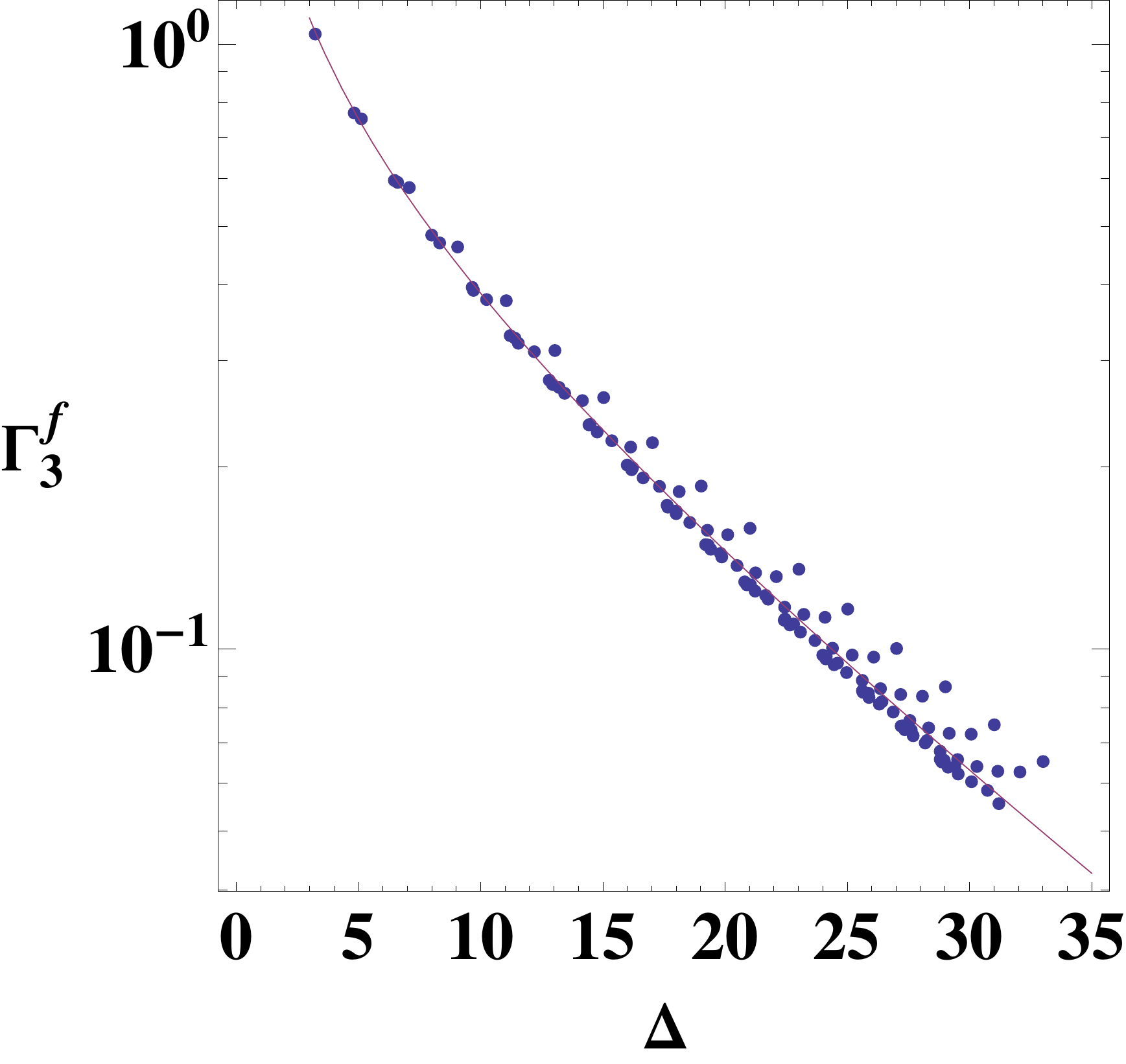}}
\quad
{\includegraphics[width=.48\textwidth]{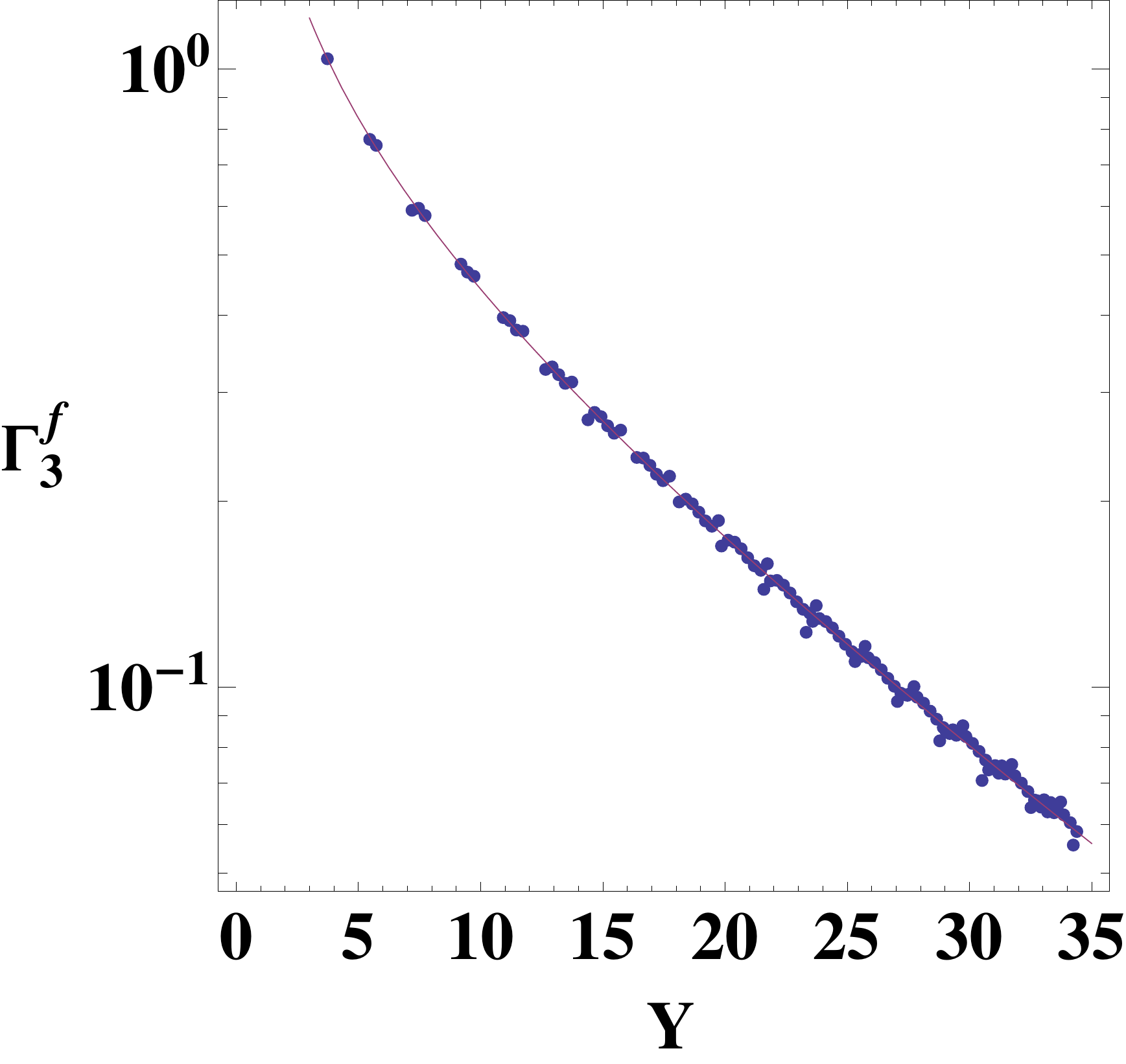}}
\caption{Comparison of the three-point correlation $\Gamma_3^{\rm f}$ with
the fit using the $\Delta$
law (left) and the $Y$ law (right) for $\beta=0.423$ isosceles triangles
with the angles less than $2\pi / 3$
and $Y < 35$.}
\label{fig:fits_Gamma3f}
\end{figure}

\begin{figure}[tbp]
\centering
{\includegraphics[width=.48\textwidth]{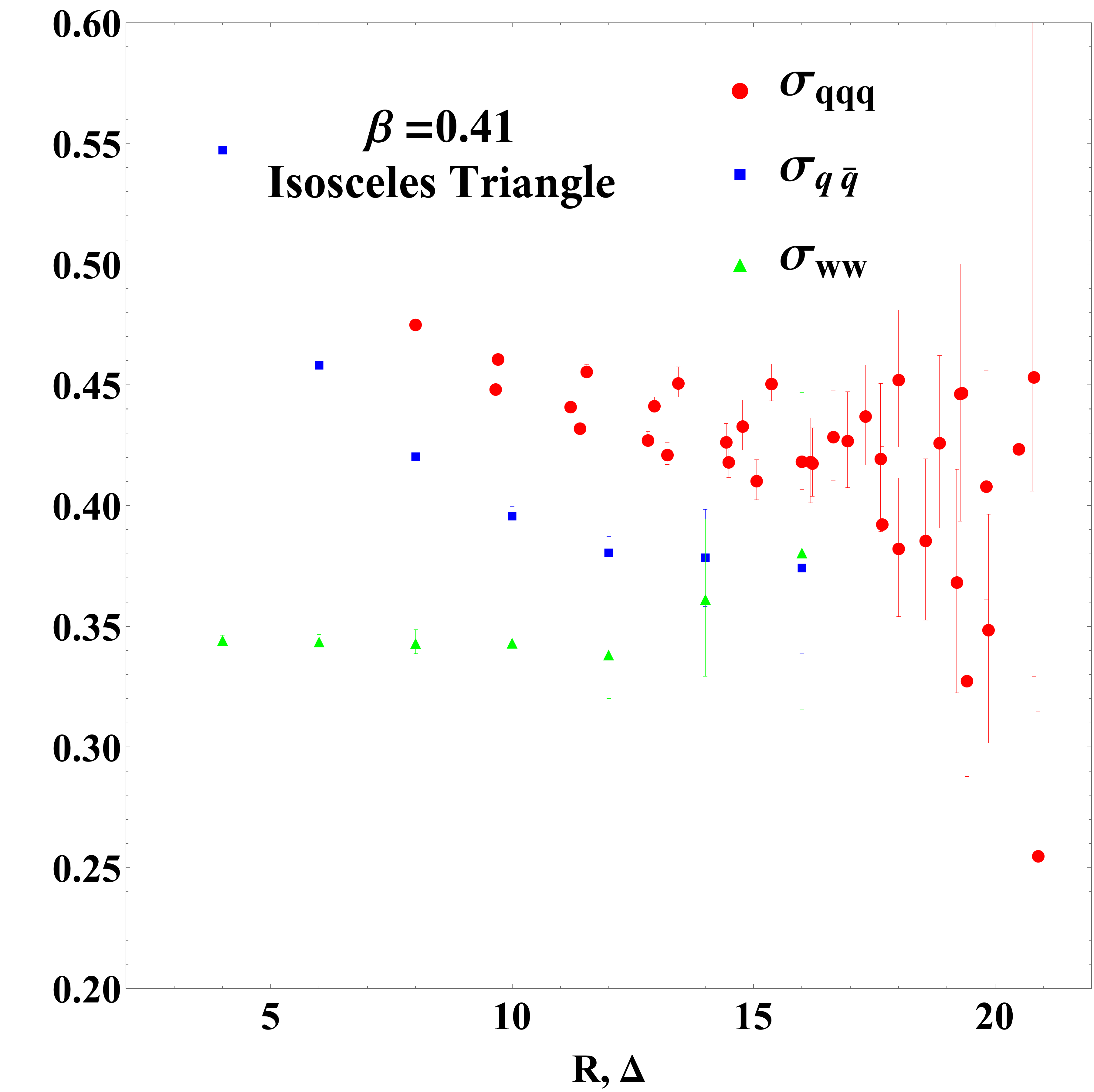}}
\quad
{\includegraphics[width=.48\textwidth]{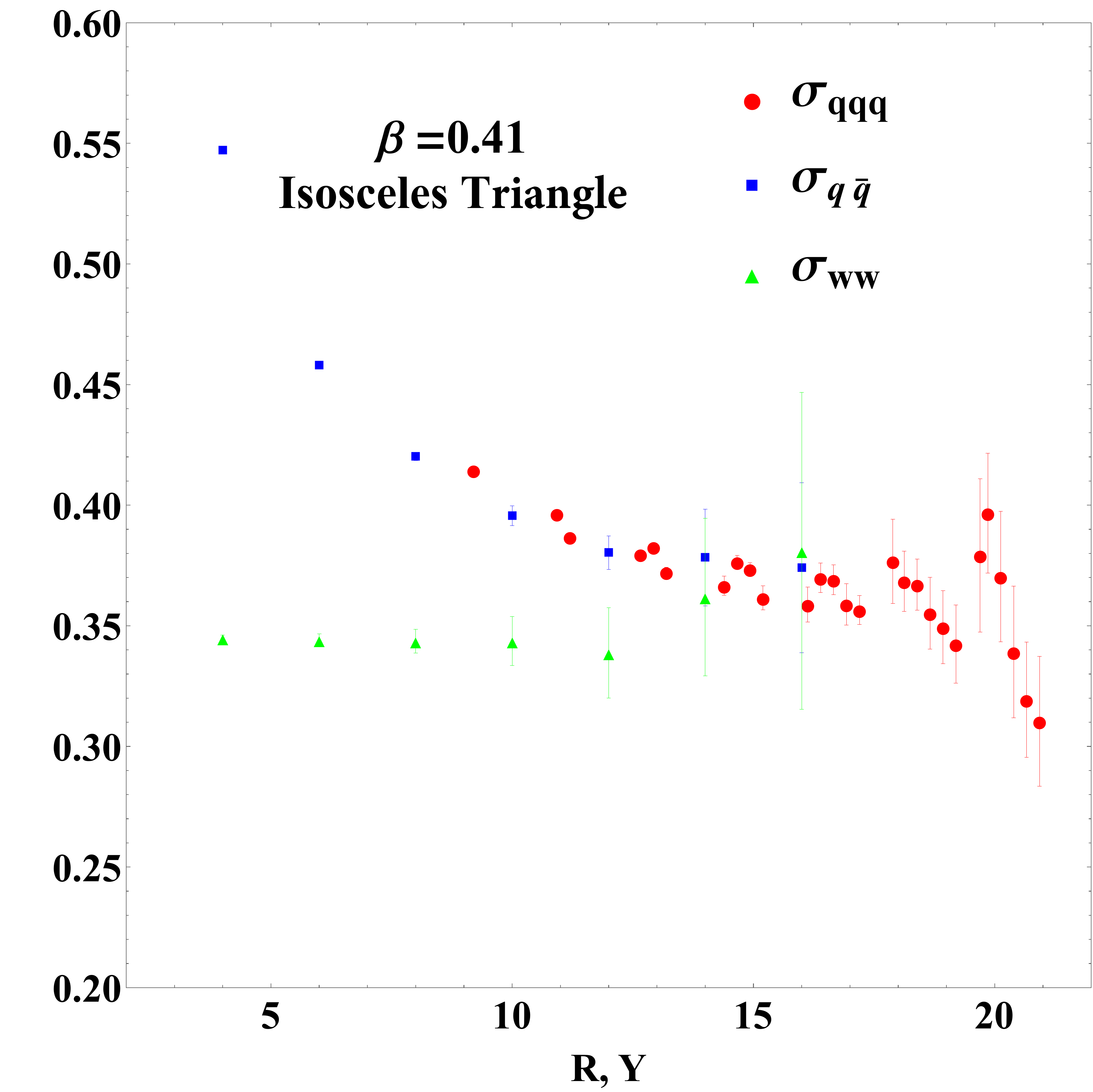}}
\caption{$\sigmaqqeff$ versus $R$, $\sigmaww^{\rm eff}$ versus $R$ and
  $\sigmaB^{\rm eff}$ versus $\Delta$ (left) and $Y$ (right) at $\beta=0.41$ on
  a $128\times 128$ lattice, for the isosceles geometry.}
\label{fig:128_Iso_0.41}
\end{figure}

\begin{figure}[tbp]
\centering
{\includegraphics[width=.48\textwidth]{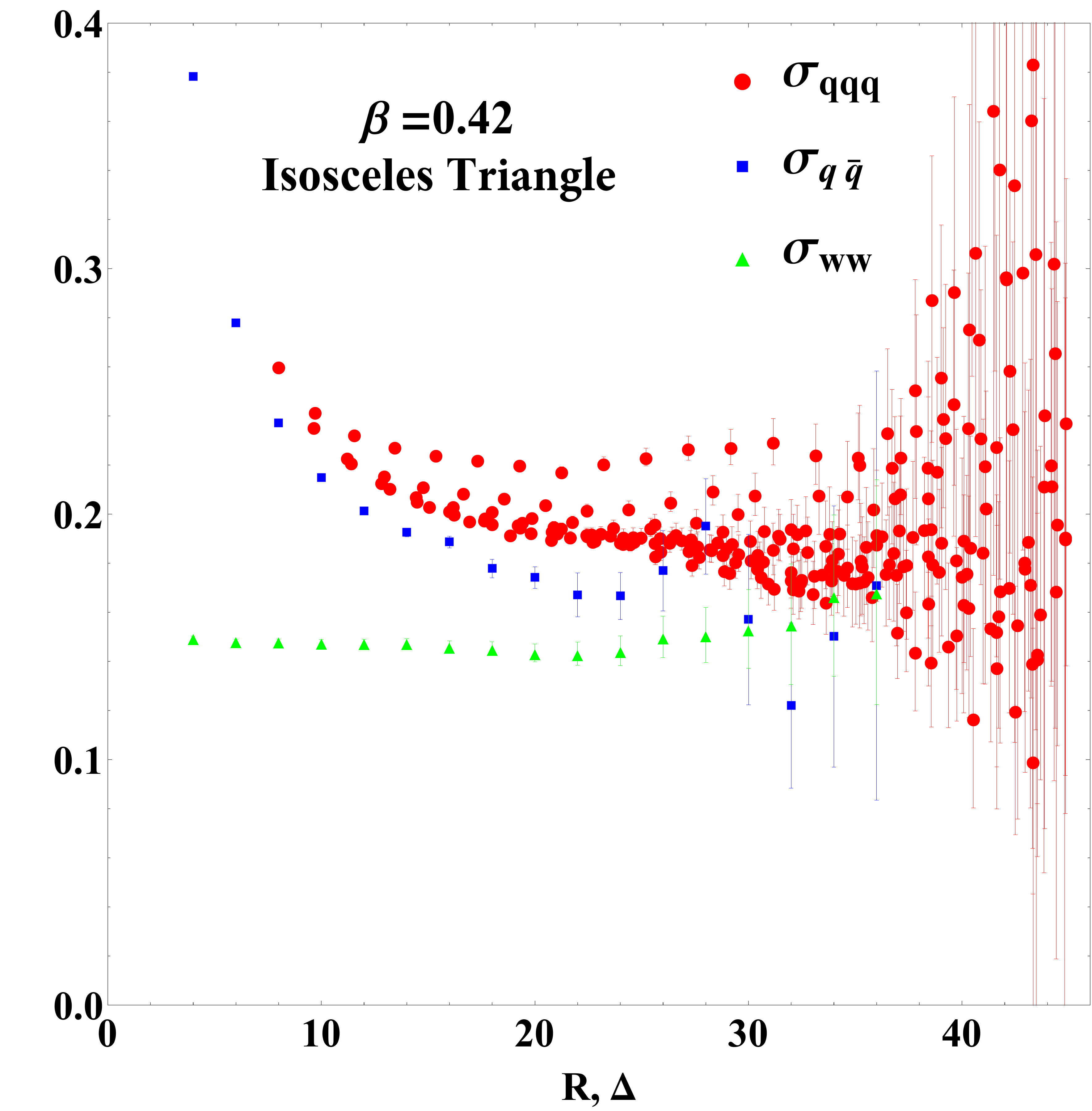}}
 \quad
{\includegraphics[width=.48\textwidth]{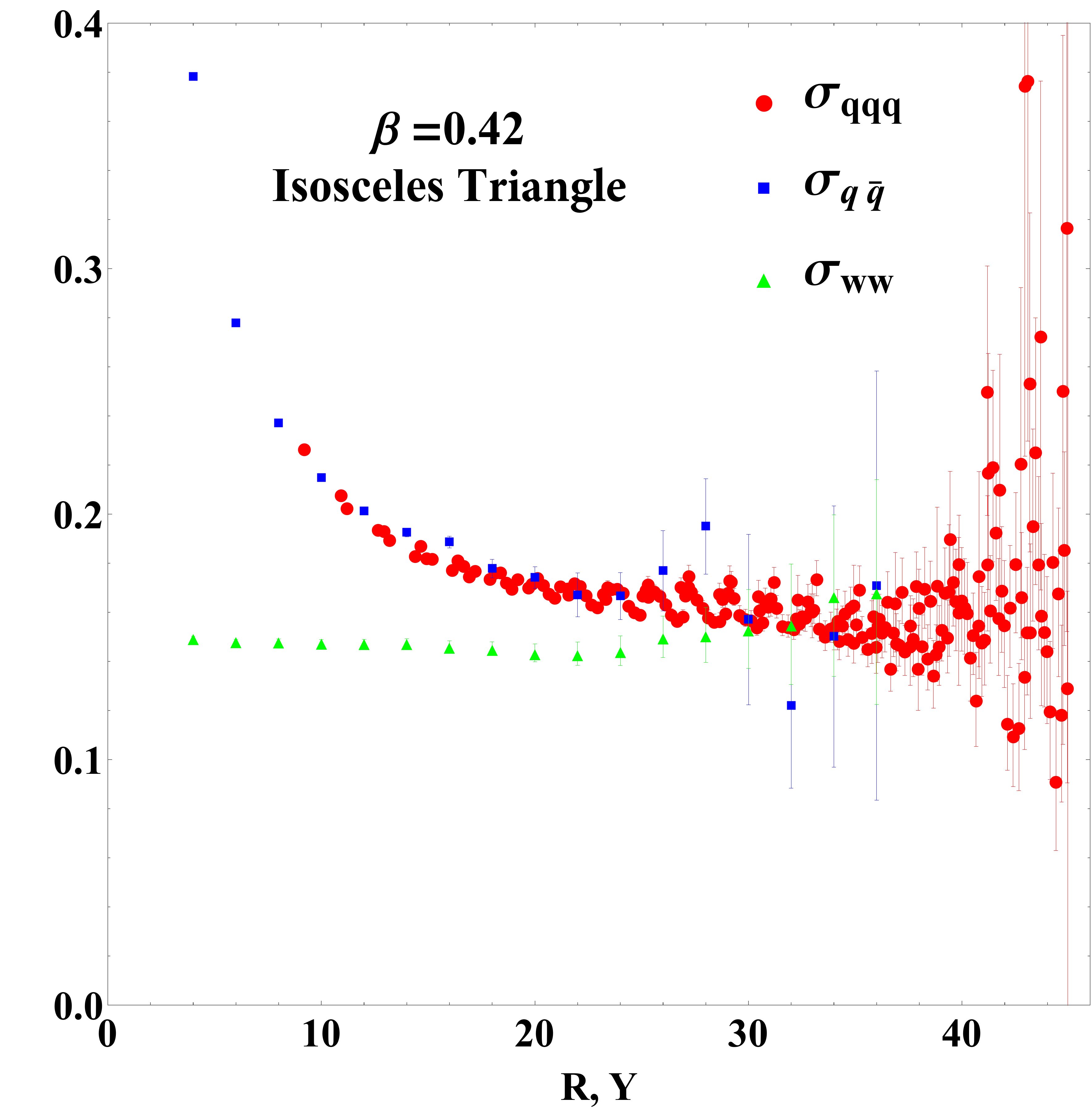}}
\caption{Same as Fig.~\ref{fig:128_Iso_0.41} at $\beta=0.42$.}
\label{fig:128_Iso_0.42}
\end{figure}

\begin{figure}[tb]
\centering
{\includegraphics[width=.48\textwidth]{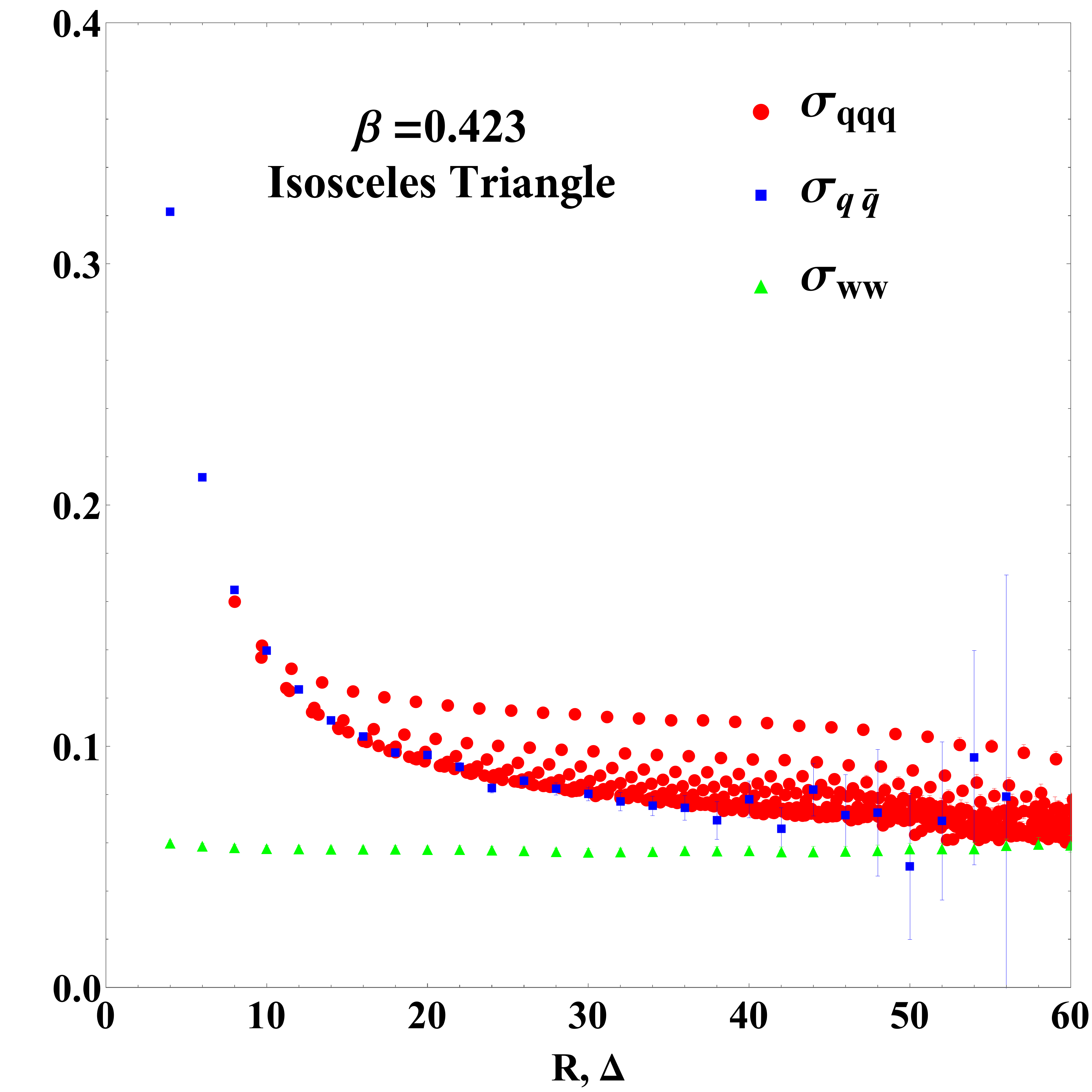}}
\quad
{\includegraphics[width=.48\textwidth]{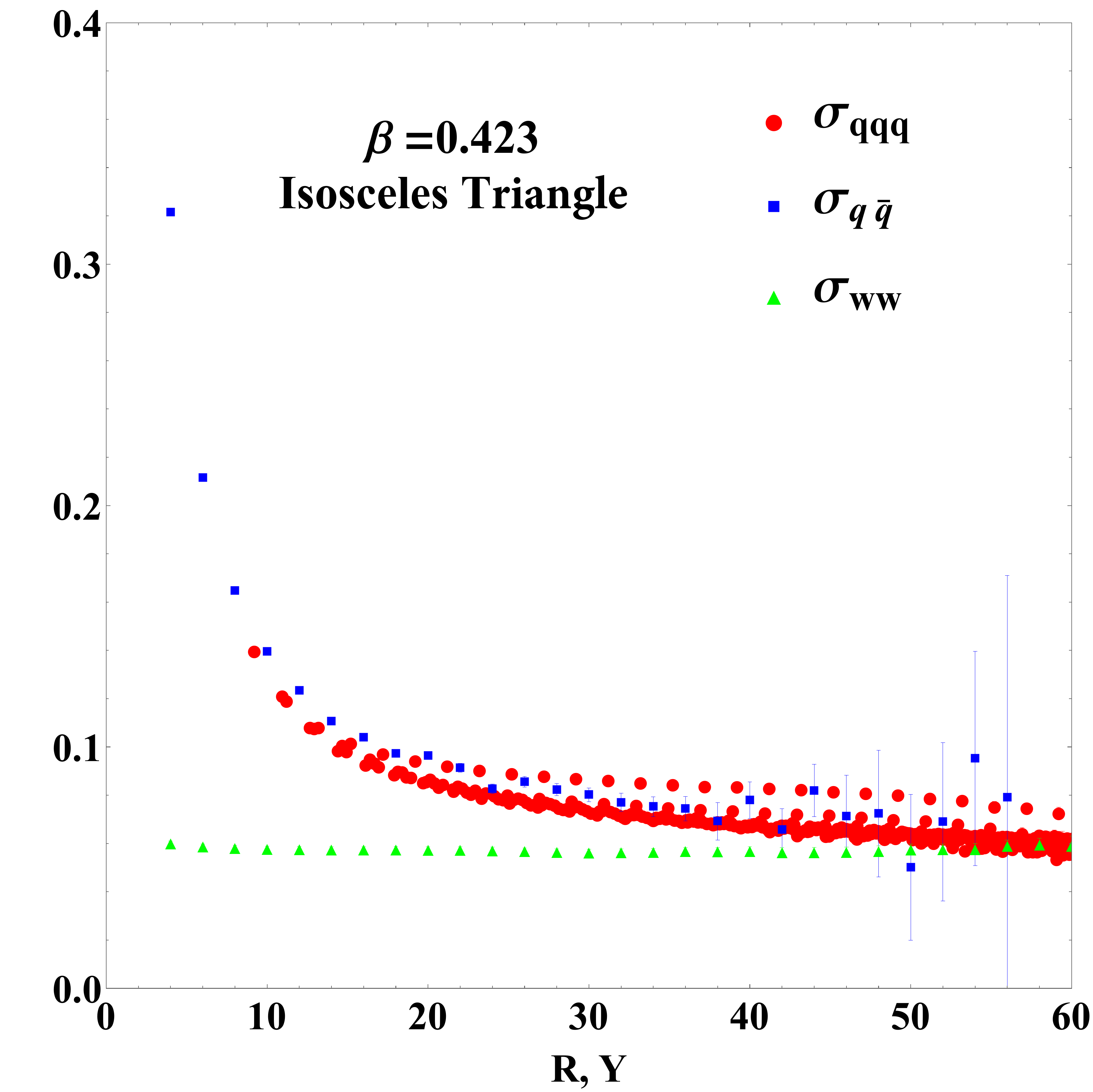}}
\caption{Same as Fig.~\ref{fig:128_Iso_0.41} at $\beta=0.423$.}\label{fig:128_Iso_0.423}
\end{figure}

First we studied the dependence of $\Gamma_3^{\rm f}$ on the geometry,
considering $\Delta$ and $Y$ laws. In Fig.~\ref{fig:scaled_Gamma3f} we see that
if we consider $\sigma$ to be proportional to $\sigmaww$ we get a
reasonable collapse for all $\beta$ values except the largest one
($\beta = 0.424$), which might be too close to the critical point for our
lattice size, $L = 128$. Still this does not allow us to discriminate between
the two laws.

We turned therefore to fits with the two laws of the three-point correlation
function for small ($\sigmaww Y < 2$) triangles, excluding those
having an angle larger than $2 \pi / 3$.
The results of these fits for $\beta = 0.423$ are shown in
Fig.~\ref{fig:fits_Gamma3f}.
In this case the fit with the $\Delta$ law gives
$A = 2.319(14),\ \sigma = 0.0631(4),\ \eta=0.503(4),\ \chiR = 28.6$,
while the fit with the $Y$ law gives
$A = 2.502(17),\ \sigma = 0.0573(3),\ \eta=0.506(5),\ \chiR = 4.67$. 
For all values of $\beta$ the $Y$ law performed better than the $\Delta$ law
(giving smaller $\chiR$), the difference getting more and more clear for
larger values of $\beta$.

To extract $\sigmaB$ we followed a procedure similar to the one used for
$\sigmaqq$. We supposed, according to~(\ref{corrfunc3_fund_largeN_small_beta})
and~(\ref{YandLambda_LargeN}), the decay law
\begin{equation}
\label{Gamma3_dlaw}
\Gamma_3^{\rm f}(R)=B\, \frac{e^{-\sigmaB\, R}}{R^\eta} \;.
\end{equation}
In this subsection $R$ is a distance parameter depending on the geometry, that
could be $\Delta$ or $Y$, respectively for the $\Delta$ and $Y$ laws, given
in Eqs.~(\ref{delta_law}) and~(\ref{Y_law}). In this proposed fitting function
we excluded the second term, corresponding to the $\Lambda$ law, since, for
the size of triangles considered in the fitting procedure, we could
hardly distinguish it from the first one, corresponding to the
$Y$ law~\footnote{Recently, a new method was suggested to reveal the
distinction between the $\Delta$ and $Y$ laws, based on the use of
hyperspherical three-body variables~\cite{Leech:2018lqu}.}

We calculated the following ratio
\begin{equation}
\label{sigmacalcc}
\sigmaB^{\mathrm{eff}} = -\dfrac{1}{R_1 - R_2} \ln \left[\dfrac{\Gamma_3^{\rm f}(R_1)}{\Gamma_3^{\rm f}(R_2)}\right] 
\end{equation}
for the pairs of triangles $(b+2, h+2)$ and $(b,h)$ for the isosceles geometry
and $(a_1 + 2, a_2 + 2)$ and $(a_1, a_2)$ for the right-angled geometry,
where $R_1$ and $R_2$ are the distance parameters of the two triangles.
The ratio is equal to
\begin{equation}
\label{gamma_eqq}
\sigmaB^{\mathrm{eff}} = -\dfrac{1}{R_1 - R_2} \ln \left[\dfrac{\Gamma_3^{\rm f}(R_1)}{\Gamma_3^{\rm f}(R_2)}\right] =
 \sigmaB + \dfrac{\eta}{R_1-R_2} \ln \left[ \dfrac{R_1}{R_2}\right] \;,
\end{equation}
where, for sufficiently large distances we can assume $R_1 \sim R_2\sim R$ and
get 
\begin{equation}
\label{gamma_3_sugamma3}
\sigmaB^{\mathrm{eff}} =  -\dfrac{1}{R_1 - R_2} \ln \left[\frac{\Gamma_3^{\rm f}(R_1)}{\Gamma_3^{\rm f}(R_2)}\right]  \overset{R \gg 1}{\simeq}
 \sigmaB + \dfrac{\eta}{R} \;.
\end{equation}

It turned out that for part of the triangle pairs $\sigmaB^{\mathrm{eff}}$ and
its jackknife error estimate could not be extracted reliably, due to one of
the correlation functions being too close to zero.
We removed from the study all the triangle pairs in which for at least one of
the triangles, at least one of the jackknife samples gave negative correlation.
The actual number of the triangle pairs for which the extraction of
$\sigmaB^{\mathrm{eff}}$ was possible, strongly depends on the value of $\beta$;
for example, for the isosceles geometry we had 274 triangle pairs
for $\beta=0.41$, 558 for $\beta=0.42$ and 783 for $\beta=0.423$.

After extracting $\sigmaB^{\mathrm{eff}}$, we plotted it directly against the
half-perimeter $\Delta$ and against the sum $Y$ of the distances of the triangle
vertices from the Fermat-Torricelli point. We overlapped these plots
with the plots of $\sigmaqqeff$ and $\sigmaww^{\mathrm{eff}}$ versus $R$
(Figs.~\ref{fig:128_Iso_0.41}-\ref{fig:128_Iso_0.423}).
We see that on the plots for the $\Delta$ law the values of
$\sigmaB^{\mathrm{eff}}$ fail to collapse into a single line, while the collapse
is much better for the $Y$ law for all $\beta$ values we studied. The
residual spread of the points can be at least partially explained by different
triangle pairs having different $R_1 - R_2$ values, which are not distinguished
on these plots. Another observation that supports the $Y$ law is that the
collapse line for the $\sigmaB^{\mathrm{eff}}$ closely matches the line
of $\sigmaqq^{\mathrm{eff}}$, which suggests that not only the sigma values
entering the two- and the three-point correlation are the same if we consider
the $Y$ law, but also that the parameters $\eta$ are similar in these cases.

\begin{figure}[tb]
\centering
{\includegraphics[width=.48\textwidth]{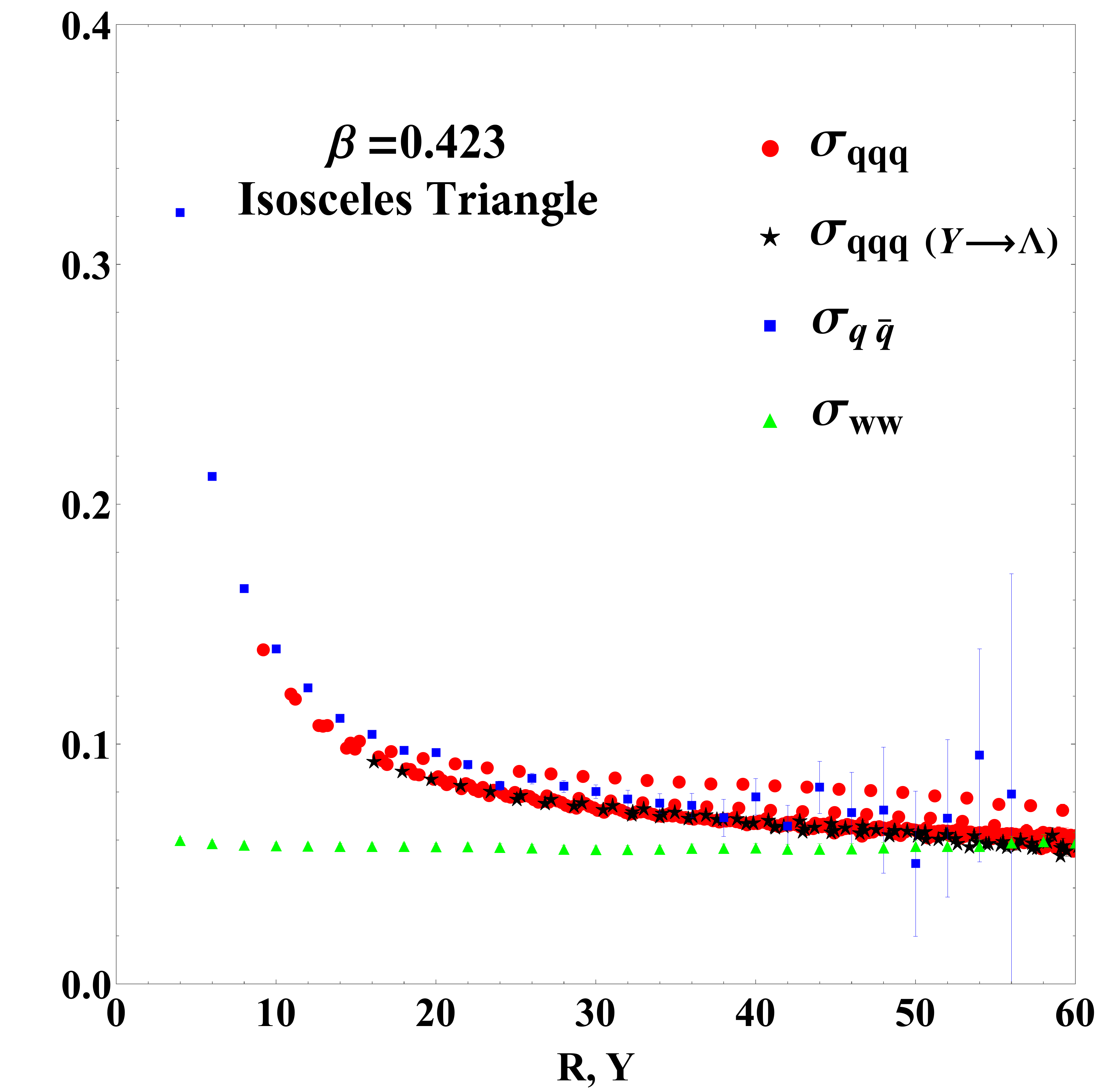}}
\caption{Same as Fig.~\ref{fig:128_Iso_0.423} (right) with data points for
  triangles having an angle larger than $2 \pi / 3$
  explicitly marked as $\sigmaB (Y \rightarrow \Lambda)$.}
\label{fig:128_YtoV}
\end{figure}

It is worth mentioning that in our study the results of the extraction of
$\sigmaB^{\mathrm{eff}}$ are compared for triangles that have strongly different
geometries: there are triangles that have similar $Y$ distance, but some of
them can have a small base and a large height, while others can have small
heights and large bases. In particular, we did not exclude triangles with
angles larger than $2 \pi / 3$ from the study of $\sigmaB^{\mathrm{eff}}$, despite
the fact that for them the Fermat-Torricelli point coincides with one of the
vertices, thus leading to a different dependence of $Y$ on $h$ and $b$. The
fact that even these ``extreme'' triangles obey the $Y$ law is explicitly
demonstrated in Fig.~\ref{fig:128_YtoV} (instead, the most outlying data points
turn out to be the ones with smallest triangle base), 
where the caption $\sigmaB (Y \rightarrow \Lambda)$ implies that for such
triangles 
the Fermat-Torricelli point coincides with one of the vertices turning the $Y$
law into the $\Lambda$ law. As can be seen from
Figs.~\ref{fig:128_Iso_0.41}-\ref{fig:128_Iso_0.423} these differences in
geometry give negligible corrections to the values of $\sigmaB^{\mathrm{eff}}$ up
to $\beta = 0.423$, providing us with an additional point in support of the
$Y$ law. 

\subsection{Extraction of critical index $\nu$ from the scaling of the
  two-point string tension} \label{nu_from_sigma}

\begin{figure}[tb]
\centering
	\includegraphics[width=0.5\textwidth]{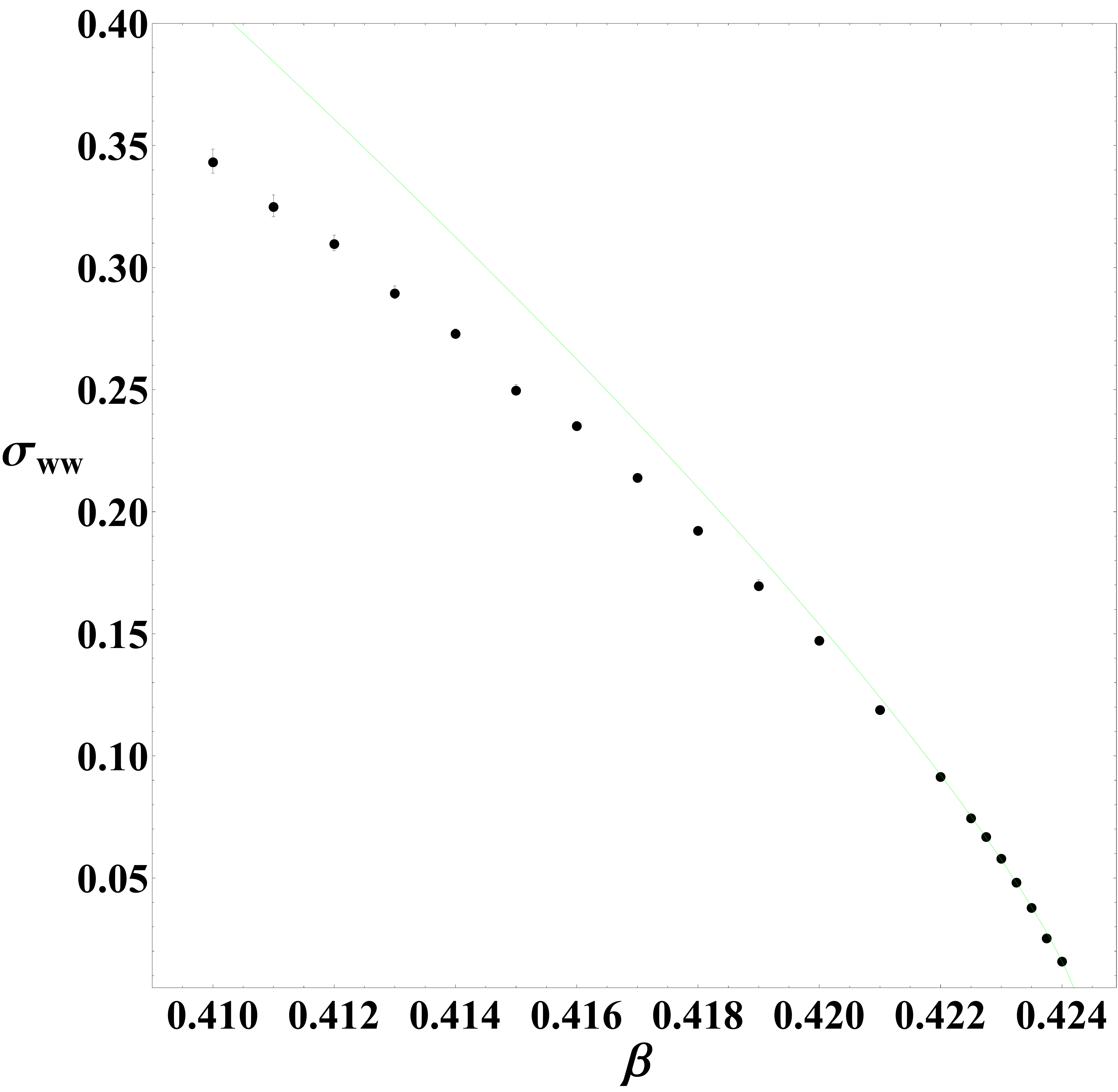}
	\caption{$\sigma_{\rm ww}$ versus $\beta$. The solid green line gives
          the result of the fit with the analytic form
          $A{(\beta_c-\beta)^{\nu}}$.}
        \label{fig:sigma_wall_vs_beta}
\end{figure}

We used the values of the string tension obtained from the wall-wall
correlation function close to the critical point to extract the critical
index $\nu$.

The values of $\sigma_{\rm ww}$, as well as the result of the fit
in the region ${0.422 < \beta < \beta_{\rm c}}$ with the scaling function
\begin{equation}
\sigma = A (\beta_{\rm c} - \beta)^{\nu} \ ,
\label{sigma_crit_scaling}
\end{equation}
\begin{equation*}
  A = 12.5(2.8),\ \ \ \beta_c = 0.424255(34),\ \ \ \nu = 0.806(37), \ \ \
  \chi^2_r \approx 3.69\ ,
\end{equation*}
are shown in the Fig.~\ref{fig:sigma_wall_vs_beta}. The value of the critical
index $\nu$ obtained in this way is compatible with both the critical
index $\nu = 5/6$ of the three-state Potts model and with our previous
estimate in~(\ref{beta_pc_scaling}). We note, however, that the scaling region
in this case is extremely narrow, and the value of $\nu$ is quite sensitive
to the inclusion of the points outside this region.

\subsection{Adjoint correlations in the confinement phase}
\label{gamma_adj_conf}

\begin{figure}[tb]
\centering
\includegraphics[width=0.48\textwidth]{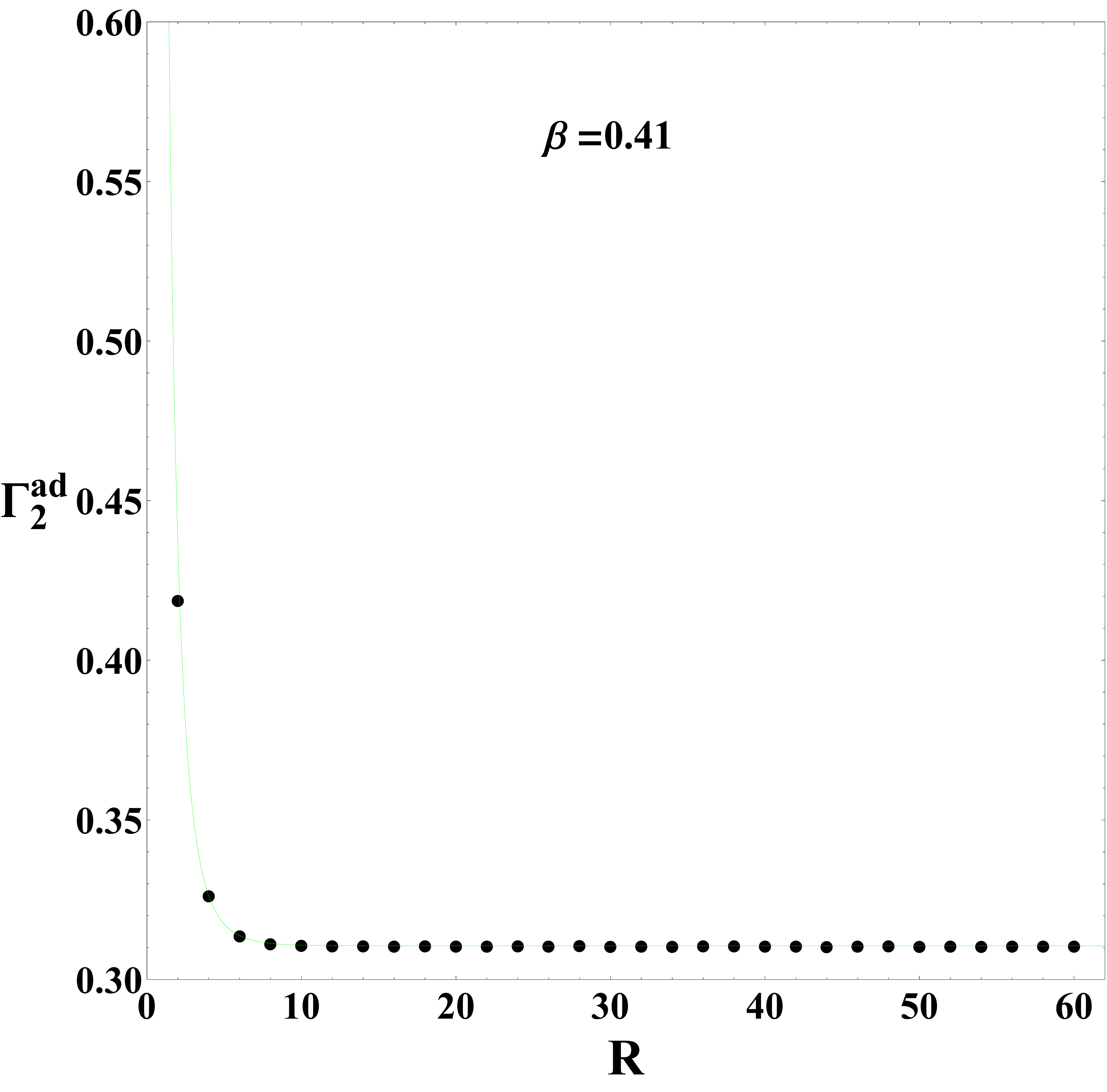}
\includegraphics[width=0.48\textwidth]{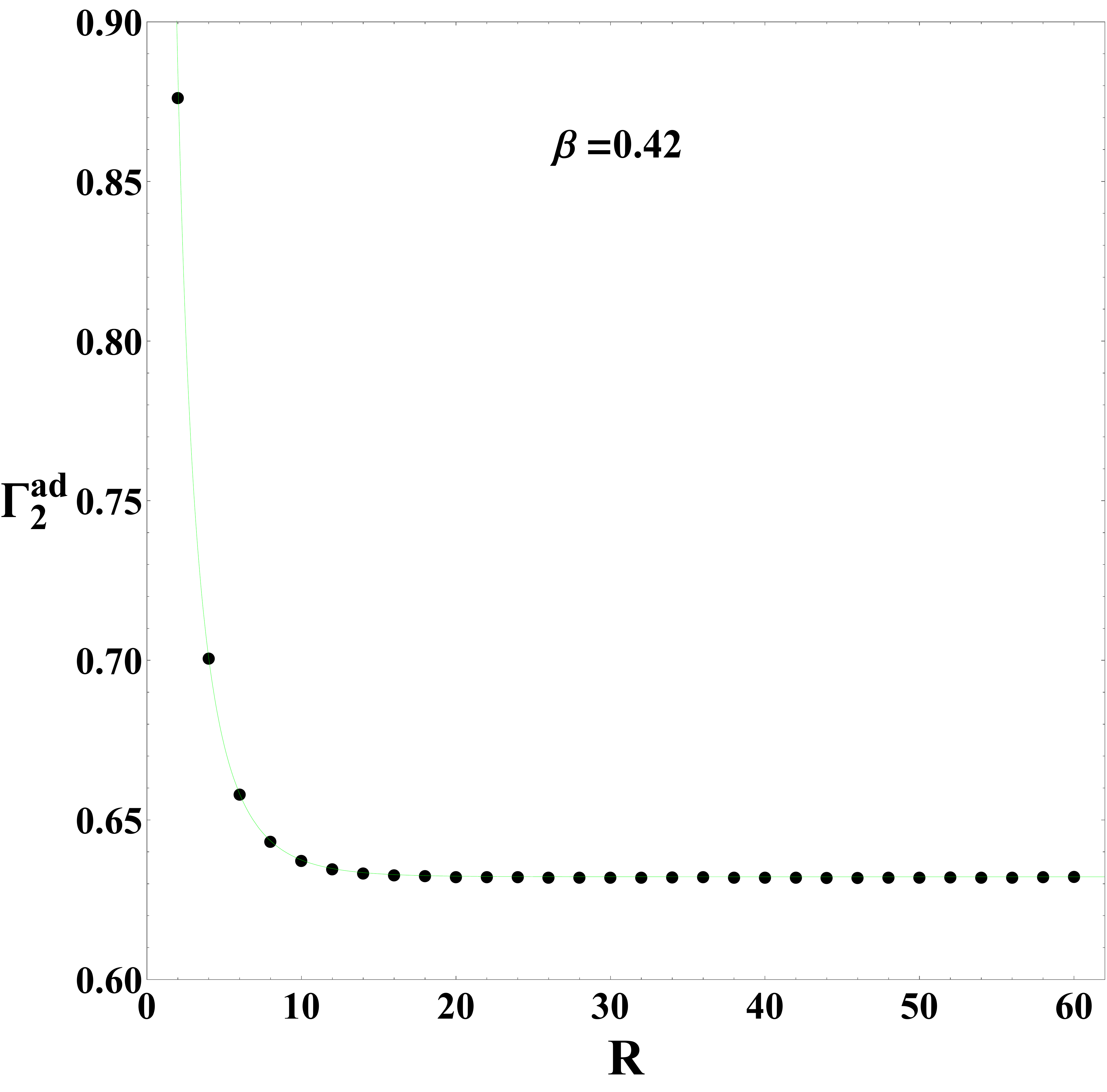}
\caption{$\Gamma_2^{{\rm ad}}$ versus $R$ at $\beta = 0.41$ (left) and
  $\beta = 0.42$ (right). The solid green line gives the result of the fit
  with the function in Eq.~\eqref{corrfunc2_adj_small_beta}.}
\label{fig:Gamma_2_AD_versus_distance}
\end{figure}

We have performed measurements of the two- and three-point correlation functions
in the adjoint representation, defined in Eqs.~(\ref{corrfunc2_adj_def})
and~(\ref{corrfunc3_adj_def}), at some values of $\beta$ below the critical
one.

Following formulae~(\ref{corrfunc2_adj_largeN_small_beta})
and~(\ref{corrfunc3_adj_largeN_small_beta}) and replacing in them
the massive Green function $G(\beta, R)$ with its asymptotic behaviour, 
we got the following models:
\begin{align}
 \label{corrfunc2_adj_small_beta}
 \Gamma_2^{{\rm ad}}(R) &= M_{\rm ad}^2 + A \frac{\exp [-2 \sigma R]}{R^{2\eta}} \;.\\
\label{corrfunc3_adj_small_beta}
 \Gamma_3^{{\rm ad}}(\{ x_i \} ) &=
A^3 \prod_{i=1}^3 \frac {\exp{ [- \sigma |x_i - x_{i+1}|] }}{|x_i - x_{i+1}|^\eta}  \nonumber \\
   &\phantom{= X} + M_{\rm ad} A^2 \sum_{i=1}^3 \frac {\exp{ [- 2 \sigma |x_i - x_{i+1}|] }}{|x_i - x_{i+1}|^{2\eta}} + M_{\rm ad}^3 \;.
\end{align}
Note that $\Gamma_3^{{\rm ad}}(\{ x_i \})$ exhibits the $\Delta$-law decay
after subtracting terms proportional to (powers of) the magnetization.

\begin{figure}[tb]
\centering
\includegraphics[width=0.48\textwidth]{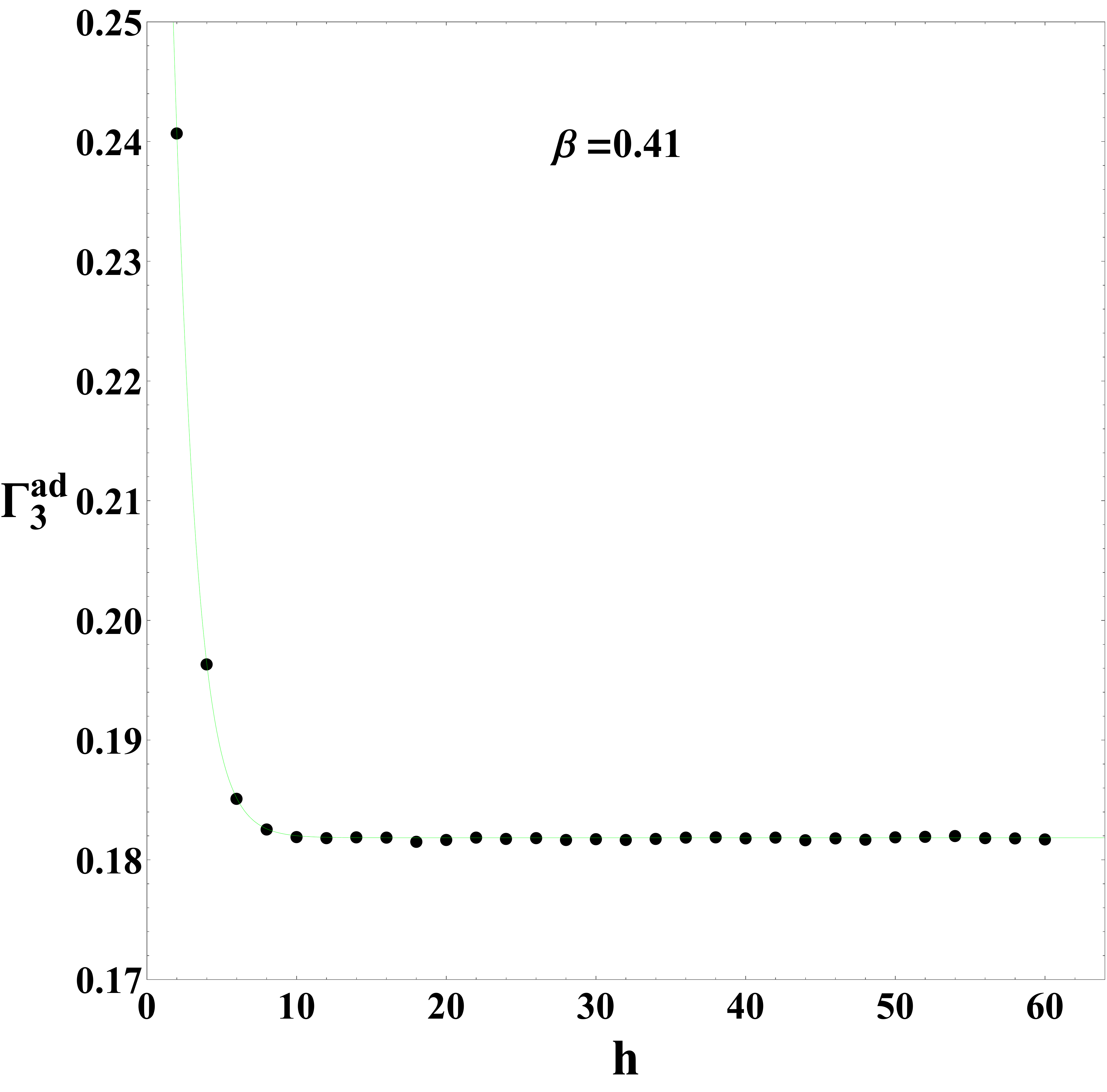}
\includegraphics[width=0.48\textwidth]{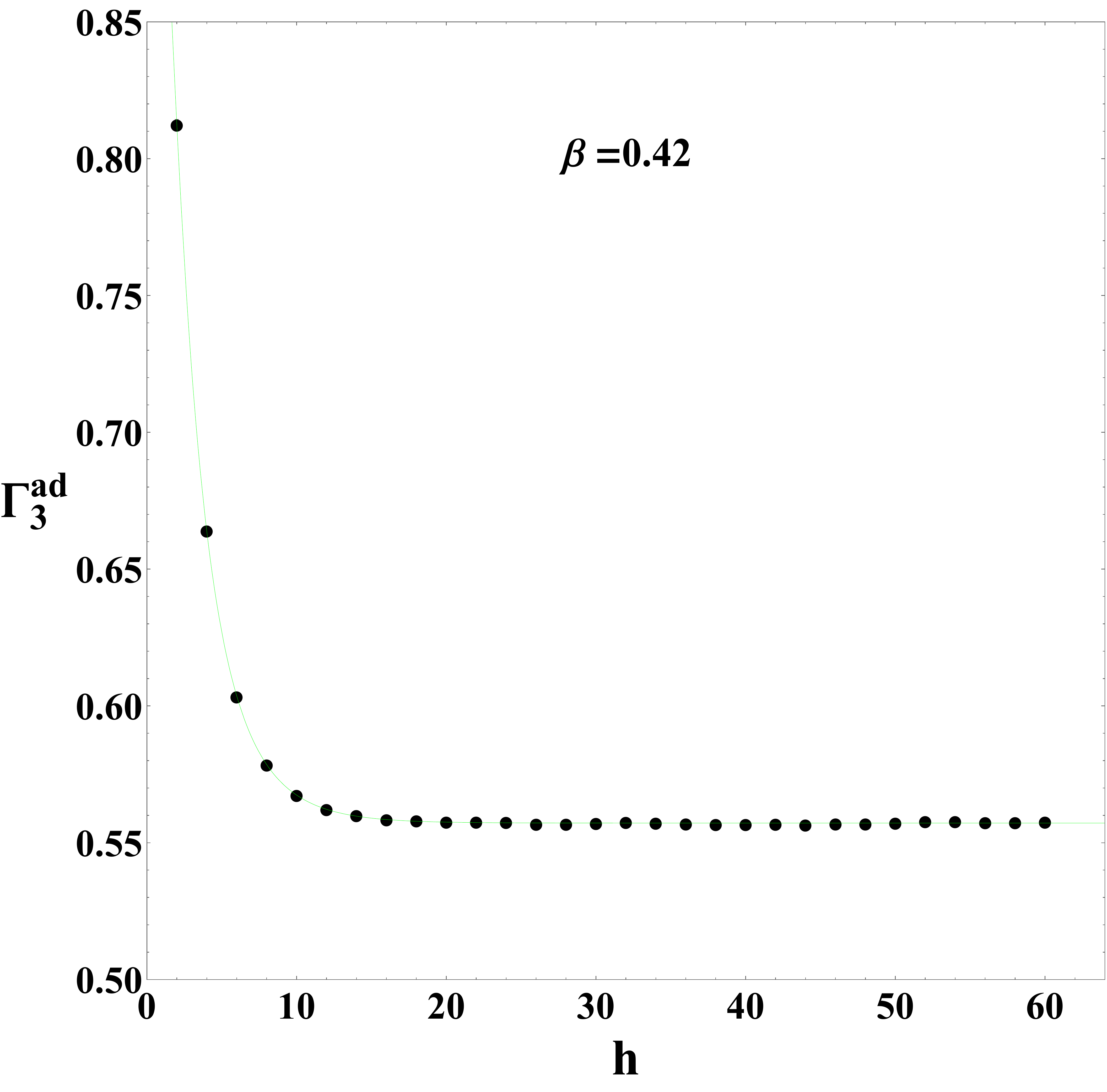}
\caption{$\Gamma_3^{{\rm ad}}$ versus the height of triangles with base $b=4$
  and $\beta = 0.41$ (left) and $\beta = 0.42$ (right). The solid green
  line gives the result of the fit with the function in
  Eq.~\eqref{corrfunc3_adj_small_beta}.}
\label{fig:Gamma_3_AD_versus_height}
\end{figure}

The results of the fitting of the adjoint correlations to the models in
Eqs.~(\ref{corrfunc2_adj_small_beta}) and~(\ref{corrfunc3_adj_small_beta})
are given in Table~\ref{table:fits_Gamma_3_adj_below_crit} (see also
Figs.~\ref{fig:Gamma_2_AD_versus_distance}
and~\ref{fig:Gamma_3_AD_versus_height}).

\begin{table}[htb]
\centering
\begin{tabular}{ | c | c | c | c | c | c | c | }
    \hline \hline
$\beta$  & $b$ & $M$           & $A$         & $\sigma$   & $\eta$     & $\chiR$ \\
\hline \hline
  0.41   &     & 0.557288(39)  & 0.743(12)   & 0.284(34)  & 0.57(10)   & 0.11  \\
\cline{2-7}
         &  2  & 0.55102(39)   & 0.8567(90)  & 0.308(16)  & 0.419(53)  & 0.094  \\
         &  4  & 0.55637(15)   & 0.777(22)   & 0.348(21)  & 0.278(80)  & 0.21   \\
         &  6  & 0.55716(10)   & 0.742(61)   & 0.330(33)  & 0.32(15)   & 0.27   \\
         &  8  & 0.557269(75)  & 0.65(11)    & 0.354(43)  & 0.17(22)   & 0.25   \\
\hline \hline
  0.415  &     & 0.645366(83)  & 0.792(18)   & 0.235(26)  & 0.499(79)  & 0.023 \\
\cline{2-7}
	 &  2  & 0.63984(75)   & 0.878(13)   & 0.244(22)  & 0.400(73)  & 0.047 \\
         &  4  & 0.64379(29)   & 0.825(30)   & 0.247(24)  & 0.376(94)  & 0.051 \\
         &  6  & 0.64501(21)   & 0.791(67)   & 0.255(30)  & 0.34(14)   & 0.036 \\
         &  8  & 0.64526(17)   & 0.77(13)    & 0.250(40)  & 0.36(22)   & 0.062 \\
\hline \hline
  0.42   &     & 0.79511(12)   & 0.870(16)   & 0.134(16)  & 0.531(53)  & 0.0078 \\
\cline{2-7}
	 &  2  & 0.79391(79)   & 0.907(11)   & 0.143(13)  & 0.452(48)  & 0.048 \\
         &  4  & 0.79353(24)   & 0.880(24)   & 0.140(14)  & 0.450(59)  & 0.064 \\
         &  6  & 0.79433(30)   & 0.885(50)   & 0.136(16)  & 0.474(83)  & 0.040 \\
         &  8  & 0.79487(26)   & 0.90(10)    & 0.132(23)  & 0.51(14)   & 0.040 \\
\hline \hline
\end{tabular}
\caption{Parameters extracted from the fits of the $\Gamma_2^{\mathrm{ad}}$ and
  $\Gamma_3^{\mathrm{ad}}$
 at some given $\beta < \beta_{\mathrm{c}}$. For each value of $\beta$ the
 first line contains the result of the fit of $\Gamma_2^{\mathrm{ad}}$ to
 \eqref{corrfunc2_adj_small_beta},
 and the next lines contain the result of the fit of the values of
 $\Gamma_3^{\mathrm{ad}}$
 obtained for the isosceles triangles with fixed base $b$ to \eqref{corrfunc3_adj_small_beta}.}
\label{table:fits_Gamma_3_adj_below_crit}
\end{table}

Unusually low $\chi^2_r$ values in the fits arise due to treating the
measurements of the correlations at different distances as independent
despite being obtained from the same set of measurements. This fact makes
the error estimates of the fit parameters unreliable. Also, the estimation of
the $\eta$ value is inaccurate, since it describes short-range corrections
to the exponential decay, on which only a few points in the fitting range
have impact. This fact is especially visible from the covariance between
$\eta$ and $\sigma$ which is very close to $-1$.

Despite that, the fact that the parameters $\sigma$ and $M$ show some degree of
stability when going from the description of the two-point correlation to the
three-point one, and also the compatibility of the results for $\sigma$ with
the $\sigmaqq$ and $\sigmaww$ values given in Table~\ref{table:128_sigma2_new},
support the validity of the suggested descriptions for the adjoint correlations.

We would like to stress that the values of $M_{\rm ad}$ extracted from the
fits do agree with direct measurements of this quantity as defined in
Eq.~(\ref{magnetization_ad_def}). 

\section{Correlation functions in deconfinement phase}

Using the same approach adopted in the subsection~\ref{gamma_adj_conf} for
the description of adjoint correlations in the confinement phase, 
we get the following models for the correlations in the deconfinement phase 
from Eqs.~(\ref{corrfunc2_fund_largeN_large_beta})-(\ref{corrfunc3_adj_largeN_large_beta}):
\begin{align}
\label{corrfunc2_fund_large_beta}
\Gamma_2^{{\rm f}}(R) &= M_f^2 \exp \left[\alpha \dfrac{\exp[-m R]}{R^{\eta}}\right]
\ , \\
\label{corrfunc3_fund_large_beta}
\Gamma_3^{{\rm f}}(\beta , \{ x_i \} )  &=
 M_f^3 \exp \left[ \alpha \sum_{i=1}^3 \dfrac{\exp[-m |x_i - x_{i+1}|]}{|x_i - x_{i+1}|^{\eta}} \right]  \ , \\
\label{corrfunc2_adj_large_beta}
\Gamma_2^{{\rm ad}}(R) &= (M_{\rm ad} + 1)^2 \exp \left[4 \alpha \dfrac{\exp[-m R]}{R^{\eta}} \right] - 2 M_{\rm ad} - 1 \ , \\
\label{corrfunc3_adj_large_beta}
\Gamma_3^{{\rm ad}}(\{ x_i \} ) &=
(M_{\rm ad} + 1)^3 \exp \left[ 4 \alpha \sum_{i=1}^3 \dfrac{\exp[-m |x_i - x_{i+1}|]}{|x_i - x_{i+1}|^{\eta}} \right]
 \nonumber \\
 &- \sum_{i=1}^3 (M_{\rm ad} + 1)^2 \exp \left[4 \alpha \dfrac{\exp[-m |x_i - x_{i+1}|]}{|x_i - x_{i+1}|^{\eta}} \right] + 3 M_{\rm ad} + 2 \ ,
\end{align}
where $m$ can be interpreted as the chromoelectric screening mass.

\begin{table}[htb]
\centering
\begin{tabular}{ | c | c | c | c | c | c | c | }
    \hline \hline
$\beta$  & $b$ & $M$           & $\alpha$    & $m$   & $\eta$     & $\chiR$ \\
\hline \hline
  0.425  &     & 1.15180(36)   & 0.4186(45)  & 0.1031(76) & 0.527(29) & 0.028 \\
\cline{2-7}
         &  2  & 1.17711(55)   & 0.2368(36)  & 0.1089(80) & 0.510(34)  & 0.051 \\
         &  4  & 1.16585(43)   & 0.2600(80)  & 0.1055(96) & 0.531(48)  & 0.028 \\
         &  6  & 1.16034(48)   & 0.287(17)   & 0.100(11)  & 0.572(70)  & 0.0093 \\
         &  8  & 1.15729(53)   & 0.314(31)   & 0.095(14)  & 0.61(10)   & 0.0036 \\
\hline \hline
  0.43   &     & 1.4678468(36) & 0.18067(16) & 0.3663(25) & 0.7054(77) & 0.00012 \\
\cline{2-7}
         &  2  & 1.474944(26)  & 0.11717(43) & 0.3905(62) & 0.574(21)  & 0.0029 \\
         &  4  & 1.469304(13)  & 0.1292(12)  & 0.3754(66) & 0.610(26)  & 0.0019 \\
         &  6  & 1.468205(12)  & 0.1490(32)  & 0.3541(85) & 0.728(40)  & 0.0014 \\
         &  8  & 1.467957(11)  & 0.1714(88)  & 0.348(13)  & 0.808(71)  & 0.0012 \\
\hline \hline
  0.435  &     & 1.582348(16)  & 0.1364(26)  & 0.560(77)  & 0.71(23)   & 0.0039 \\
\cline{2-7}
	 &  2  & 1.586050(80)  & 0.0908(24)  & 0.568(44)  & 0.58(14)   & 0.012 \\
         &  4  & 1.582808(45)  & 0.0997(67)  & 0.542(49)  & 0.67(18)   & 0.0077 \\
         &  6  & 1.582422(37)  & 0.122(20)   & 0.512(61)  & 0.87(27)   & 0.0047 \\
         &  8  & 1.582353(30)  & 0.194(61)   & 0.436(73)  & 1.37(37)   & 0.0052 \\
\hline \hline
  0.44   &     & 1.6581470(11) & 0.11482(22) & 0.722(13)  & 0.687(37)  & 0.0020 \\
\cline{2-7}
	 &  2  & 1.660580(28)  & 0.0744(10)  & 0.789(40)  & 0.34(12)   & 0.0077 \\
         &  4  & 1.658353(11)  & 0.0754(49)  & 0.780(63)  & 0.34(23)   & 0.0088 \\
         &  6  & 1.6581876(83) & 0.069(18)   & 0.89(13)   & 0 $\pm$ 0.55 & 0.0085 \\
         &  8  & 1.6581549(54) & 0.090(53)   & 0.80(17)   & 0.36(91)   & 0.0042 \\
\hline \hline
\end{tabular}
\caption{Parameters extracted from the fits of $\Gamma_2^{\mathrm{f}}$ and
  $\Gamma_3^{\mathrm{f}}$ at $\beta > \beta_{\mathrm{c}}$. For each value
  of $\beta$ the first line contains the result of the fit of
  $\Gamma_2^{\mathrm{f}}$ to \eqref{corrfunc2_fund_large_beta},
  and the next lines contain result of the fit of the values of
  $\Gamma_3^{\mathrm{f}}$ obtained for isosceles triangles with fixed base $b$
  to \eqref{corrfunc3_fund_large_beta}.}
\label{table:fits_Gamma_3_fund_above_crit}
\end{table}

\begin{table}[htb]
\centering
\begin{tabular}{ | c | c | c | c | c | c | c | }
    \hline \hline
$\beta$  & $b$ & $M$           & $\alpha$    & $m$   & $\eta$   & $\chiR$ \\
\hline \hline
  0.425  &     & 1.52849(31)   & 0.04314(83) & 0.086(16) & 1.097(63) & 0.0062 \\
\cline{2-7}
         &  2  & 1.54873(77)   & 0.03227(73) & 0.100(15) & 1.008(59) & 0.022 \\
         &  4  & 1.53420(47)   & 0.0348(16)  & 0.093(17) & 1.043(81) & 0.0087 \\
         &  6  & 1.53086(50)   & 0.0385(35)  & 0.084(20) & 1.11(12)  & 0.0046 \\
         &  8  & 1.52966(53)   & 0.0422(64)  & 0.079(23) & 1.17(16)  & 0.0049 \\
\hline \hline
  0.43   &     & 1.9422411(77) & 0.0385(23)  & 0.335(17) & 1.079(93) & 0.019 \\
\cline{2-7}
         &  2  & 1.95648(20)   & 0.02654(42) & 0.369(19) & 0.891(64)  & 0.018 \\
         &  4  & 1.94449(13)   & 0.0287(14)  & 0.345(32) & 0.97(13)   & 0.016 \\
         &  6  & 1.94279(11)   & 0.0343(38)  & 0.316(40) & 1.15(19)   & 0.021 \\
         &  8  & 1.94237(10)   & 0.0458(89)  & 0.275(44) & 1.43(24)   & 0.017 \\
\hline \hline
  0.435  &     & 2.159588(31)  & 0.03716(58) & 0.546(69) & 0.96(21)   & 0.026 \\
\cline{2-7}
	 &  2  & 2.17020(25)   & 0.02495(70) & 0.530(61) & 0.89(20)   & 0.051 \\
         &  4  & 2.16073(12)   & 0.0270(35)  & 0.51(11)  & 0.96(42)   & 0.033 \\
         &  6  & 2.15976(76)   & 0.033(12)   & 0.49(16)  & 1.16(69)   & 0.023 \\
         &  8  & 2.159588(60)  & 0.063(28)   & 0.38(12)  & 1.87(67)   & 0.028 \\
\hline \hline
  0.44   &     & 2.3244414(44) & 0.03673(14) & 0.727(26)  & 0.840(76)  & 0.00080 \\
\cline{2-7}
	 &  2  & 2.33305(39)   & 0.0234(13)  & 0.74(14)   & 0.64(43)   & 0.011 \\
         &  4  & 2.32509(12)   & 0.0236(25)  & 0.729(67)  & 0.65(25)   & 0.017 \\
         &  6  & 2.324522(75)  & 0.0245(64)  & 0.801(92)  & 0.45(38)   & 0.0087 \\
         &  8  & 2.324454(56)  & 0.060(48)   & 0.56(21)   & 1.7(29)    & 0.015 \\
\hline \hline
\end{tabular}
\caption{Parameters extracted from the fits of $\Gamma_2^{\mathrm{ad}}$ and
  $\Gamma_3^{\mathrm{ad}}$ at $\beta > \beta_{\mathrm{c}}$. For each value
  of $\beta$ the first line contains the result of the fit of
  $\Gamma_2^{\mathrm{ad}}$ to \eqref{corrfunc2_adj_large_beta},
  and the next lines contain result of the fit of the values of
  $\Gamma_3^{\mathrm{ad}}$ obtained for isosceles triangles with fixed base $b$
  to \eqref{corrfunc3_adj_large_beta}.}
\label{table:fits_Gamma_3_adj_above_crit}
\end{table}

\begin{figure}[tb]
\centering
\includegraphics[width=0.49\textwidth]{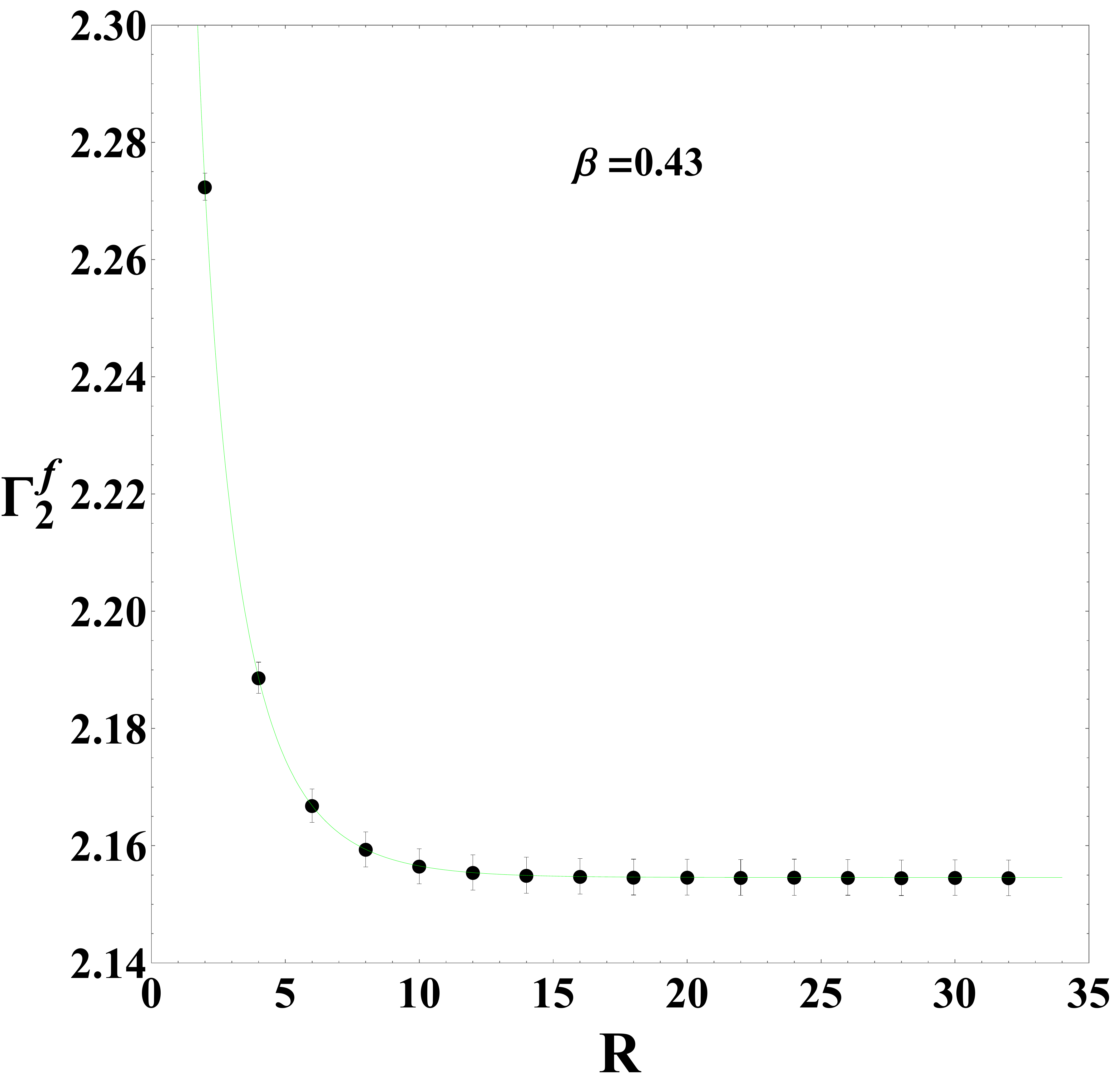}
\includegraphics[width=0.49\textwidth]{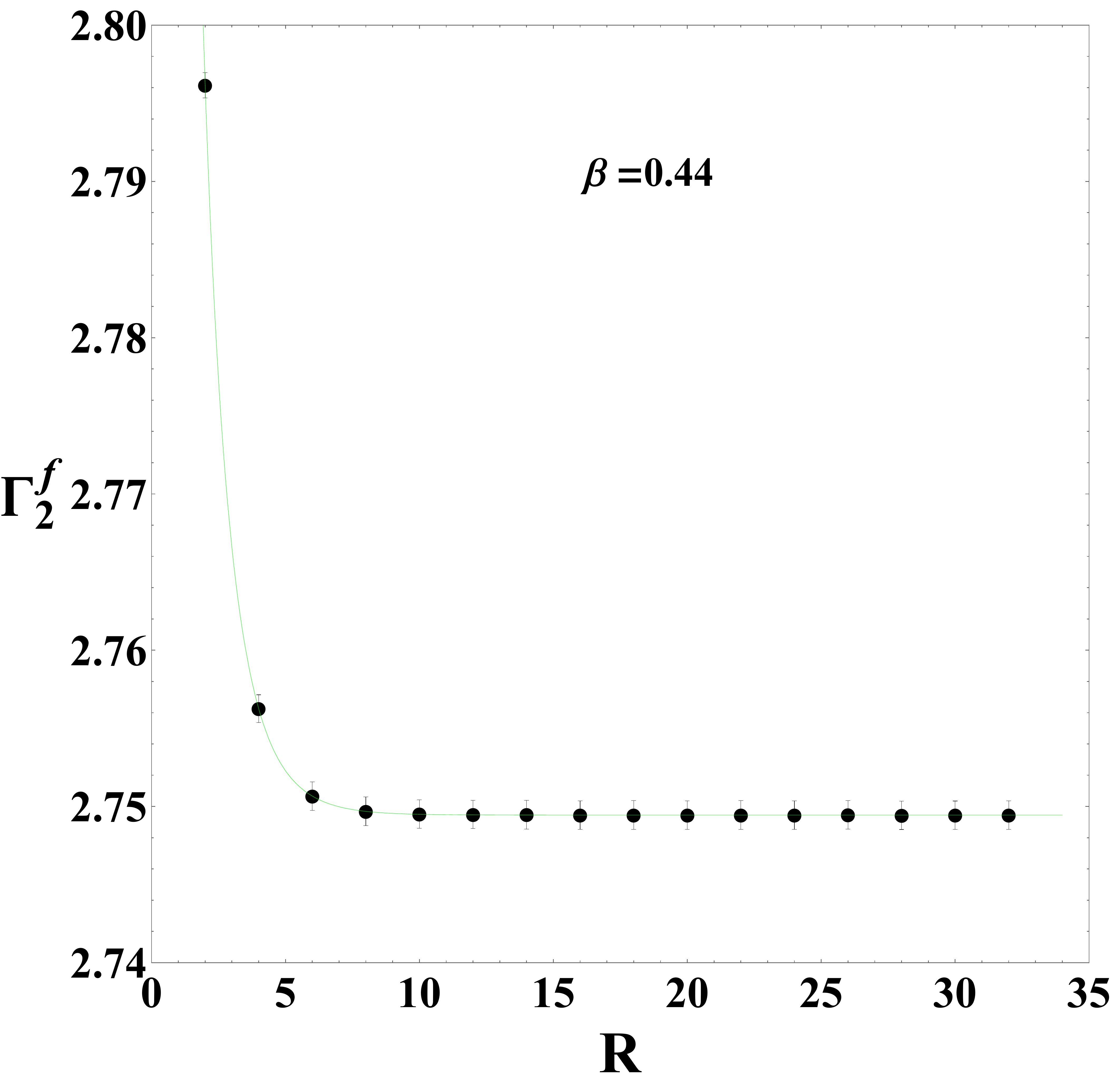}
\caption{$\Gamma_2^{{\rm f}}$ versus $R$ at $\beta = 0.43$ (left) and
  $\beta = 0.44$ (right). The solid green line gives the result of the fit
  with the function in Eq.~\eqref{corrfunc2_fund_large_beta}.}
\label{fig:Gamma_2_F_0.43_0.44_versus_distance}
\end{figure}

\begin{figure}[tb]
\centering
\includegraphics[width=0.49\textwidth]{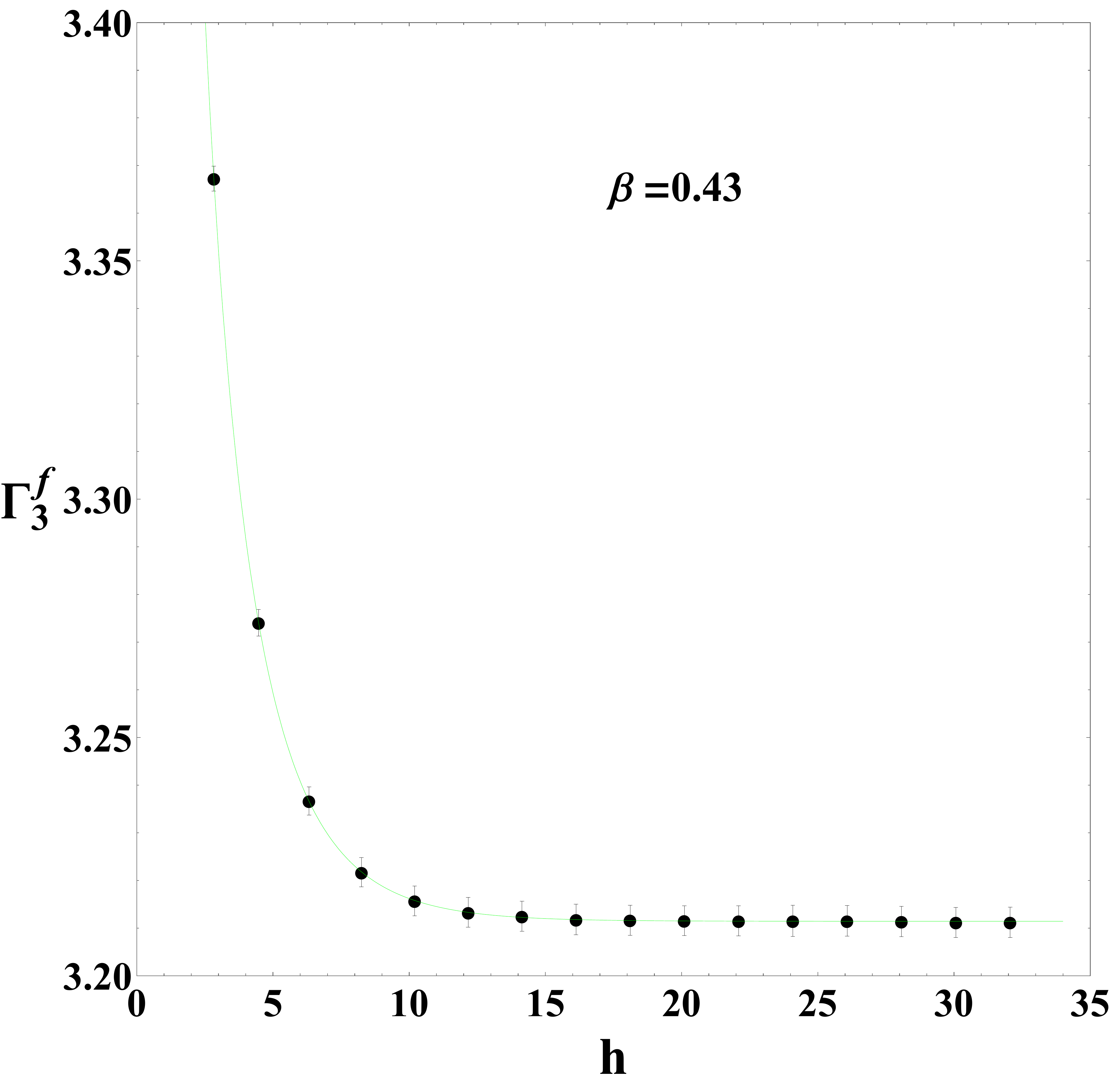}
\includegraphics[width=0.49\textwidth]{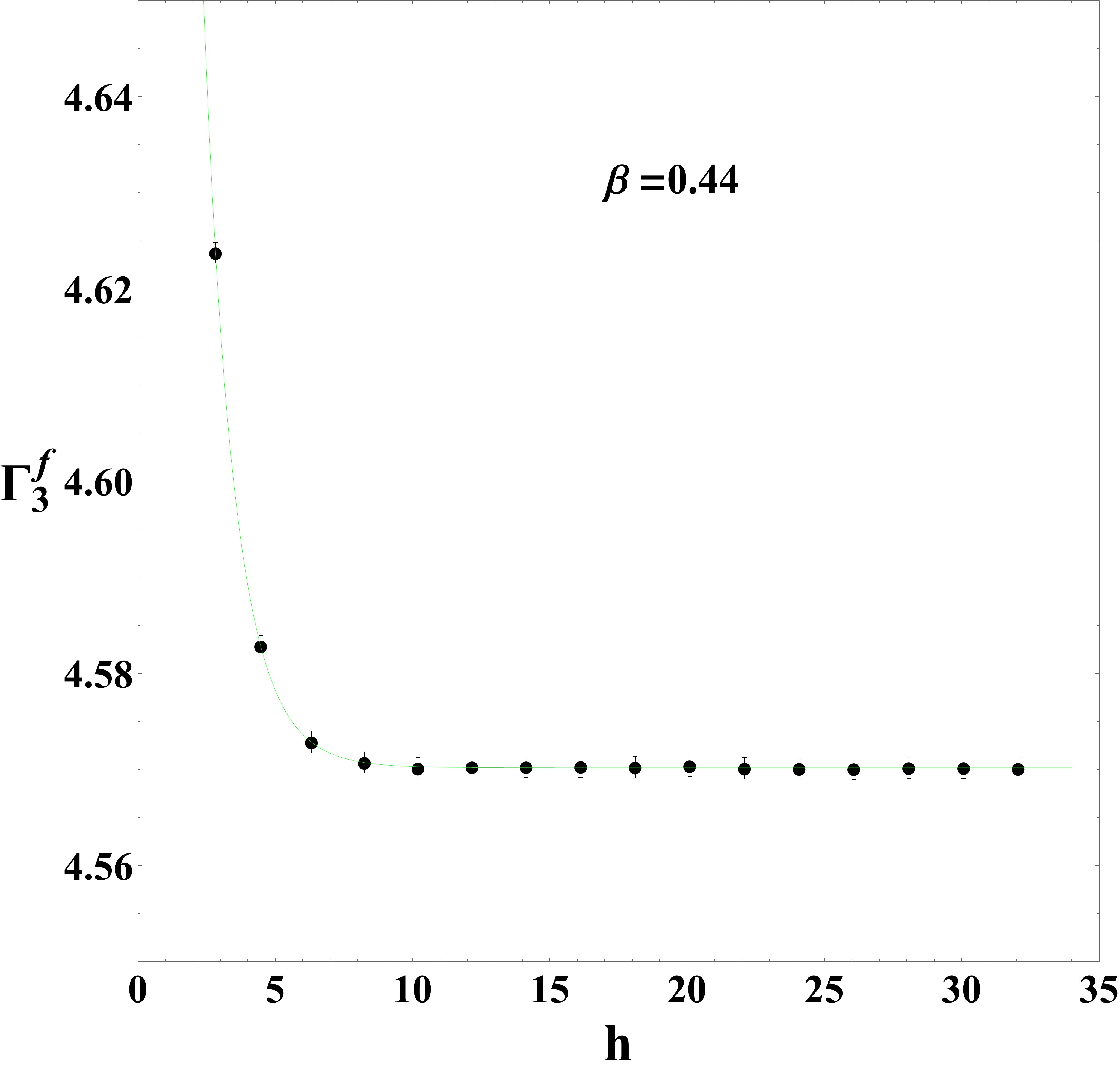}
\caption{$\Gamma_3^{{\rm ad}}$ versus the height of triangles with
  base $b=4$ at $\beta = 0.43$ (left) and $\beta = 0.44$ (right).
  The solid green line gives the result of the fit with the function in
  Eq.~\eqref{corrfunc3_fund_large_beta}.}
\label{fig:Gamma_3_F_0.43_0.44_versus_height}
\end{figure}

\begin{figure}[tb]
\centering
\includegraphics[width=0.48\textwidth]{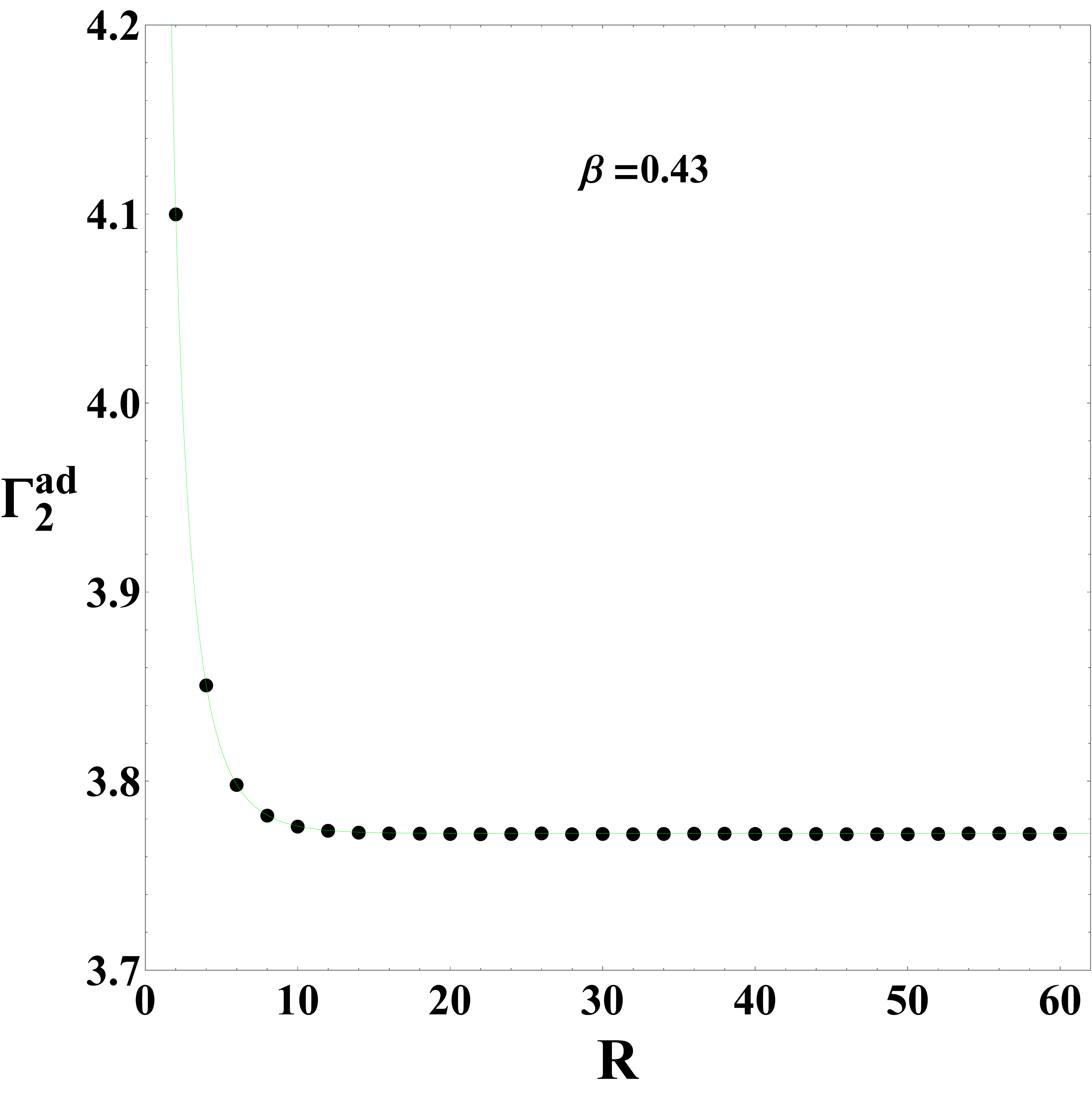}
\includegraphics[width=0.49\textwidth]{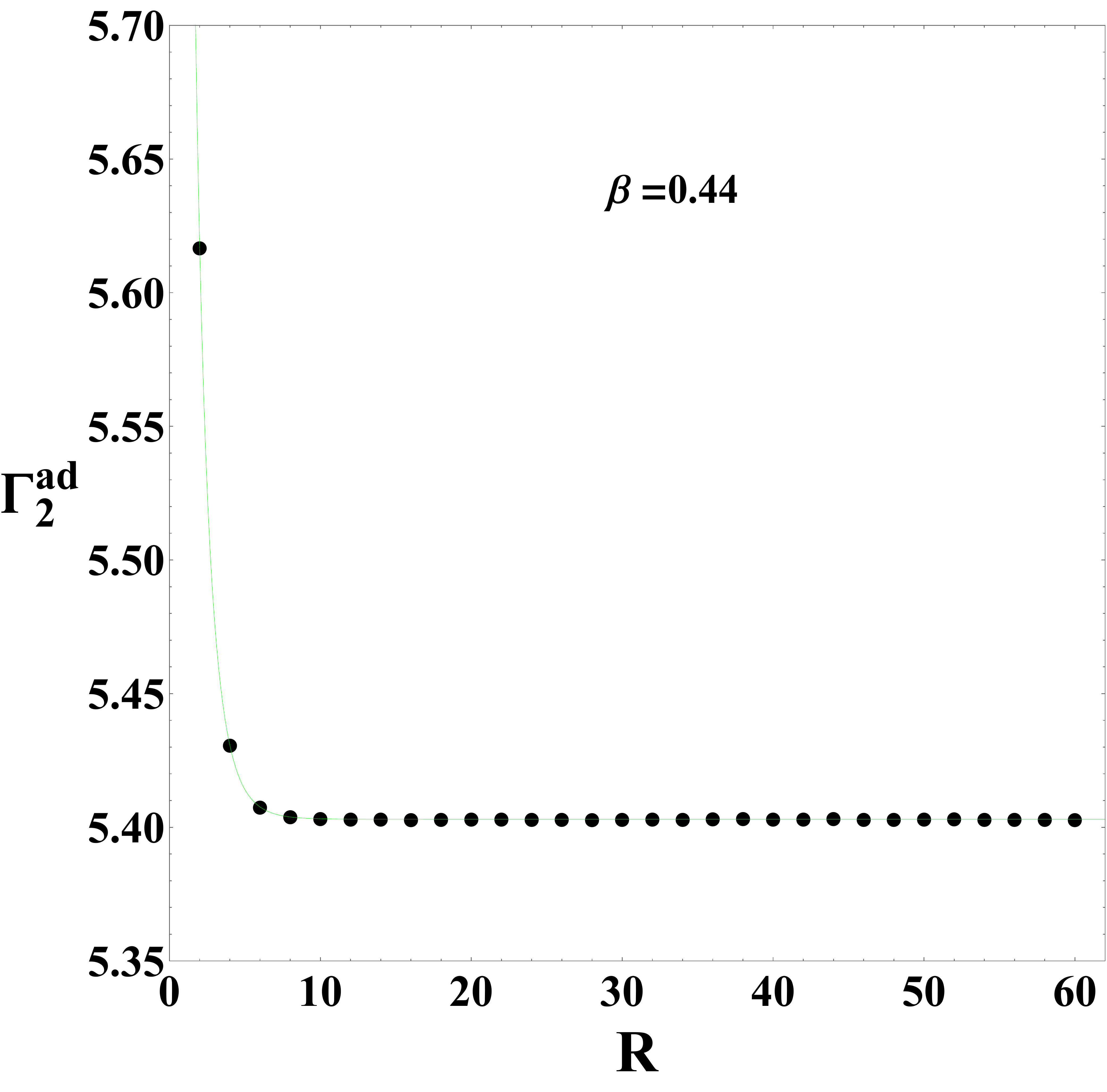}
\caption{$\Gamma_2^{{\rm ad}}$ versus $R$ at $\beta = 0.43$ (left) and
  $\beta = 0.44$ (right).
  The solid green line gives the result of the fit with the function in
  Eq.~\eqref{corrfunc2_adj_large_beta}.}
\label{fig:Gamma_2_AD_0.43_0.44_versus_distance}
\end{figure}

\begin{figure}[tb]
\centering
\includegraphics[width=0.48\textwidth]{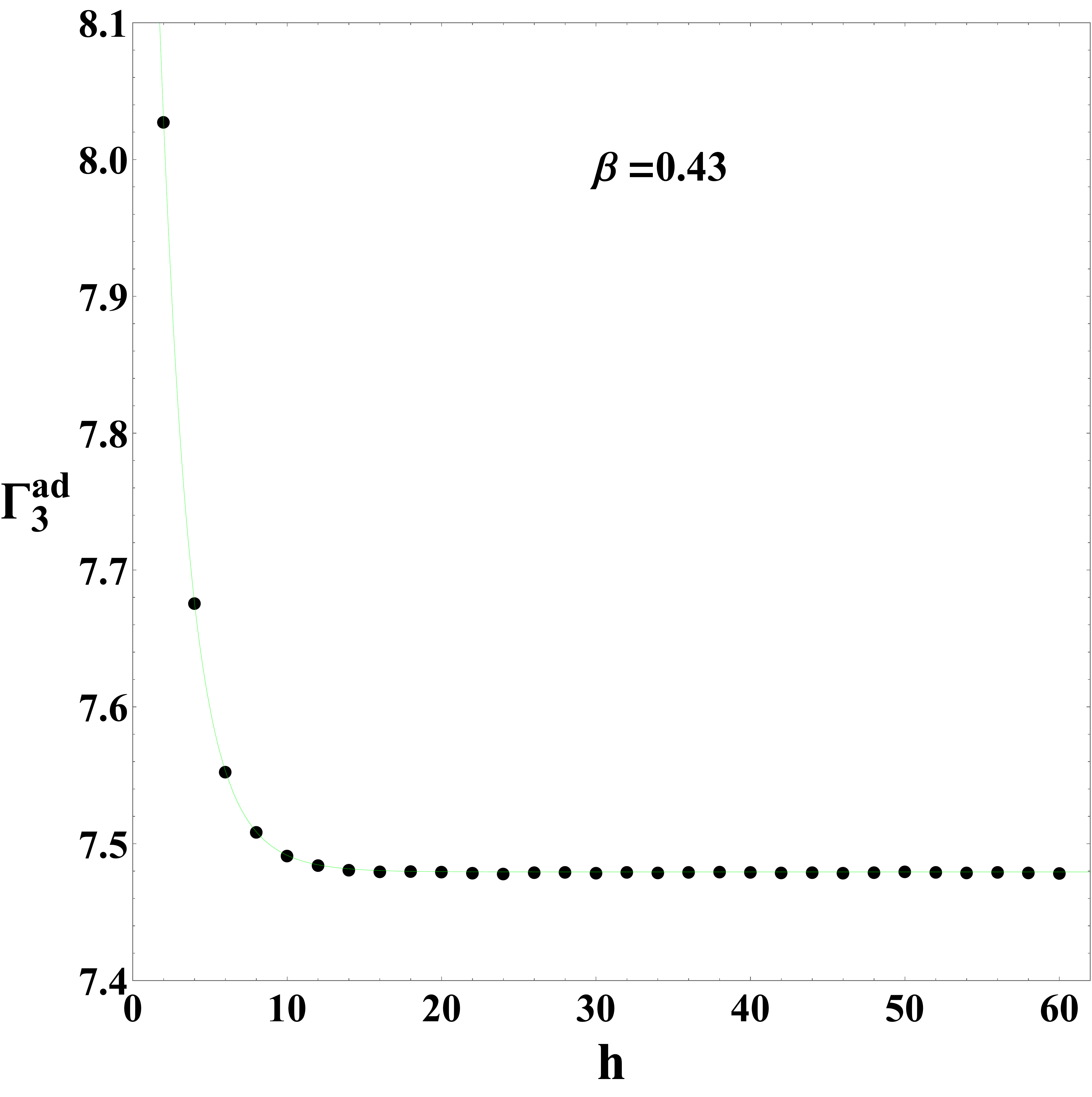}
\includegraphics[width=0.49\textwidth]{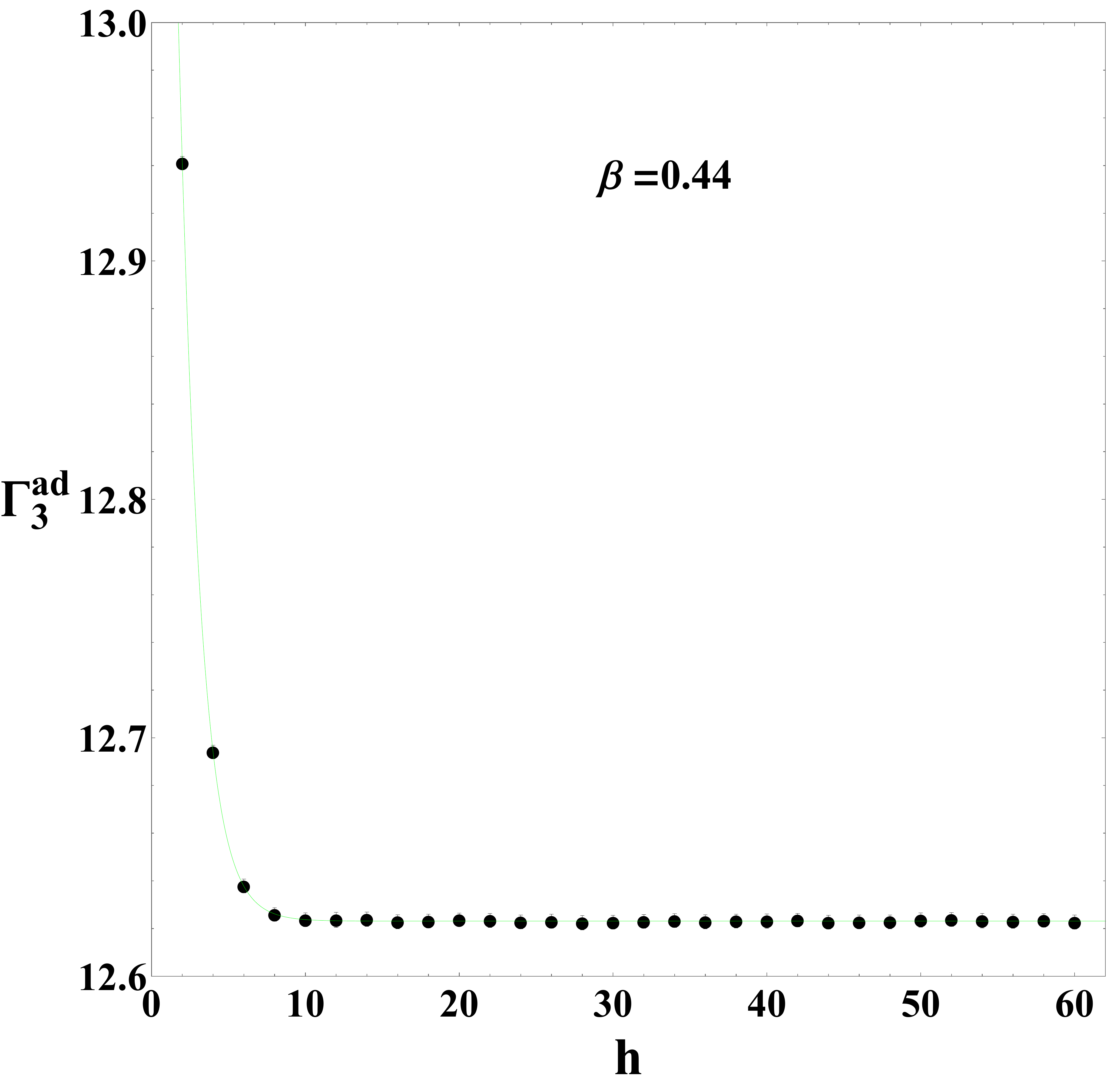}
\caption{$\Gamma_3^{{\rm ad}}$ versus the height of triangles with base $b=4$
  at $\beta = 0.43$ (left) and $\beta = 0.44$ (right).
  The solid green line gives the result of the fit with the function in
  Eq.~\eqref{corrfunc3_adj_large_beta}.}
\label{fig:Gamma_3_AD_0.43_0.44_versus_height_above}
\end{figure}

We used these formulae to describe the correlation functions measured above
$\beta_{\rm c}$.
The remarks given at the end of subsection \ref{gamma_adj_conf} apply here as
well. The results of the fits are gathered in
Table~\ref{table:fits_Gamma_3_fund_above_crit} for the fundamental correlations
(see also Figs.~\ref{fig:Gamma_2_F_0.43_0.44_versus_distance}
and~\ref{fig:Gamma_3_F_0.43_0.44_versus_height})
and in Table~\ref{table:fits_Gamma_3_adj_above_crit} for the adjoint ones
(see also Figs.~\ref{fig:Gamma_2_AD_0.43_0.44_versus_distance}
and~\ref{fig:Gamma_3_AD_0.43_0.44_versus_height_above}).
We see that the values of $M$ and $m$ extracted from the
fits for two-point and three-point correlations are compatible. Moreover, the
$m$ values extracted from the fits of the fundamental and adjoint
correlations are also compatible between themselves
(the same should not apply to the $M$ values, since they represent the average
magnetization in two different representations). This supports the validity of
the formulae~(\ref{corrfunc2_fund_large_beta})-(\ref{corrfunc3_adj_large_beta})
for the description of the correlation functions in the deconfinement phase.

Also in this case, we found that the values of $M_{\rm ad}$ extracted from the
fits do agree with direct measurements of this quantity as defined in
Eq.~(\ref{magnetization_ad_def}). 

In the deconfinement phase the value of $m$ becomes the inverse correlation
length for the connected part of the correlation, as can be seen from the
Taylor expansion of the outer exponent in Eq.~(\ref{corrfunc2_fund_large_beta}).
When we approach the critical point from above the value of $m$ should vanish
as 
\begin{equation}
m = A_{\rm dec} (\beta - \beta_{\rm c})^{\nu} \ .
\label{m_crit_scaling}
\end{equation}
The ratio $A_{\rm dec} / A$, where $A$ is the amplitude for the scaling of $\sigma$ in the deconfinement 
phase given by Eq.~(\ref{sigma_crit_scaling}), equals 2.657 for $2d$ $Z(3)$ universality class~\cite{universal_ratios}. 
Using this universal ratio and the parameters obtained in subsection~\ref{nu_from_sigma}, we have 
calculated the prediction that Eq.~(\ref{m_crit_scaling}) gives for the values of $m$. 
It turns out that for $\beta = 0.425$ the predicted value $m = 0.100(35)$ is in good agreement with 
the values of $m$ in Tables~\ref{table:fits_Gamma_3_fund_above_crit}, \ref{table:fits_Gamma_3_adj_above_crit},
while for larger values of $\beta$ this agreement becomes worse ($m = 0.52(15)$ for $\beta = 0.43$).
This might be explained by the scaling holding only in a narrow region around $\beta_{\rm c}$, 
similarly to Fig.~\ref{fig:sigma_wall_vs_beta}.

\section{Summary}

In this study we performed an extensive analysis of a two-dimensional
effective $SU(3)$ Polyakov loop model. Differently from other approaches
of the same kind, in our effective model the basic degrees of
freedom are traces of $SU(3)$ matrices and not $Z(3)$ spins. The
partition function is therefore integrated with a group-invariant measure.
The motivation for this choice is that it can help to catch some important  
features of the $(2+1)$-dimensional $SU(3)$ lattice gauge theory at finite
temperature that escape approaches based on the center degrees of freedom. For
example, one cannot define adjoint correlation functions in $Z(3)$ models. 
Our main goal was to examine the behaviour of the three-quark potential below
the critical point and to distinguish between possible scenarios for
the three-point correlation function decay: the $\Delta$ law and the $Y$ law.
Considering the triangle whose vertices are the positions of the three sources,
the $\Delta$ potential depends on the perimeter of this triangle, while
the $Y$ potential depends on the sum $Y$ of the distances of its vertices
from the Fermat-Torricelli point. Other studies accomplished in the paper 
include investigation of the two- and three-point correlation functions 
in the adjoint representation both below and above critical point and 
calculation of critical indices in the vicinity of the deconfinement phase 
transition. 

Our main findings can be summarized as follows: 
\begin{itemize}
\item 
Similarly to the pure gauge $SU(3)$ LGT and $Z(3)$ spin model, 
the leading contribution to the three-point fundamental correlation 
function in the strong coupling region is described by the $Y_l$ law, as
explained in section 2.5. Exact $Y$ law is restored as soon as the rotational
symmetry is also restored. 
\item 
From the study of the large-$N$ limit of the model we also obtained the
general form for the two- and three-point correlations in fundamental and
adjoint representations both above and below the critical point. 
The analytical results suggest that the fundamental three-point correlation 
behaves as in Eq.~(\ref{YandLambda_LargeN}), {\it i.e.} it is described by 
a combination of $Y$ and $\Lambda$ laws if all angles of the triangle 
are less than $2\pi/3$ or by only the $\Lambda$ law if any of the angles 
is larger than $2\pi/3$. We have not found analytical support in favour 
of $\Delta$ law (one possibility is that this law is suppressed in the
large-$N$ limit). The comparison with the results of numerical simulations shows
that these forms can indeed be used to describe the behaviour of the
corresponding correlations. 
\item 
  The critical behaviour across the deconfinement transition supports the
  universality conjecture that this model is in the universality class of the
  two-dimensional $Z(3)$ spin model. In particular, we have determined the
  critical indices $\nu$ and $\eta$ from finite size scaling (the index $\nu$
  has been also evaluated directly from the two-point correlation). Their
  values agree well with the values in the Potts model. 
\item 
  The fact that the assumption of $Y$ law gives much better collapse of the
  effective string tension to a single curve, when different locations 
  of the three sources are taken with the same value of $Y$, indicates
  that in this effective model the $Y$ law is preferred.
  We have found the agreement of the effective string tensions for the two- and
  three-point correlation functions. Moreover, all string tensions appearing in 
  the two-quark potential and in the three-quark potential with $Y$ or $\Delta$
  law agree up to uncertainties.
  This result is also supported by the analytic study of the model in the
  large-$N$ limit. 
  Since our results are supportive of the $Y$ law even for small triangles, we
  do not observe a smooth crossover from $\Delta$ to $Y$ law as conjectured
  in~\cite{forcrand1, caselle}. For triangles with one of the angles larger
  than $2\pi/3$ the three-point fundamental correlation function follows
  the $\Lambda$ law.
\item 
  In the deconfinement phase the screening masses for fundamental two- and
  three-point correlations also coincide up to numerical errors. 
\item 
  Adjoint correlations share a similar pattern in the deconfinement phase, 
  namely the corresponding screening masses are consistent with each other for
  two- and three-point correlators. 
  Moreover, they seem to coincide in the deconfined phase with the fundamental
  screening masses.
\item 
  In the confined phase the connected part of the adjoint two-point correlation
  equals the square of the fundamental one after subtraction of magnetization.
  The connected part of the adjoint three-point correlation is consistent with
  the $\Delta$ law below the critical point and agrees with the large-$N$
  predictions. 
\end{itemize}

This work can be straightforwardly extended to the case of three dimensions,
which is certainly more relevant from the physical point of view, though
being technically more involved. Another possible extension is to
consider the three-quark system in different color channels, both in the
confined and in the deconfined phase (see, {\it e.g.},
Ref.~\cite{Huebner:2004qf}).

\vspace{0.5cm}

{\bf \large Acknowledgements}

\vspace{0.2cm}

We gratefully acknowledge useful discussions with Leonardo Cosmai.
Numerical simulations have been performed on the ReCaS Data Center of
INFN-Cosenza. V.C. acknowledges financial support from the INFN HPC{\_}HTC
and NPQCD projects. A.P. acknowledges financial support from the
INFN NPQCD project. O.B. also thanks INFN (Fondo FAI) for financial support.


\vspace{0.5cm}

\clearpage

\begin{thebibliography}{99}

%
\bibitem{sommer_84} R.~Sommer, J.~Wosiek, Phys. Lett. {\bf B149} (1984) 497.
%
\bibitem{sommer_86} R.~Sommer, J.~Wosiek, Nucl. Phys. {\bf B267} (1986) 531.
%
\bibitem{bali} G.~Bali, Phys. Rep. {\bf 343} (2001) 1.
%
\bibitem{takahashi} T.T.~Takahashi, H.~Suganuma, Y.~Nemoto, H.~Matsufuru,
Phys. Rev. {\bf D65} (2002) 114509.
%
\bibitem{forcrand3} Ph.~de Forcrand, C.~Alexandrou, A.~Tsapalis,
Phys. Rev. {\bf D65} (2002) 054503.
%
\bibitem{forcrand1} C.~Alexandrou, Ph.~de Forcrand, O.~Jahn, Nucl. Phys. B
Proc. Suppl. {\bf 119} (2003) 667.
%
\bibitem{forcrand2} Ph.~de Forcrand, O.~Jahn, Nucl. Phys. {\bf A755} (2005) 475.
%
\bibitem{polikarpov1} V.~Bornyakov, H.~Ichie, Y.~Mori, D.~Pleiter, M.~Polikarpov,
G.~Schierholz, T.~Streuer, H.~Stuben, T.~Suzuki, Phys. Rev. {\bf D70} (2004) 054506.
%
\bibitem{polikarpov2} V.~Bornyakov, P.~Boyko, M.~Chernodub, M.~Polikarpov,
hep-lat/0508006.
%
\bibitem{caselle} M.~Caselle, G.~Delfino, P.~Grinza, O.~Jahn, N.~Magnoli,
J. Stat. Mech.: Theory and Experiment {\bf 2006.03} (2006) P03008.
%
\bibitem{sato} T.~Sato, V.~Dmitra\v{s}inovi\v{c}, M.~\v{S}uvakov,
Phys. Rev. {\bf D80} (2009) 054501.
%
\bibitem{Brambilla:2009cd}
N.~Brambilla, J.~Ghiglieri and A.~Vairo, Phys.\ Rev.\ {\bf D81} (2010) 054031
and references therein.
%
\bibitem{bakry} A.S.~Bakry, X.~Chen, P.-M.~Zhang,
Phys. Rev. {\bf D91} (2015) 114506.
%
\bibitem{Andreev:2015iaa}
O.~Andreev, Phys.\ Lett.\ {\bf B756} (2016) 6.
%
\bibitem{Andreev:2015riv}
O.~Andreev, Phys.\ Rev.\ {\bf D93} (2016) 105014.
%
\bibitem{koma} Y.~Koma, M.~Koma, Phys. Rev. {\bf D95} (2017) 094513.
%
\bibitem{Leech:2018lqu}
J.~Leech, M.~\v{S}uvakov and V.~Dmitra\v{s}inovi\v{c},
Acta Phys.\ Polon.\ Supp.\  {\bf 11} (2018) 435.
%
\bibitem{sun_effaction} J.~Langelage, S.~Lottini, O.~Philipsen, JHEP {\bf 1102} (2011) 057.
%
\bibitem{svetitsky} B.~Svetitsky, L.~Yaffe, Nucl. Phys. {\bf B210} (1982) 423.
%
\bibitem{teper_08} J.~Liddle, M.~Teper, {\em The deconfining phase transition in
$D=2+1$ $SU(N)$ gauge theories}, arXiv:0803.2128v1 [hep-lat].
%
\bibitem{2dun_bkt} O.~Borisenko, V.~Chelnokov, F.~Cuteri, A.~Papa,
Phys. Rev. {\bf E94} (2016) 012108. 
%
\bibitem{spin_flux1} C.~Gattringer, Nucl. Phys. {\bf B850} (2011) 242.
%
\bibitem{dual_lgt} O.~Borisenko, V.~Chelnokov, S.~Voloshyn,
{\em Duals of $U(N)$ LGT with staggered fermions},
EPJ Web Conf., 175 (2018) 11021.
%
\bibitem{dual_sunspin} O.~Borisenko, V.~Chelnokov, S.~Voloshyn,
{\em Duals of non-abelian lattice spin models}, in preparation.
%
\bibitem{dual_largeN_exact} O.~Borisenko, V.~Chelnokov, S.~Voloshyn,
{\em Exact solution of the Polyakov loop models in the large $N$ limit}, in preparation. 
%
\bibitem{largeN_mean-field} C.~H.~Christensen, Phys. Lett. {\bf B714} (2012) 306.
%
\bibitem{universal_ratios} G.~Delfino, J.L.~Cardy, Nucl. Phys. {\bf B519} (1998) 551.
%
\bibitem{Huebner:2004qf}
K.~H\"ubner, O.~Kaczmarek, F.~Karsch and O.~Vogt,
{\em Free energies of static three quark systems},
Strong and Electroweak Matter 2004 (2005) 371-375,
doi:10.1142/9789812702159\_0057, hep-lat/0408031. 

\end{thebibliography}
\end{document}